\documentclass[a4paper,11pt]{article}
\usepackage{aas_macros}
\usepackage{jcappub}

\usepackage{amsmath}
\usepackage{bm}
\usepackage{xcolor}
\usepackage[utf8]{inputenc}
\usepackage{hyperref}
\usepackage{url}
\usepackage{graphicx}
\usepackage{amssymb}
\usepackage[amssymb]{SIunits}
\usepackage{float}  % to fix the figure positions
\usepackage{multirow}
\usepackage{bbold}

\newcommand{\beq} {\begin{equation}}
\newcommand{\eeq} {\end{equation}}
\newcommand{\bal} {\begin{aligned}}
\newcommand{\eal} {\end{aligned}}
\newcommand{\vk} {\bm{k}}
\newcommand{\vx} {\bm{x}}
\newcommand{\vtheta} {\bm{\theta}}
\newcommand{\vl} {\bm{\ell}}

\title{
Astrophysics \& Cosmology from Line Intensity Mapping vs Galaxy Surveys
}

\author[a,b]{Emmanuel Schaan}
\author[a,b,c]{and Martin White}
\affiliation[a]{Lawrence Berkeley National Laboratory, One Cyclotron Road, Berkeley, CA 94720, USA}
\affiliation[b]{Berkeley Center for Cosmological Physics, UC Berkeley, CA 94720, USA}
\affiliation[c]{Department of Physics, University of California, Berkeley, CA 94720, USA}
\emailAdd{eschaan@lbl.gov}
\emailAdd{mwhite@berkeley.edu}

\abstract{
Line intensity mapping (LIM) proposes to efficiently observe distant faint galaxies and map the matter density field at high redshift.
Building upon the formalism in a companion paper, 
we first highlight the degeneracies between cosmology and astrophysics in LIM. 
We discuss what can be constrained from measurements of the mean intensity and redshift-space power spectra.
With a sufficient spectral resolution, the large-scale redshift-space distortions of the 2-halo term can be measured, helping to break the degeneracy between bias and mean intensity.
With a higher spectral resolution, measuring the small-scale redshift-space distortions disentangles the 1-halo and shot noise terms.
Cross-correlations with external galaxy catalogs or lensing surveys further break degeneracies.
We derive requirements for experiments similar to SPHEREx, HETDEX, CDIM, COMAP and CONCERTO.

We then revisit the question of the optimality of the LIM observables, compared to galaxy detection, for astrophysics and cosmology.
We use a matched filter to compute the luminosity detection threshold for individual sources.
We show that LIM contains information about galaxies too faint to detect, in the high-noise or high-confusion regimes.
We quantify the sparsity and clustering bias of the detected sources and compare them to LIM, showing in which cases LIM is a better tracer of the matter density.
We extend previous work by answering these questions as a function of Fourier scale, including for the first time the effect of cosmic variance, pixel-to-pixel correlations, luminosity-dependent clustering bias and redshift-space distortions.
}

\begin{document}
\maketitle
\flushbottom

\section{Introduction}

The high redshift Universe is one of the frontiers for astrophysics and cosmology.
Tracking the evolution of galaxy populations from early times to today contains a wealth of astrophysical information about galaxy formation and evolution \cite{Breysse16, Yue15, Silva17}, and the processes which reionized the Universe \cite{Lidz09, Gong12, Lidz11}.
It spans a large fraction of the observable Universe, enabling high precision measurements of the evolution of structure in the Universe, testing the properties of $\Lambda$CDM \cite{Dinda18, Carucci17} and the masses of the neutrinos.
Because the matter distribution is more linear at higher redshift, its simpler and more accurate modeling should also allow us to better reconstruct the initial conditions of the Universe, and test the presence of primordial non-Gaussianities \cite{Moradinezhad19, Fonseca18, Munoz15}.

% define IM
Intensity mapping (IM) is emerging as a promising way to efficiently probe these last swaths of the Universe, and thus learn about astrophysics and cosmology  \cite{Kovetz17,Kovetz19}.
It saves on observing time and angular resolution by not attempting to detect individual galaxies, and instead inferring their properties from their combined atomic and molecular line emissions.
The IM field is still in its infancy, but growing fast with a few, low-significance detections to date.  
For instance, the COPSS experiment reported a detection of CO IM \cite{Keating16}, while Planck detected [C{\sc ii}] IM from Planck \cite{Pullen18}. 
If successful, IM may revolutionize our understanding of galaxy evolution and the large-scale structure in the Universe.

% But the sensitivity to both is also a source of degeneracies
A major challenge of the IM approach is to disentangle the signal of interest from dominant foregrounds.
Even in the absence of foregrounds, the simultaneous sensitivity of IM to astrophysics and cosmology also constitutes a hurdle, causing degeneracies between the two.

% We use our formalism from companion paper to explore this.
In this paper, we apply the formalism derived in \cite{paper1} to clarify the degeneracies between astrophysics and cosmology from IM \cite{Mao08, Bernal19}, focusing on a number of representative optical and infrared lines and experiments.
Using quasi-linear scales \cite{Castorina19} or higher order information (e.g.\ from the bispectrum) may help \cite{Beane18}.
%
% experiments we consider
As an exemplar of optical and infrared IM, we consider the SPHEREX survey \cite{Dore14, Dore16, Dore18}, a satellite capable of mapping structure in the H$\alpha$, H$\beta$, [O{\sc ii}], [O{\sc iii}] and Lyman-$\alpha$ lines.
We also consider H$\alpha$ (656.28~nm) and [O{\sc iii}] (495.9~nm and 500.7~nm) IM with HETDEX and Lyman-$\alpha$ (121.6~nm) IM with CDIM, as examples of intensity mapping in different regimes.
In the far infrared (IR) and sub-millimeter, a number of experiments are targeting the [C{\sc ii}] (158~$\mu$m) line \cite{Lagache18,Parshley18} or the ladder of rotational CO lines (2.6~mm for CO 1-0) \cite{Keating16,Cleary16,Bower16,Crites17}.
As examples of far IR IM, we focus on CO from COMAP \cite{Cleary16, Li16} and [C{\sc ii}] from CONCERTO \cite{Dumitru19, Serra16, Concerto20}.
Many more lines and IM experiments exist. 
We restricted ourselves to only a few ones, selected to showcase different limiting regimes within LIM.

% LIM vs galaxy detection
In this paper, we also revisit the comparison between the IM approach and the more traditional galaxy detection, for astrophysics and cosmology.
For astrophysics, the question of the sensitivity of LIM to galaxies too faint to detect was tackled in \cite{Silva17}, focusing on the LIM mean intensity. 
We extend this work by considering the various components of the LIM power spectrum too.
For cosmology, the comparison of LIM observables and galaxy detection as tracers of the matter density field was tackled in a formal and intuitive way in \cite{Cheng19},
focusing on a single pixel from an intensity map and under some simplifying approximations.
The formalism we build in the companion paper~\cite{paper1} is perfectly suited to extend this study.
We thus compare LIM and galaxy detection as tracers of the matter density field,  as a function of Fourier scale, including the effects of cosmic variance, pixel-to-pixel correlations, luminosity-dependent clustering bias and redshift-space distortions (RSD).

% Outline
We finish with a brief outline of the paper. 
We apply our formalism to answer some of the key questions about astrophysics and cosmology from LIM in \S\ref{sec:whatcanwelearnfromlim}.
This section identifies the degeneracies between astrophysical and cosmological parameters in LIM, and presents simple requirements and forecasts for LIM surveys.
We discuss the effect of the degeneracies which arise from the limited angular resolution, and the importance of spectral resolution in breaking these degeneracies by measuring the RSD.
We quantify the sparsity of halos and galaxies in LIM, a key quantity for line deconfusion methods.
In \S\ref{sec:lim_vs_galaxies}, we derive a 3D matched filter to determine the detection threshold for individual galaxies in intensity maps. We then compare the catalog of detected galaxies to the IM itself, both as probes of faint galaxies and as tracers of the matter density field.
We summarize our results in \S\ref{sec:conclusions}.

% cosmo parameters
As in the companion paper~\cite{paper1}, we assume throughout a flat $\Lambda$CDM cosmology from ref.~\cite{PlanckParams16}, 
with $\Omega_\text{CDM}=0.267$, $\Omega_b = 0.0493$, $H_0 = 67.12\,$km/s/Mpc, $A_S = 2.3\times 10^{-9}$, $n_s=0.9624$, $N_\text{eff}=3.046$ with massless neutrinos.
All distances, volumes and wavevectors are quoted in comoving units, and typically in $h^{-1}$Mpc or $h\,{\rm Mpc}^{-1}$ units unless otherwise specified.
Halo masses refer to virial masses in $h^{-1}M_\odot$, defined as the mass enclosed within the virial radius, where the density is a factor $\Delta_\text{vir crit}(z) = 18\pi^2 + 82\left[\Omega_m(z)-1\right] - 39\left[\Omega_m(z)-1\right]^2$ higher than the critical density $\rho_\text{crit}(z)$ \cite{Bryan98}.

%%%%%%%%%%%%%%%%%%%%%%%%%%%%%%%%%%%%%%%%%%%%%%%%%%%%%%%%%%
%%%%%%%%%%%%%%%%%%%%%%%%%%%%%%%%%%%%%%%%%%%%%%%%%%%%%%%%%%
\section{Multi-line halo model: refresher}
\label{sec:reminder_formalism}

In this section, we briefly summarize the halo model results derived in ref.~\cite{paper1}. More discussion and interpretation of these results can be found in \S2 of that paper.
We start from the multi-line conditional luminosity function (CLF; \cite{Yang03}) $\phi\left( L_1,..., L_n | m \right)$ such that
$\phi\left( L_1, ..., L_n | m \right) dL_1 ... dL_n$ 
is the mean number of galaxies in one halo of mass $m$
with luminosity $L_1, ..., L_n$ in lines $1, ..., n$.
From the CLF, one can recover the usual (unconditional) luminosity function $\Phi$,
such that $\Phi(L_1, ..., L_n) dL_1 ... dL_n$ is the mean number density of galaxies with luminosities $L_1, ..., L_n$ in lines $1, ..., n$, regardless of host halo mass.
The relation between $\Phi$ and $\phi$ is:
\beq
\Phi(L_1, ..., L_n)
=
\int dm\ n(m)
\phi(L_1, ..., L_n | m),
\eeq

The mean intensity, $\bar{I}_j$, in line $j$ is determined by the first moment of the univariate CLF, $\phi\left( L_j | m \right)$:
\beq
\bar{I}_j
=
\frac{1}{4\pi \nu_j^0}
\frac{c}{H(z)}
\int dL_j\ \Phi(L_j) L_j
\label{eq:mean_intensity_lf}
\eeq
where $\nu_j^0$ is the rest-frame frequency of line $j$.

The intensity power spectrum, a function of the comoving wave vector modulus $k$, cosine of the angle to the line of sight $\mu$ and redshift $z$, is made of three contributions:
\begin{equation}
    P_{i,j}(k,\mu,z) = P_{i,j}^\text{2-halo}(k, \mu, z) +
    P_{i,j}^\text{1-halo}(k, \mu, z) +
    P_{i,j}^\text{shot}(z).
\label{eqn:Ptotal}
\end{equation}

The 2-halo term is given by
\beq
P_{i,j}^\text{2-halo}(k, \mu, z)
=
\bar{I}_i \bar{I}_j
\left( b_i + F \mu^2 \right)
\left( b_j + F \mu^2 \right)\; 
P_\text{lin}.
\label{eqn:P2halo}
\eeq
Here, the intensity bias is
\beq
b_j(k, \mu, z)
\equiv \mathcal{L}_j^{-1}
\int
dm\ n(m)\;
L_j(m)
\;
b(m) 
\;
u(k,m)
e^{-k^2 \mu^2\sigma_d^2(m) / 2}
.
\label{eq:effective_bias}
\eeq
In this expression, $L_j$ is the mean line $j$ luminosity for a halo of mass $m$:
\beq
L_j(m)
\equiv
\int dL_j\ \phi(L_j|m) L_j,
\eeq
and $\sigma_d = \sigma_{v\ \text{1D}}/aH$ is the dispersion of the spurious displacement in redshift space due to the random line-of-sight motion within halos $\sigma_{v\ \text{1D}}$, $n(m)$ is the halo mass function, $u(k,m)$ is the (normalized) halo profile and $\mathcal{L}_j \equiv \int dm \ n(m) L_j(m)$ is the luminosity density in line $j$.
The line-of-sight velocity dispersion is taken to be that of a singular isothermal sphere of mass $m$ and radius $r_\text{vir}$, i.e.\ $\sigma_{v\ \text{1D}}^2 = \mathcal{G} m / 2 r_\text{vir}$.
The effective growth rate of structure is computed as 
\beq
F(k, \mu, z)
\equiv
f\;
\int dm \; n(m)\;
\left( \frac{m}{\bar{\rho}} \right)
u(k,m) 
e^{-k^2 \mu^2 \sigma_d^2(m) /2}
\label{eq:effective_growth_rate}
\eeq
such that $F\to f$ as $k\to 0$.
The 1-halo term can be written
\beq
P_{i,j}^\text{1-halo}(k, \mu, z)
=
\bar{I}_i \bar{I}_j
\frac{U_{i,j}^2(k, \mu, z)}{\bar{n}^\text{h eff}_{i,j}}
\label{eq:p1h}
\eeq
where the effective mean number density of halos, $\bar{n}^\text{h eff}_{i,j}$, properly counts the halos by taking into account their luminosities in lines $i$ and $j$:
\beq
\bar{n}^\text{h eff}_{i,j}
=
\frac{
\left( \int dm \; n(m) L_i(m)  \right)
\left( \int dm \; n(m) L_j(m)  \right)
}
{\int dm \; n(m) L_i(m) L_j(m)}.
\label{eqn:nheff}
\eeq
and the effective squared halo profile $U_{i,j}^2$ is defined by
\beq
U_{i,j}^2(k, \mu, z)
\equiv
\frac{
\int dm \; n(m)
\left| u(k,m) \right|^2
e^{-k^2\mu^2\sigma^2}
L_i(m) L_j(m)
}
{
\int dm \; n(m)
L_i(m) L_j(m)
}.
\eeq

Finally, the shot noise term can be written
\beq
P_{i,j}^\text{shot}(z)
=
\frac{\bar{I}_i \bar{I}_j}{\bar{n}^\text{gal eff}_{i,j}}
\eeq
in terms of an effective number density of galaxies that properly takes into account the distribution of galaxy luminosities in lines $i$ and $j$:
\beq
\frac{1}{\bar{n}^\text{gal eff}_{i,j}}
=
\frac{
\int dL_i dL_j\ \Phi(L_i, L_j) L_i L_j
}
{
\left( \int dL_i \ \Phi(L_i) L_i  \right)
\left( \int dL_j \ \Phi(L_j) L_j  \right)
}
.
\label{eqn:ngaleff}
\eeq

%%%%%%%%%%%%%%%%%%%%%%%%%%%%%%%%%%%%%%%%%%%%%%%%%%%%%%%%%%
%%%%%%%%%%%%%%%%%%%%%%%%%%%%%%%%%%%%%%%%%%%%%%%%%%%%%%%%%%
\section{What can we learn from LIM?}
\label{sec:whatcanwelearnfromlim}

The LIM observables are a combination of astrophysical and cosmological parameters.
What can we learn from LIM about cosmology and astrophysics?
In this section, we derive simple measurement requirements (calibration, spatial and spectral resolution, survey area and depth) and identify the degeneracies that can be broken and those that cannot (see also ref.~\cite{Bernal19}).

We perform forecasts for experiments similar to SPHEREx \cite{Dore14, Dore16, Dore18}, CDIM \cite{Cooray16}, HETDEX \cite{Hill08, Hill16}, COMAP \cite{Cleary16, Li16} and CONCERTO \cite{Dumitru19, Serra16} in Fig.~\ref{fig:tradeoff_a_halpha_spherex}-\ref{fig:tradeoff_a_cii_concerto}.
The specifications assumed here are approximate, and we list them in Table~\ref{tab:exp_specs}.
For each experiment, the detector noise calculation is described in Appendix.~\ref{app:experimental_detector_noise}.
In practice, we compute the LIM observables at the redshifts where the line luminosity functions from \cite{Sobral13, Colbert13, Mehta15, Cochrane17, Popping16, Cassata11} have been observed. 
In some cases, we thus extrapolate the noise power spectra outside the fiducial redshift range quoted in Table~\ref{tab:exp_specs}.
\begin{table}[H]
\centering
\begin{tabular}{ |p{2cm} p{3cm} p{1.5cm} p{1.5cm} p{3cm} p{1cm}|}
\hline
Line & Experiment & PSF  & $z$ & Observed           & $\mathcal{R}$\\
     &            &      &     & $\nu$ or $\lambda$ &              \\
\hline
\hline
H$\alpha$, [O{\sc iii}] & SPHEREx-like & $6''$ & $0-6$ & $0.75-4.8\mu$m & 40 \\
\hline
Ly-$\alpha$ & SPHEREx-like & $6''$ & 5-9 & $0.75-1.2\mu$m & 150 \\
\hline
H$\alpha$ & CDIM-like & $1''$ & $5-8$ & $0.7-6.7\mu$m & 300 \\
\hline
Ly-$\alpha$ & HETDEX-like & $3''$ & $1.9-3.5$ & $0.35-0.55\mu$m & 800 \\
\hline
CO & COMAP-like & $3'$ & $2.4-3.4$, $5.8-7.8$ & $26-34$ GHz & 800 \\
\hline
[C{\sc ii}] & CONCERTO-like & $0.24'$ & $4.5-8.5$ & $200-360$ GHz & $300$ \\
\hline
\end{tabular}
\caption{
Summary of lines and experimental configurations explored in this paper
to forecast RSD measurements, and compare the intensity mapping approach to the catalog of individually detected galaxies.
We follow \cite{Dore14, Dore16, Dore18} for the SPHEREx-like specifications,
\cite{Cooray16} for CDIM,
\cite{Hill08, Hill16} for HETDEX,
\cite{Cleary16, Li16} for COMAP
and \cite{Dumitru19, Serra16} for CONCERTO.
In this paper, we do not exactly follow the specified redshift ranges, and instead compute power spectra at the redshifts where our luminosity functions are available.
}
\label{tab:exp_specs}
\end{table}

%%%%%%%%%%%%%%%%%%%%%%%%%%%%%%%%%%%%%%%%%%%%%%%%%%%%%%%%%%
\subsection{Astrophysics \& Cosmology: degeneracies}
\label{subsec:astro_cosmo_deg}

%%%%%%%%%%%%%%%%%%%%%%%%%%%%%%%%%%%%%%%%%%%%%%%%%%%%%%%%%%
\subsubsection{RSD disentangles power spectrum components}
\label{sec:degeneracies_rsd}

As we have shown, the redshift-space LIM power spectrum contains a wealth of information on astrophysics and cosmology, encoded in the parameters ($\bar{I}_i$, $b_i$, $\bar{n}_{i,j}^{\rm h, eff}$, $\bar{n}_{i,j}^{\rm gal, eff}$, $U_{i,j}^2$ $P_m$, $f$).
For an ideal experiment, with a perfect calibration, angular and spectral resolutions, and sensitivity, the various observables we have considered can all be related to these parameters. 
Schematically one would measure:
\begin{equation}
\left\{
\bal
&\text{Mean intensity} \quad && \bar{I}_i\\
&\text{2-halo power spectrum} && P^\text{2h}_{i,j} = \bar{I}_i \bar{I}_j (b_i + f\mu^2) (b_j + f\mu^2) P_m\\
&\text{1-halo power spectrum} && P^\text{1h}_{i,j} = \bar{I}_i \bar{I}_j U^2_{i,j} / \bar{n}_{i,j}^\text{h eff}\\
&\text{Shot noise power spectrum} && P^\text{shot}_{i,j} = \bar{I}_i \bar{I}_j / \bar{n}_{i,j}^\text{gal eff}\\
\eal
\right.
\label{eq:param_degeneracies}
\end{equation}
where we have ignored the $k$ and $\mu$ dependence of $F$ and $b$ in the two-halo term, and assuming we have data at large enough scales that $F\to f$ is a good approximation.

The mean intensity contains astrophysical information about the mean number density of sources and their mean luminosity.
It also normalizes the power spectrum.
However, measuring the mean intensity requires an absolute calibration of the instrument, which may not be available.
In that case, the dependence of the power spectrum on $\bar{I}_i \bar{I}_j$ in Eq.~\eqref{eq:param_degeneracies} above cannot be taken out unless one assumes external constraints on $P_m$ or cross-correlates with an external sample.

The 2-halo, 1-halo and shot-noise terms are not measured separately. Instead, only their sum is observable.
The 2-halo term is generically dominant on large scales though, making it easy to distinguish.
Its $\mu$-dependence ($\propto (f\mu)^{0,1,2}$) can be expressed as \cite{Kaiser87,H98}
\begin{equation}
  P^\text{2h}_{i,j} = \bar{I}_i \bar{I}_j \left[b_i b_j + f(b_i+b_j)\mu^2 + f^2\mu^4\right] P_m,
\end{equation}
just like in galaxy surveys.
Assuming a known growth rate of structure, $f$,
the ratios of the monopole to the quadrupole or to the hexadecapole yield the values of the biases.
Furthermore, on these large scales, we have shown in \cite{paper1} that there is no decorrelation between intensity maps from different lines.
Taking ratios of large-scale power spectra, e.g.\ $P^\text{2h}_{i,j} / P^\text{2h}_{i,i}$, thus provides all the intensity ratios $\bar{I}_j / \bar{I}_i$.
This way, even in the absence of absolute calibration of the mean intensities, the ratios of mean intensities can be recovered from the redshift-space 2-halo power spectrum.  If further information on the normalization of $P_m$ is assumed (e.g.\ from external measurements), then all $\bar{I}_j$ can again be obtained.
However, if the spectral resolution is insufficient to measure the $\mu$-dependence of the 2-halo term, 
then biases and ratios of mean intensities can no longer be estimated separately (see also Appendix~B in \cite{paper1}).
Assuming a known cosmology, only the ratios $(\bar{I}_j b_j) / (\bar{I}_i b_i)$ can be recovered.

In principle, the 1-halo and shot noise terms can be distinguished in two ways.
First, on extremely small scales across the line of sight ($k_{\perp}\sim 10-100\,h\,{\rm Mpc}^{-1}$), the turn over in the halo profile $U^2_{i,j}$ means that only the shot noise remains dominant.
Second, on very small scales along the line of sight ($k_\parallel \sim 1-10\,h\,{\rm Mpc}^{-1}$), the FOG effect suppresses the 2-halo and 1-halo terms, such that again only the shot noise remains.
In practice, these scales are often inaccessible due to the instrumental angular and spectral resolutions. 
Furthermore, we do not expect our halo model to describe these very small scales accurately.
Distinguishing the 1-halo and shot noise terms would require modeling a number of additional effects, which we neglected here, that would be important on these very small scales.
Distinguishing the 1-halo and shot noise terms is thus difficult in practice.
This is unfortunate, as the 1-halo and shot noise contain different astrophysical information:
\beq
\left\{
\bal
&P^{1h} \propto \int dm\ n(m) \left( \int dL\ \phi(L|m) L \right)^2 \\
&P^\text{shot} \propto \int dm\ n(m) \int dL\ \phi(L|m) L^2 \\
\eal
\right.
.
\eeq
The 1-halo term is a measurement of the mean galaxy luminosity (first moment of the CLF), squared and averaged over host halo mass.
The shot noise is instead a measurement of the mean squared galaxy luminosity (second moment of the CLF), averaged over host halo mass.
If the 1-halo and shot noise terms are completely degenerate, this astrophysical information on the galaxy population is lost.

In summary, measuring redshift-space distortions in the power spectrum is crucial in order to extract astrophysical information from LIM.
On large scales, detecting the supercluster infall effect (or Kaiser effect \cite{Kaiser87}) enables one to separately measure the bias and mean intensity ratios.  However this necessitates a sufficiently large survey volume and sufficient spectral resolution.
On small scales, resolving the FOGs would allow us to measure separately the 1-halo and shot noise terms, which would be degenerate otherwise.
Doing so requires even higher spectral resolution than resolving supercluster infall, making it challenging in practice.
In the next subsection, we explore more quantitatively the survey requirements on volume, spectral and angular resolution, to help in guiding experimental design.

%%%%%%%%%%%%%%%%%%%%%%%%%%%%%%%%%%%%%%%%%%%%%%%%%%%%%%%%%%
\subsubsection{Cross-correlations disentangle Astrophysics \& Cosmology}
\label{sec:degeneracy_breaking_external_tracer}

We have seen that LIM suffers from a degeneracy between the amplitude of the matter power spectrum and the mean intensity.
This key degeneracy can be broken by cross-correlating the intensity map with an external tracer e.g., photometric or spectroscopic galaxy catalogs or CMB lensing.

A photometric galaxy sample with known bias $b_g$ and CMB lensing ($b_g=1$) are analogous.
On large scales, the auto and cross-spectra across the LOS are simply
\beq
\left\{
\bal
&P_{\text{LIM}, \text{LIM}} = \left( \bar{I}_\text{LIM} b_\text{LIM} \right)^2 P_\text{lin}\\
&P_{\text{LIM}, g} = \bar{I}_\text{LIM} b_\text{LIM}\ b_g P_\text{lin}\\
\eal
\right.
\eeq
Since $b_g$ is known, the ratio of auto to cross gives $\bar{I}_\text{LIM} b_\text{LIM}$.
This extracts astrophysical information from LIM, breaking the degeneracy with the amplitude of the linear power spectrum.
Cross-correlations with upcoming photometric surveys like the Vera Rubin Observatory Legacy Survey of Space and Time\footnote{\url{https://www.lsst.org/}} \cite{LSST12}, and CMB lensing surveys like Simons Observatory\footnote{\url{https://simonsobservatory.org/}} \cite{Ade19} and CMB-S4\footnote{\url{https://cmb-s4.org/}} \cite{Abazajian16} will therefore be extremely valuable.

Importantly, foregrounds in LIM can limit our ability to correlate it with these 2D fields.
Indeed, continuum foregrounds can contaminate the low $k_\parallel$ modes of the intensity map, making them unusable. Unfortunately, these low $k_\parallel$ modes are precisely the ones needed to cross-correlate with a 2D projected field.
This issue is discussed in \cite{Modi19}, and can be circumvented by reconstructing the missing low $k_\parallel$ modes \cite{Doux16, Zhu16, Li19}.

If the galaxy catalog with known bias $b_g$ is spectroscopic, the $\mu^2$ term of the power spectrum can be measured:
\beq
\left\{
\bal
&P_{\text{LIM}, \text{LIM}} = 2 \bar{I}_\text{LIM}^2 b_\text{LIM} f \mu^2 P_\text{lin}\\
&P_{\text{LIM}, g} = \bar{I}_\text{LIM} \left( b_\text{LIM} + b_g \right) f \mu^2 P_\text{lin}\\
\eal
\right.
\eeq
Given that $\bar{I}_\text{LIM} b_\text{LIM}$ is now known, the ratio of auto to cross allows to solve for $\bar{I}_\text{LIM}$ and $b_\text{LIM}$ separately.
Hence high-redshift spectroscopic surveys e.g., MegaMapper \cite{Schlegel19}, will allow to extract even more astrophysical information.

Following the same reasoning, the $\mu=0$ and $\mu^2$ terms of the auto and cross-spectra allow to extract $P_\text{lin}$ and $f$.
However, the uncertainty on $\sigma_8$ and $f$ is limited by that of $b_g$, such that LIM may not constrain them beyond the external survey.
However, once the amplitude of the LIM power spectrum is fixed this way, LIM can be used to extend the scales probed, or to add modes along the LOS, compared to the photometric or CMB lensing survey.

%%%%%%%%%%%%%%%%%%%%%%%%%%%%%%%%%%%%%%%%%%%%%%%%%%%%%%%%%%
\subsubsection{Summary of degeneracies}
\label{sec:summary_degeneracies}

The astrophysical information in LIM is encoded in $\bar{I}, b, \bar{n}^\text{h eff}$ and $\bar{n}^\text{gal eff}$. 
The astrophysical degeneracies can be summarized as follows:
\begin{itemize}
\item $\bar{n}^\text{gal eff}$ can be measured from the shot noise amplitude, if very small angular scales are probed, or on intermediate scales along the LOS where the FOGs suppress the confusing 1-halo term.
\item The ratio of the $\mu^2$ and $\mu^0$ terms in the 2-halo power spectrum yields $f/b$. Assuming the $\Lambda$CDM value for $f$ thus provides the bias $b$.
\item $\bar{I}$ can be measured directly as the mean intensity, or inferred from the 2-halo power spectrum.
For angular modes, $P^{2h}\sim (\bar{I} b)^2 P_\text{lin}$.
If $b$ is known as above, and $P_\text{lin}$ from $\Lambda$CDM, we can infer $\bar{I}$.
Otherwise, without assuming $\Lambda$CDM, ratios of $P_{i,j}^{2h}(\mu=0)$ reconstruct the ratios $\bar{I}_j b_j/\bar{I}_i b_i$.
\item Cross-correlations with an external 2D (3D) tracer with known bias $b_g$ allow to extract $\bar{I}b$ ($\bar{I}$ and $b$ separately).

\end{itemize}

For the purpose of cosmology, the quantities of interest are instead $P_\text{lin}$ (its amplitude $\sigma_8^2$ and scale-dependence) and $f$.
The cosmological degeneracies are summarized as follows:
\begin{itemize}
\item From the 2-halo angular/radial modes, only the combinations $\bar{I}^2b^2P_\text{lin}$
and
$\bar{I}^2b f P_\text{lin}$
are measured.
The amplitude of the power spectrum and the growth rate of structure are degenerate with cosmology.
\item Cross-correlations with an external tracer with known bias $b_g$ allow to extract $\bar{I}b$ (2D tracer) or $\bar{I}$ and $b$ separately (3D tracer).
This lets us extract $P_\text{lin}$ from LIM, but the precision on its amplitude is determined by the precision on $b_g$.
\item We have not treated the question of primordial non-Gaussianity here, addressed in the literature \cite{Moradinezhad19, Fonseca18, Munoz15}.
\item Combining multiple tomographic bins can further help separating astrophysics and cosmology from LIM, as shown in \cite{Heneka18}.
\end{itemize}

%%%%%%%%%%%%%%%%%%%%%%%%%%%%%%%%%%%%%%%%%%%%%%%%%%%%%%%%%%
\subsection{Survey requirements}
\label{subsec:astro_cosmo_reqs}

In this section, we derive requirements for an experiment to be able to detect redshift-space distortions.
We focus first on the large-scale supercluster infall effect, by forecasting the precision with which the power spectrum quadrupole can be measured.
We then discuss the possibility of detecting the finger-of-God effect on small scales.

%%%%%%%%%%%%%%%%%%%%%%%%%%%%%%%%%%%%%%%%%%%%%%%%%%%%%%%%%%
\subsubsection{Spectral resolution: resolving supercluster infall but not the FOGs}
\label{sec:rsd_forecasts}

\begin{figure}[h!]
\centering
\includegraphics[width=0.32\textwidth]{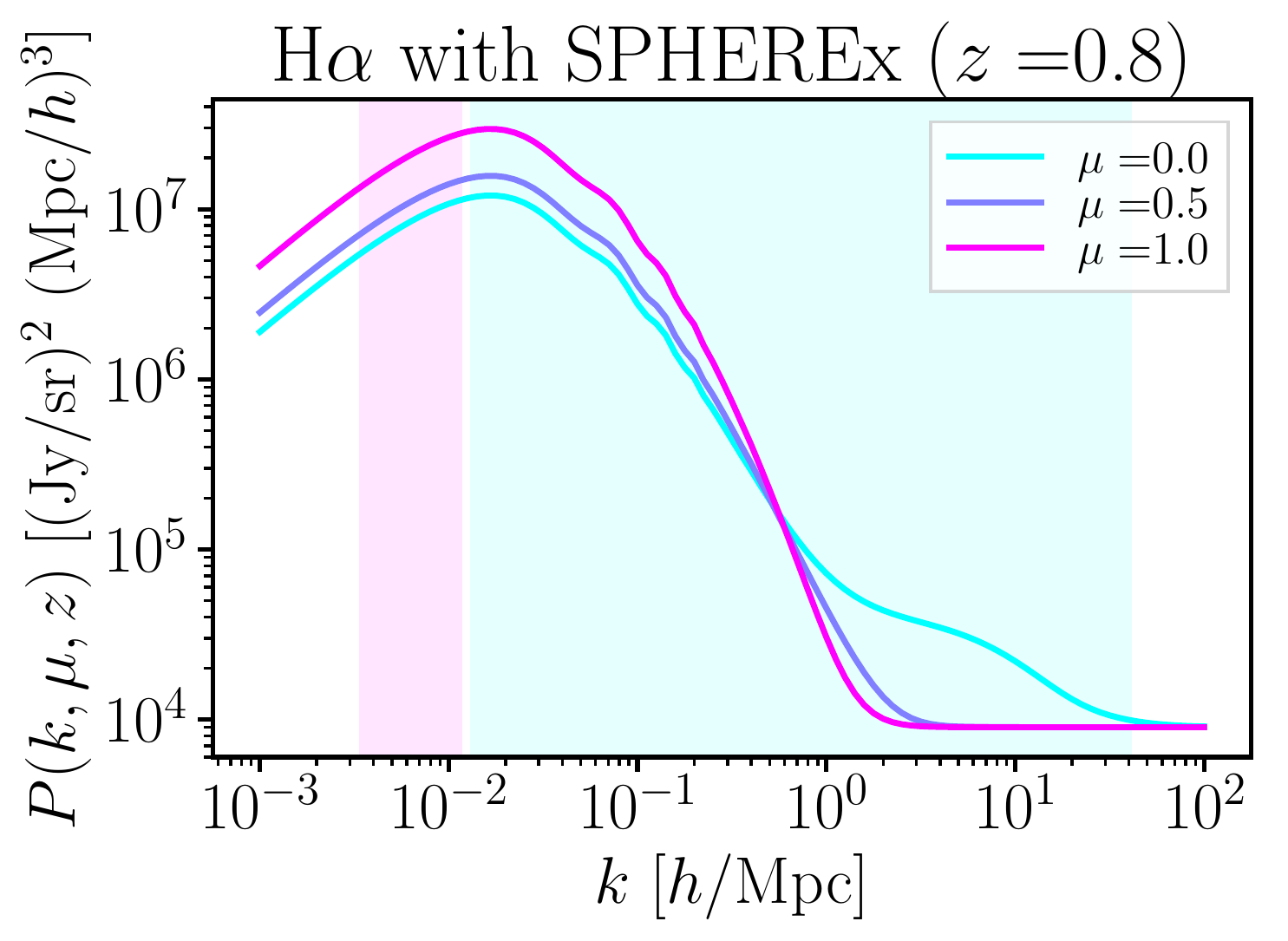}
\includegraphics[width=0.32\textwidth]{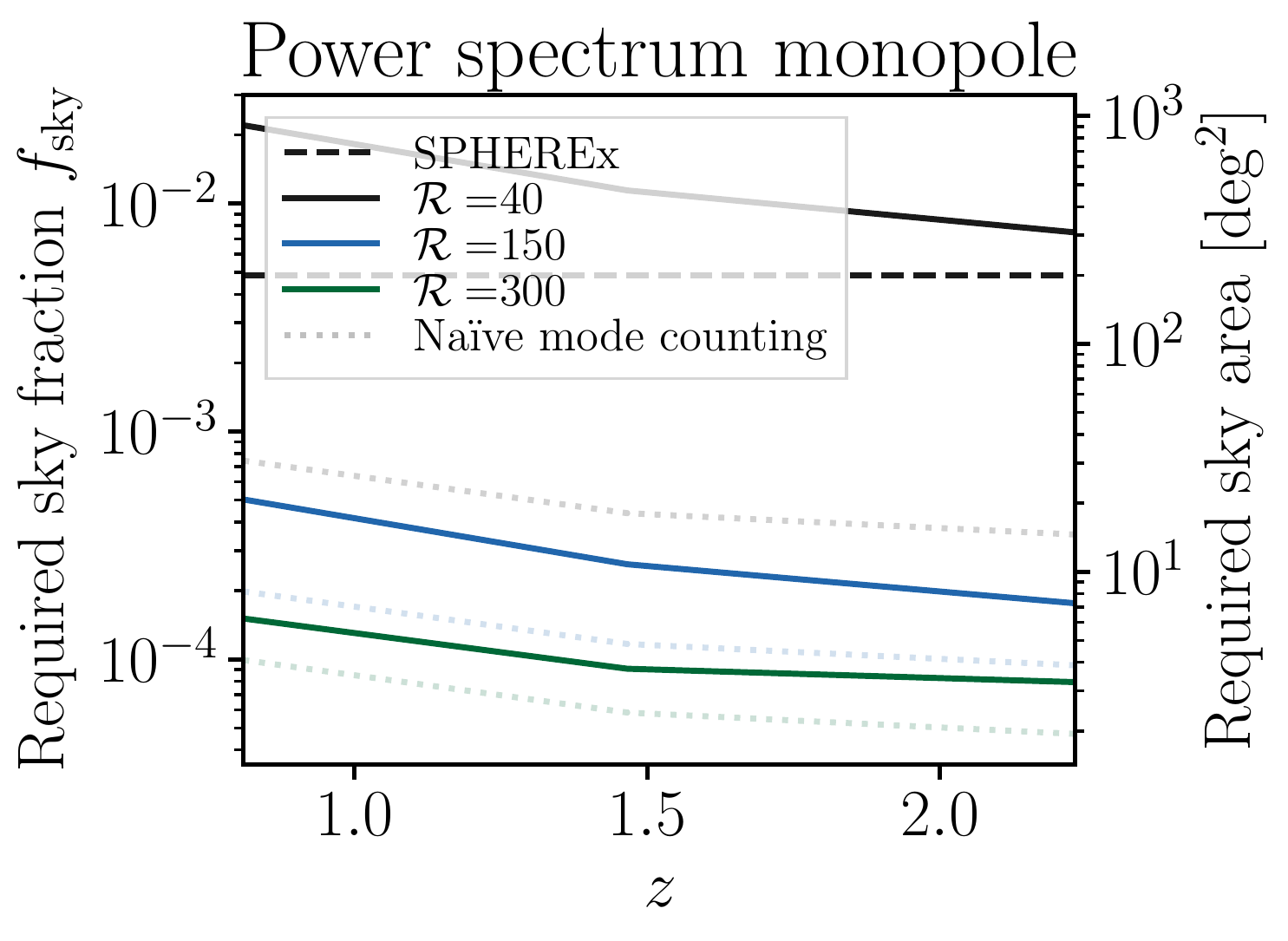}
\includegraphics[width=0.32\textwidth]{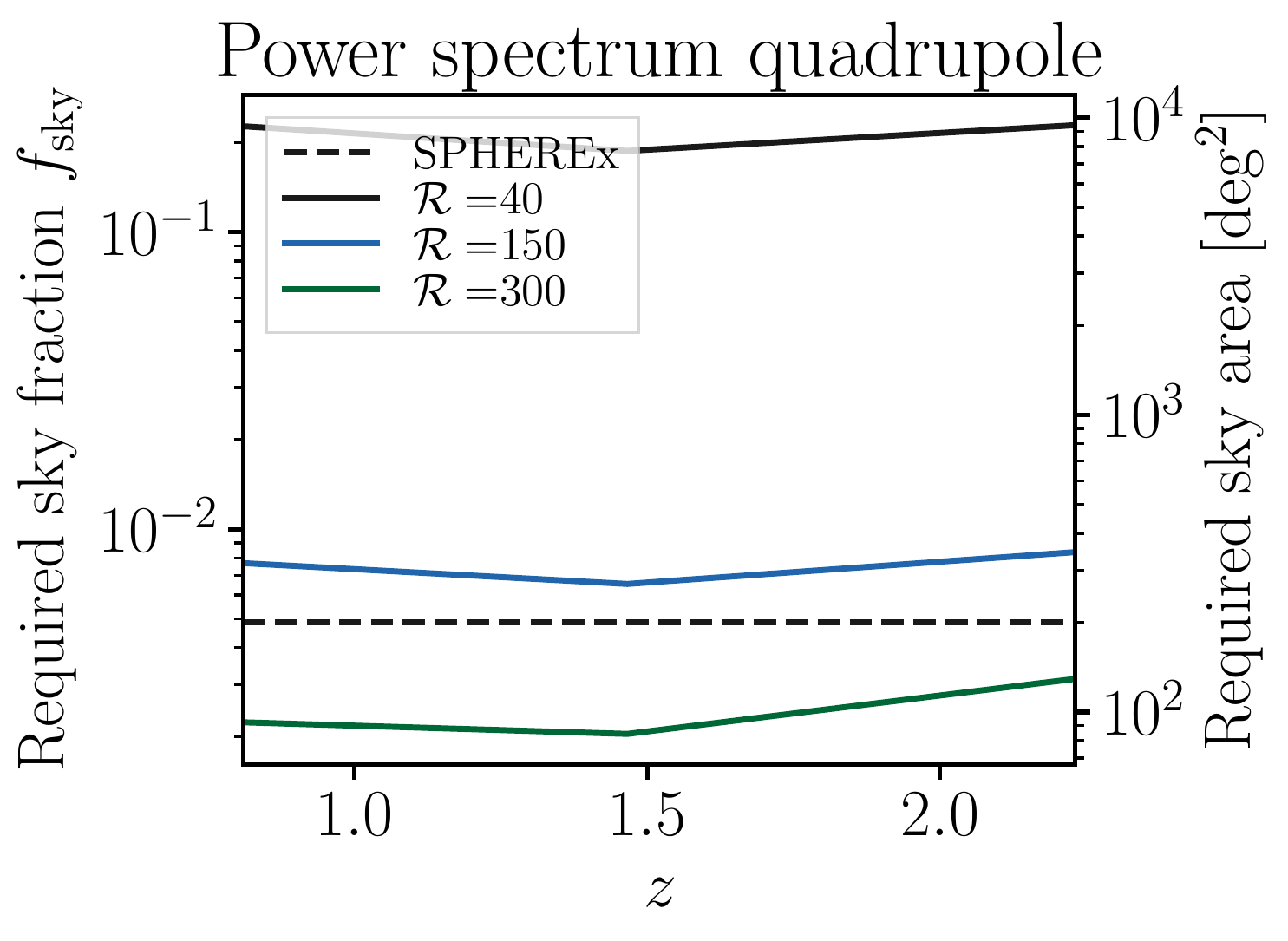}
\caption{
SPHEREx-like deep field forecast.
The left panel shows the scales probed along (pink) and across (cyan) the LOS, assuming a survey depth of $\Delta z=1$ and a PSF FWHM of $6''$.
The middle and right panels show the marginalized uncertainties on the H$\alpha$ power spectrum monopole and quadrupole, for three different spectral resolutions, keeping only modes with $|\vk|\leq0.1 h/$Mpc where the 2-halo term dominates.
The unmarginalized uncertainties (not shown) roughly agree with the simple mode counting (Eq.~\eqref{eq:tradeoff_nmodes}, dotted lines), and are up to 50 times smaller than the marginalized ones (shown).
The large degradation when marginalizing is a consequence of the very anisotropic coverage of the Fourier plane.
The lack overlap in the scales measured along and across the LOS also implies that the quadrupole is highly degenerate with the scale-dependence of the power spectrum, an effect not included here.
}
\label{fig:tradeoff_a_halpha_spherex}
\end{figure}

Because LIM probes low halo masses, the luminosity weighted galaxy bias of the sources is small (see Fig.~2 in \cite{paper1}),
% \ref{fig:b_nh})
comparable to the growth rate of structure out to $z\sim 2$ .
At these redshifts, supercluster infall \cite{Kaiser87} therefore enhances the power spectrum along the line of sight by a large factor, almost two, as shown in Fig.~\ref{fig:tradeoff_a_halpha_spherex} (left panel).
This makes LIM particularly well-suited to measure large scale redshift-space distortions.  (By contrast, luminous galaxies at high redshift tend to have large bias \cite{Wilson19} and hence a relatively isotropic clustering that makes it difficult to observe redshift-space distortions.)
As discussed above the anisotropic clustering allows us to measure the galaxy bias, assuming a known growth rate of structure.

Furthermore, Fig.~\ref{fig:tradeoff_a_halpha_spherex} shows that the finger-of-god effect suppresses the 2-halo and 1-halo terms along the line of sight on scales $k\sim 1 h\,{\rm Mpc}^{-1}$, leaving only the shot noise term.
If these very small scales can be resolved the RSD thus allows us to measure the 1-halo and shot noise terms separately on scales where they would otherwise be degenerate.

However, the spectral resolving power of the instrument, $\mathcal{R} \equiv \lambda / \sigma_\lambda = (1+z) / \sigma_z$, limits the wavevectors accessible along the line of sight.
Indeed, the spectrograph resolution effectively multiplies the Fourier space density field with the Gaussian smoothing kernel $\exp[-k^2\mu^2\sigma_\chi^2 / 2] = \exp[-k^2\mu^2 / (2 k_{\parallel\ \text{max}}^2)]$,
with $k_{\parallel\ \text{max}} = 1/\sigma_\chi^2 = a H \mathcal{R} /c$.
As shown in the right panel of Fig.~\ref{fig:tradeoff_a_halpha_spherex}, 
the SPHEREx resolving power $\mathcal{R}=40-150$ severely limits its ability to detect the FOG effect, and thus disentangle the 1-halo term and the shot noise.  Further, the range of $k$ probed along the line of sight is different from that probed across the line of sight, requiring modeling assumptions to extract any anisotropy.
Incidentally, this spectral resolution also means that the line-of-sight baryonic acoustic oscillations are barely resolved.

\begin{figure}[h!]
\centering
\includegraphics[width=0.32\textwidth]{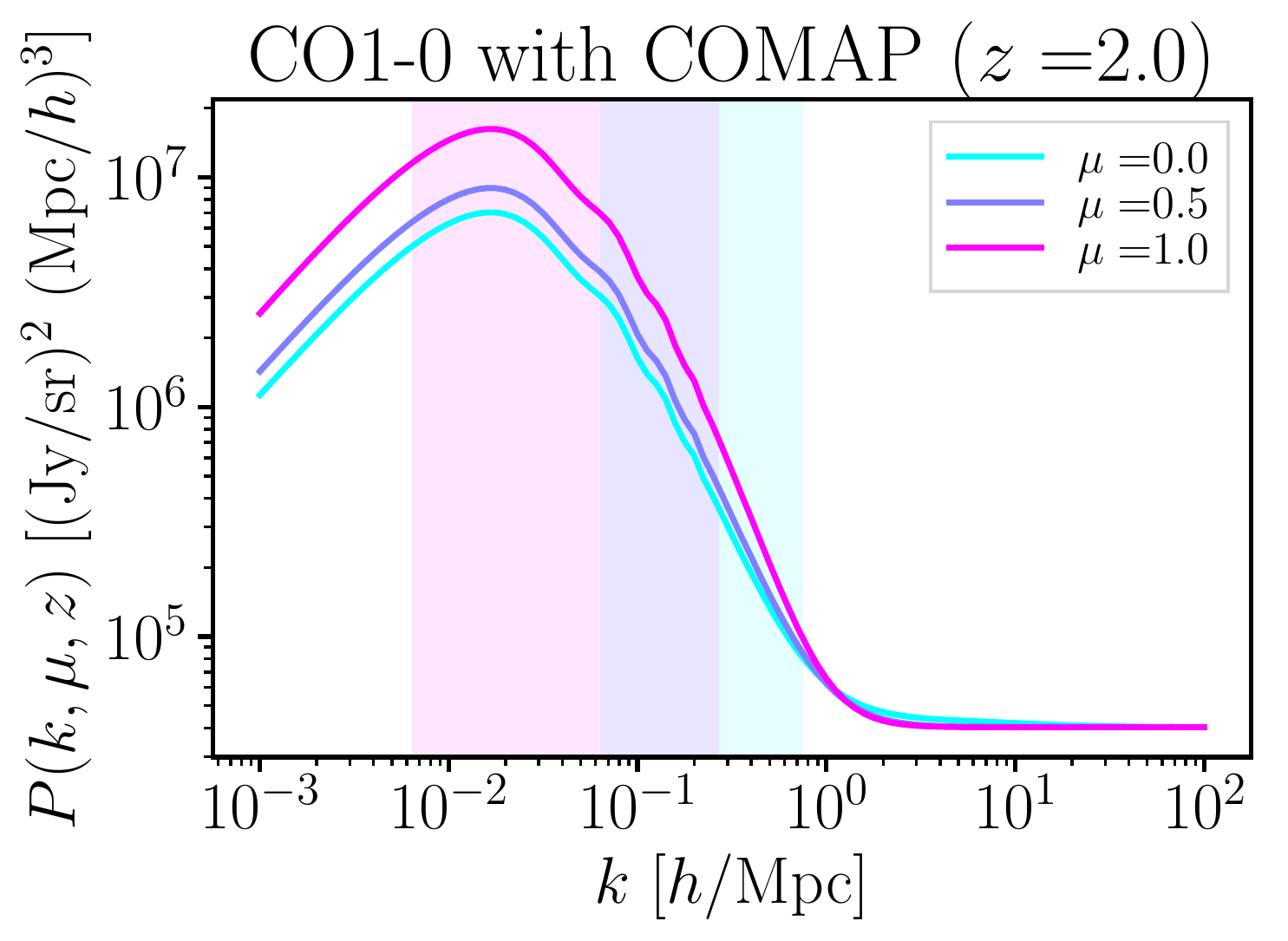}
\includegraphics[width=0.32\textwidth]{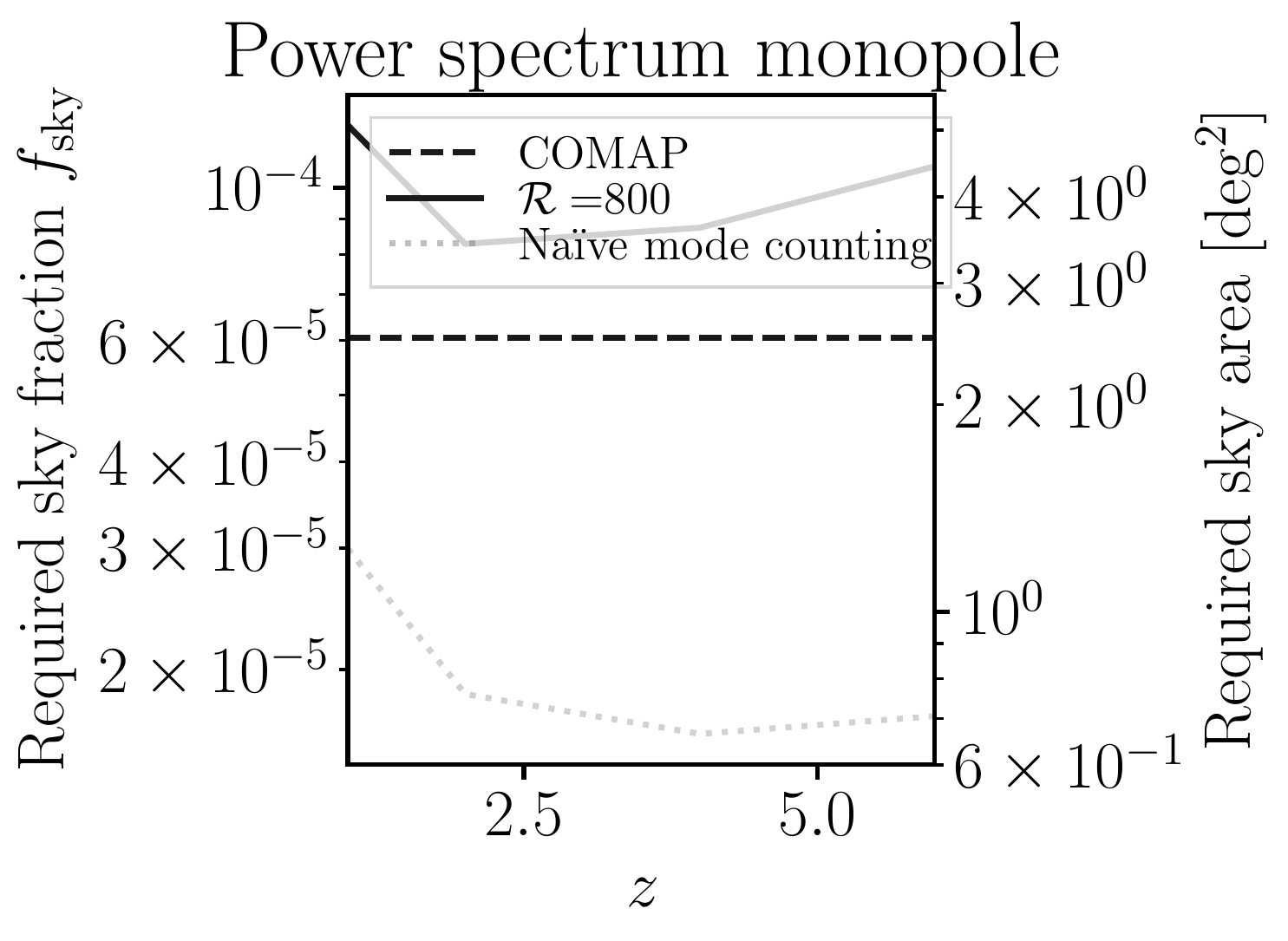}
\includegraphics[width=0.32\textwidth]{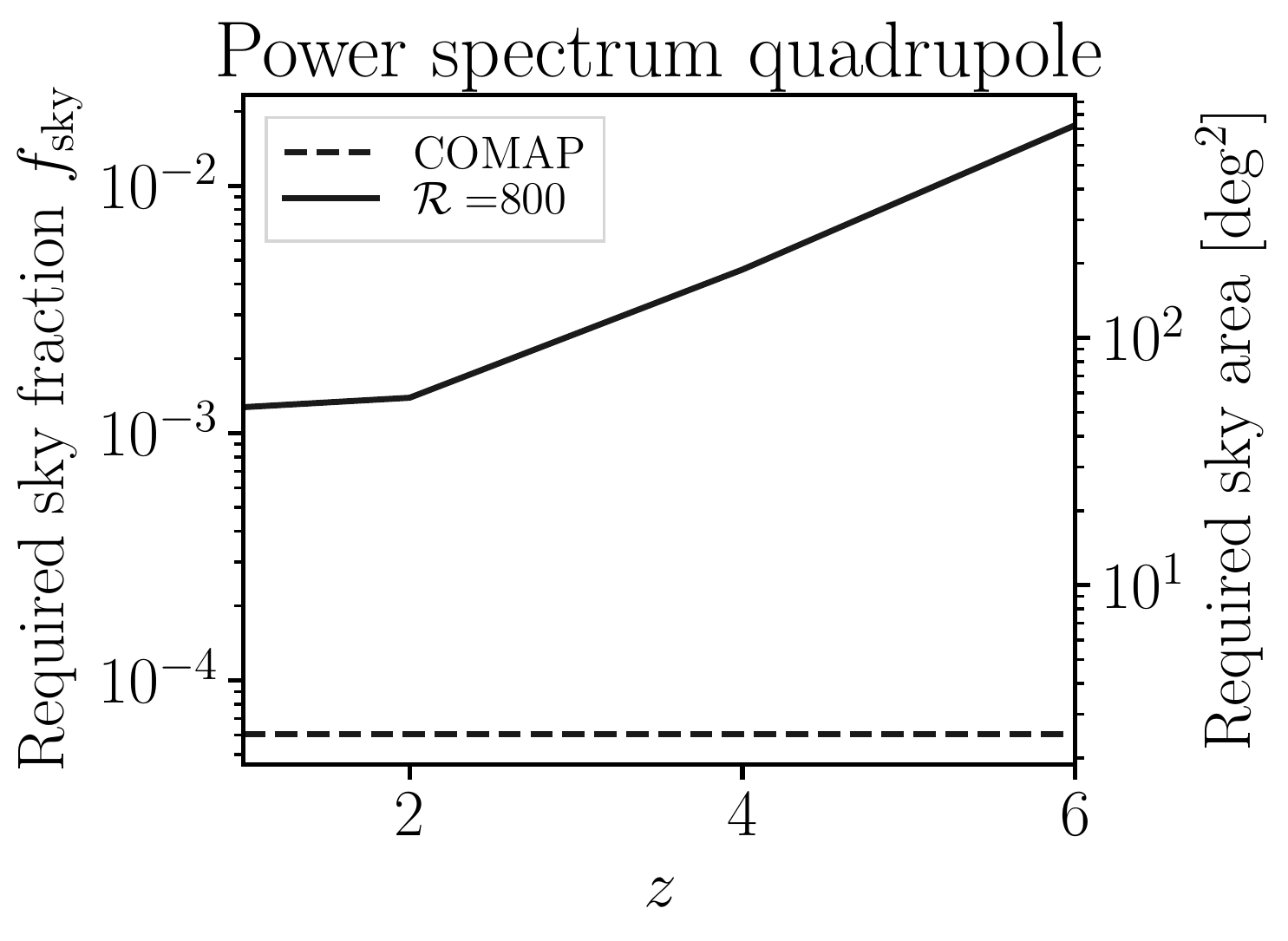}
\caption{
COMAP-like forecast.
Same as Fig.~\ref{fig:tradeoff_a_halpha_spherex}, for COMAP.
Here, the SNR in $k_\perp$ and $k_\parallel$-modes is more balanced, such that the unmarginalized uncertainties on the CO power spectrum monopole and quadrupole (not shown) agree with the marginalized ones.
Furthermore, there is a significant overlap between the scales probed along and across the LOS, making the quadrupole measurement robust to variation in the scale-dependence of the power spectrum.
The simple mode counting disagrees with the realistic monopole forecast at high redshift, where the 1-halo term contributes an important source of noise even at $k=0.1h/$Mpc.
A survey like COMAP over $\sim 100$~deg$^2$ would enable measuring the power spectrum monopole and quadrupole.
}
\label{fig:tradeoff_a_co_comap}
\end{figure}

\begin{figure}[h!]
\centering
\includegraphics[width=0.32\textwidth]{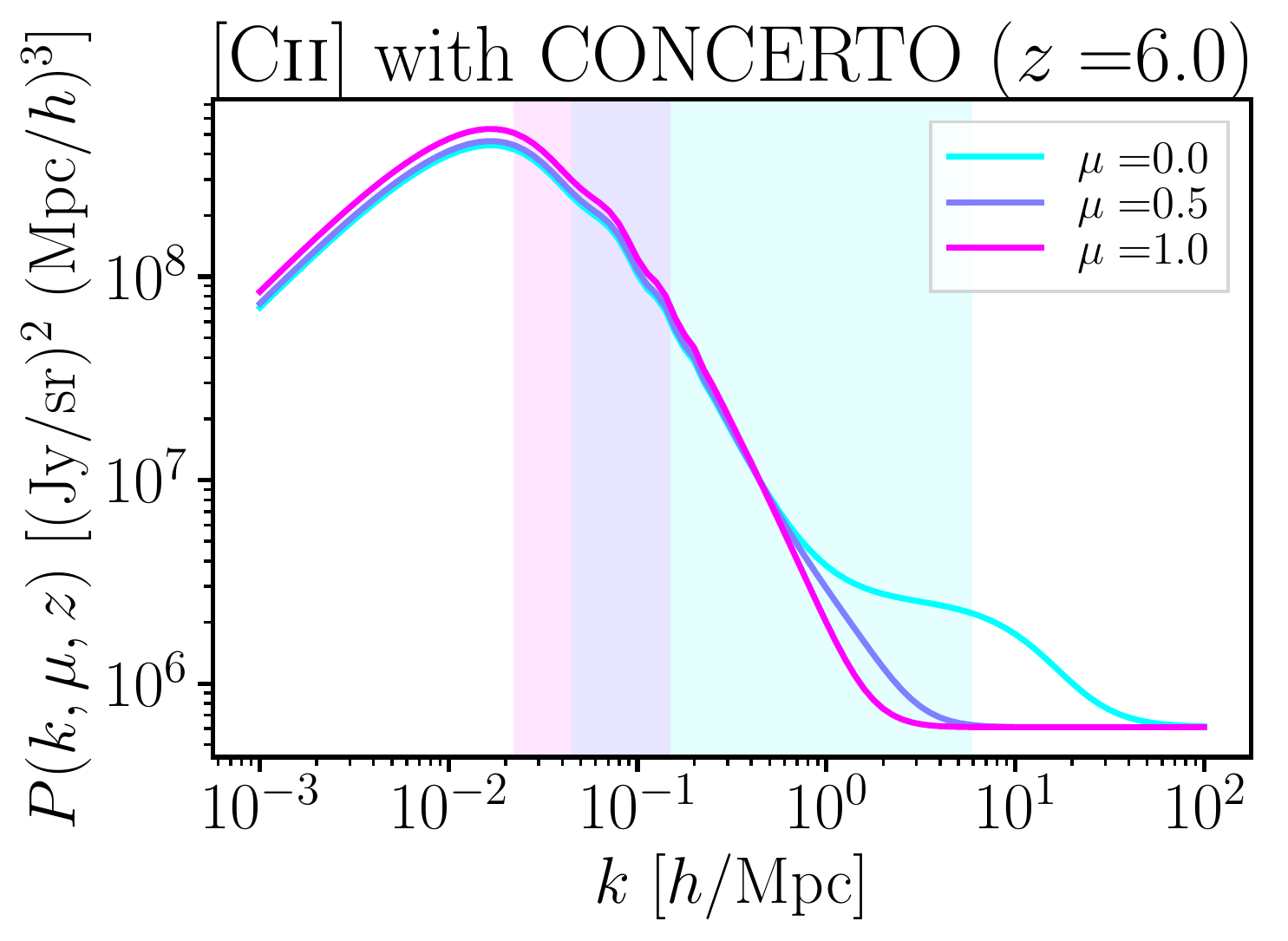}
\includegraphics[width=0.32\textwidth]{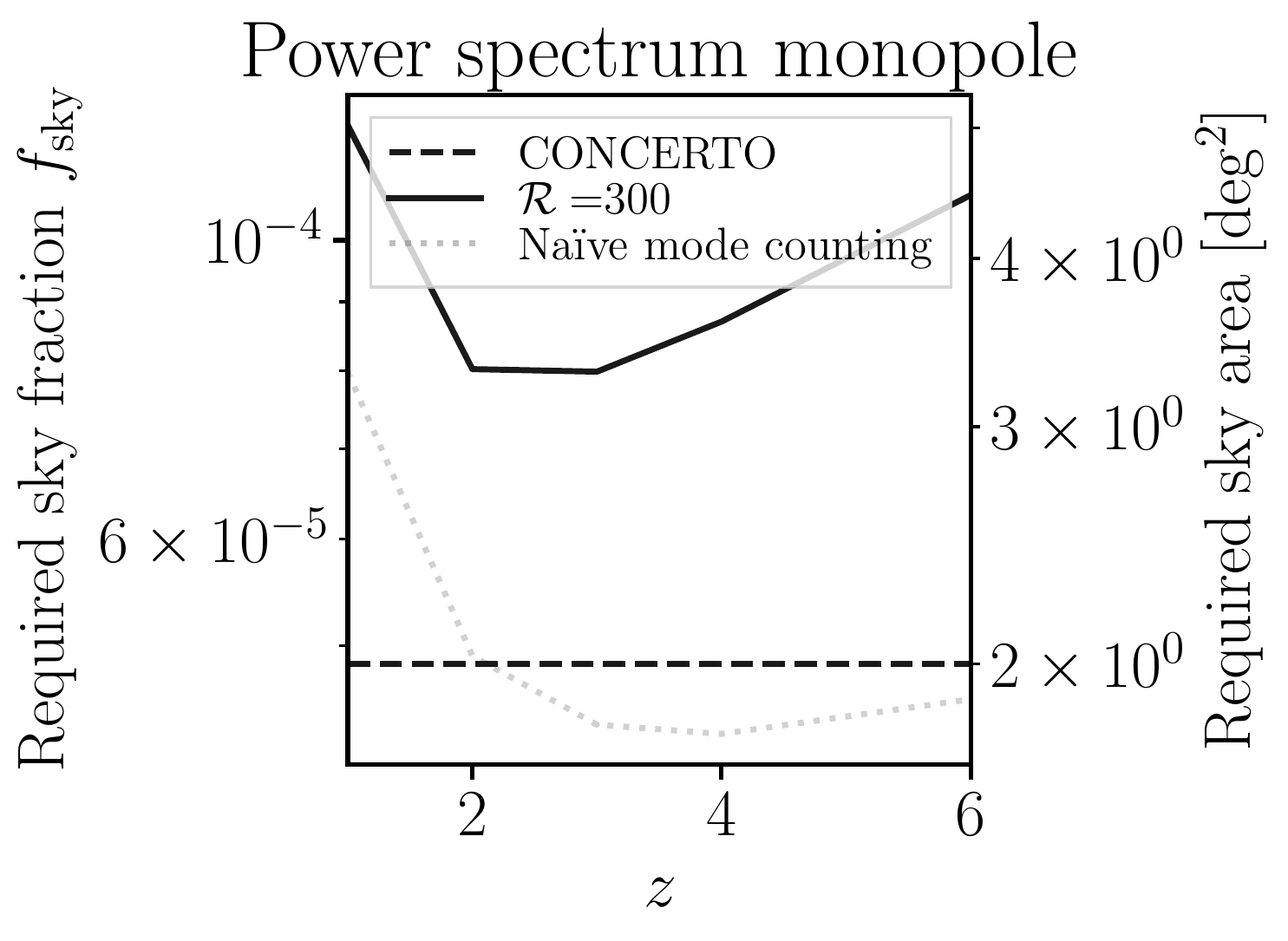}
\includegraphics[width=0.32\textwidth]{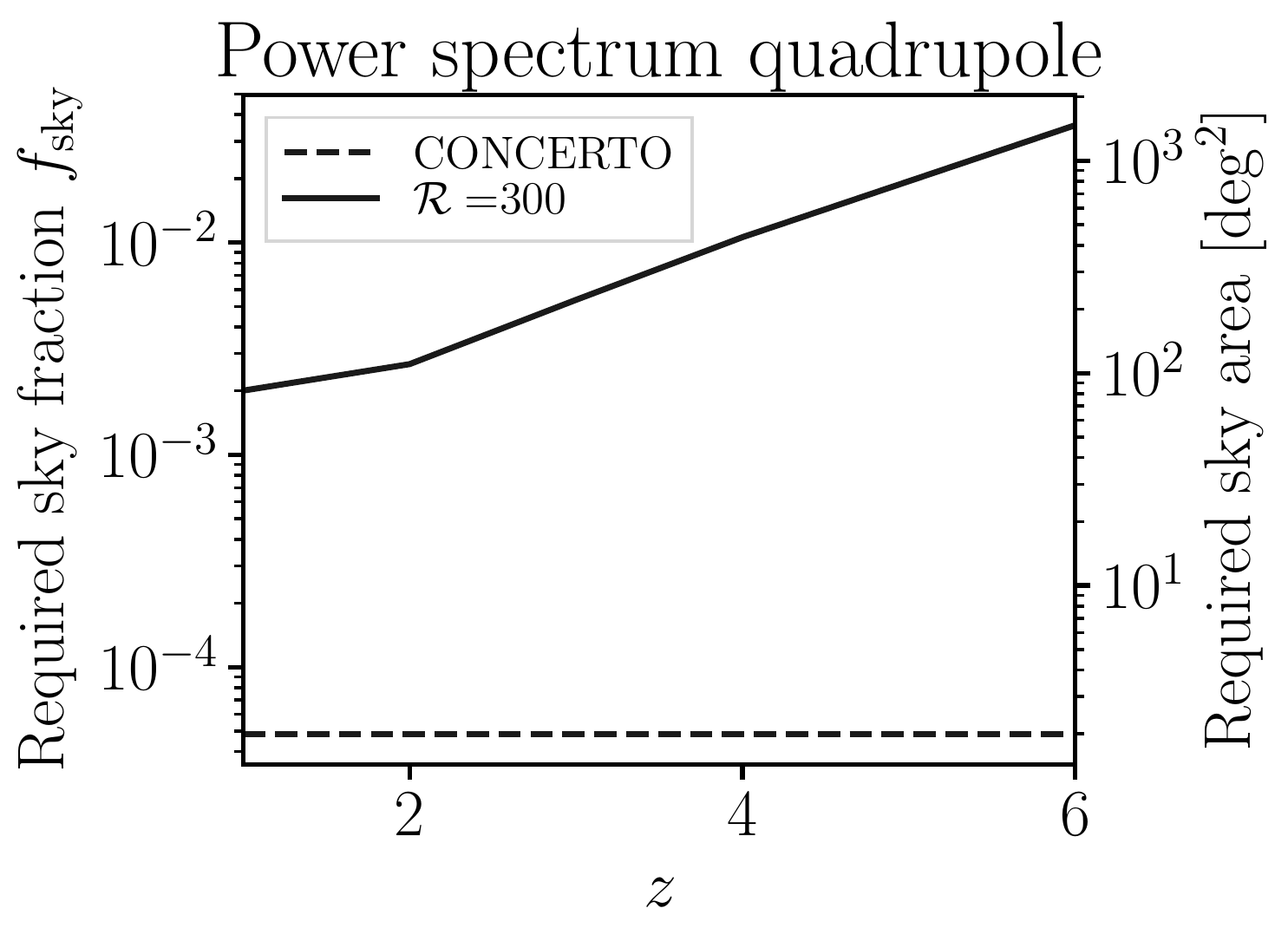}
\caption{
CONCERTO-like forecast.
The same conclusions as for COMAP (Fig.~\ref{fig:tradeoff_a_co_comap}) hold.
}
\label{fig:tradeoff_a_cii_concerto}
\end{figure}

\paragraph{How well is the power spectrum measured?}
A rough estimate of the relative uncertainty on the large-scale power spectrum, assuming Gaussian fluctuations and 2-halo domination, is $\sqrt{2/N_\text{modes}}$, where $N_\text{modes}$ is the number of independent Fourier modes accessible.
For a $\sim10\%$ precision on the 2-halo term amplitude, we therefore require $N_\text{modes}=200$.
In the flat sky geometry, this is simply the Fourier space volume, in units of the Fourier resolution element, i.e.\ $N_\text{modes} = \left(\int d^2 k_\perp / k_{\perp f}^2 \right) \times \left(\int dk_\parallel / k_{\parallel f}^2\right)$, 
where the fundamental modes are 
$k_{\perp f} = 2\pi / \sqrt{4\pi f_\text{sky} \chi^2}$
across the LOS and $k_{\parallel f} = 2\pi / \Delta \chi$ along the LOS, with $\Delta \chi$ the comoving depth of the survey volume.
This gives:
\beq
N_\text{modes} = f_\text{sky} \chi^2 
\left(k_{\perp\ \text{max}}^2 - k_{\perp\ \text{min}}^2 \right)
\ \times
\ \frac{\Delta \chi}{\pi} 
\left( k_{\parallel\ \text{max}} - k_{\parallel\ \text{min}} \right).
\eeq
In what follows, we set 
$k_{\perp\ \text{min}} = k_{\parallel\ \text{min}}=0$,
but note that continuum foregrounds will likely impose a higher value of $k_{\parallel\ \text{min}}$.
If $k_{\perp\ \text{max}}$ and $k_{\parallel\ \text{max}}$ are set to the maximum accessible values allowed by the angular and spectral resolutions, the number of modes becomes the real space survey volume in units of the angular and spectral resolution elements:
$N_\text{modes} 
= f_\text{sky} / \sigma_\text{PSF}^2 \times \Delta\chi / (\pi \sigma_\chi).$
Here, instead, we wish to roughly assess the precision of the 2-halo term measurement only. 
We thus keep $k_{\parallel\ \text{max}} = aH\mathcal{R}/c$, determined by the spectral resolution,
but set $k_{\perp \text{max}} = 0.1\,h\,{\rm Mpc}^{-1}$, where the 2-halo term still dominates, rather than the maximum allowed by the PSF.
The number of modes thus becomes:
\beq
N_\text{modes} 
= f_\text{sky} \chi^2 k_{\perp\ \text{max}}^2 \ \frac{ a H \mathcal{R}\, \Delta\chi}{\pi c}.
\label{eq:tradeoff_nmodes}
\eeq
This rough estimate shows that a poor spectral resolution can be compensated by a larger sky area\footnote{Increasing the sky area also keeps the $\mu$-distribution of the accessible Fourier modes unchanged, since it only increases the density of Fourier modes, through $k_{\perp f}$.}.
This trade off is illustrated in Fig.~\ref{fig:tradeoff_a_halpha_spherex} (left panel).
However, this na\"ive mode counting neglects the noise from the detector and from the 1-halo and shot noise terms, which reduce our ability to measure the 2-halo term.
It also neglects the redshift-space distortions, which enhance the 2-halo term.
In Figs.~\ref{fig:tradeoff_a_halpha_spherex}-\ref{fig:tradeoff_a_cii_concerto}, we compare this mode counting to the more realistic Fisher forecast for the power spectrum monopole and quadrupole below, in the absence of detector noise.
Fig.~\ref{fig:tradeoff_a_halpha_spherex} shows that the 1-halo and shot noise terms are indeed small at $k\simeq 0.1 h\,{\rm Mpc}^{-1}$ for H$\alpha$.
This is not the case for CO and [C{\sc ii}] at high redshifts (Figs.~\ref{fig:tradeoff_a_co_comap} and \ref{fig:tradeoff_a_cii_concerto}), where the 1-halo term is not negligible even at $k\simeq 0.1 h\,{\rm Mpc}^{-1}$, and the na\"ive mode counting overestimates our ability to extract the information in the 2-halo term.

\paragraph{How well is the supercluster infall effect measured?}
The constraints on 
any set of cosmological parameters from the LIM power spectrum
can be estimated using the Fisher matrix \cite{Tegmark97, White09}: 
\beq
F_{\alpha, \beta}
\equiv 
V_{\rm surv}
\int \frac{d^3\vk}{\left( 2\pi \right)^3}\
\frac{\partial P}{\partial \alpha}
\frac{\partial P}{\partial \beta}\
\frac{1}{2\left( P + N\times\text{PSF}^{-2} \times\text{SPSF}^{-2} \right)^2},
\eeq
where $P$ is the total LIM power spectrum (1-halo, 2-halo and shot noise), $N$ the instrumental noise power spectrum (set to zero here), $V_{\rm surv}$ is the survey volume, and PSF and SPSF are the Fourier-space angular and spectral point-spread functions, respectively.
Consistent with our mode counting approximation above, we replace the noise power spectrum with maximum wave vectors along and across the LOS.
The Fisher matrix thus simplifies to: 
\beq
F_{\alpha, \beta}
\simeq
\frac{V_{\rm surv}}{2}
\int_{0}^{k_{\perp\text{max}}} \frac{k_\perp dk_\perp}{2\pi}\
\int_{-k_{\parallel\text{max}}}^{k_{\parallel\text{max}}}
\frac{dk_\parallel}{2\pi}\
\frac{\partial \ln P}{\partial \alpha}
\frac{\partial \ln P}{\partial \beta}
.
\label{eq:fisher_f}
\eeq
The predicted uncertainty on parameters $\alpha$, marginalizing over the other parameters, is then given by
$\sigma_\alpha^\text{marg.} = \left( F^{-1} \right)_{\alpha, \alpha}$.
These are typically larger than the unmarginalized uncertainty
$\sigma_\alpha^\text{unmarg.} = 1 / F_{\alpha, \alpha}$.
We consider a simple angular decomposition of the power spectrum into a monopole and quadrupole, with fiducial amplitudes $A_0=A_2=1$:
\beq
P(k, \mu)
=
A_0 P_0(k) + A_2 P_2(k) \frac{1}{2}\left( 3\mu^2 - 1 \right)
\quad\text{with}\quad
\left\{
\bal
&P_0(k) \equiv \left( 1 + \frac{2}{3}\beta + \frac{1}{5}\beta^2 \right)\ I^2 b^2 P_\text{lin}(k)\\
&P_2(k) \equiv \left( \frac{4}{3}\beta + \frac{4}{7}\beta^2 \right)\ I^2 b^2 P_\text{lin}(k)\\
\eal
\right.
.
\eeq
We show forecasts for experiments similar to SPHEREx (deep fields), COMAP and CONCERTO in Fig.~\ref{fig:tradeoff_a_halpha_spherex}-\ref{fig:tradeoff_a_cii_concerto}.
We show the required sky area needed to produce a 10\% measurement of the monopole and quadrupole, and compare it to the planned survey area.

We find that the simple mode counting argument reproduces the Fisher forecast for COMAP and CONCERTO, when the 1-halo term is small. 
At high redshift, where the 1-halo term of CO and [C{\sc ii}] becomes more important relative to the 2-halo term, the simple mode counting overestimates the statistical power of the experiment.
This makes sense since the large 1-halo term then acts as an important source of noise.
For experiments similar to COMAP and CONCERTO, a survey area of $\sim 100$~deg$^2$ is sufficient to measure the monopole and quadrupole to 10\%.

In the case of SPHEREx, the marginalized uncertainties on the monopole and quadrupole can be as much as 50 times larger than the unmarginalized uncertainties.
On the contrary, the unmarginalized and marginalized uncertainties are almost identical for COMAP and CONCERTO.
This can be attributed to the coverage of the Fourier space, more anisotropic in SPHEREx.
Indeed, schematically, the power spectrum monopole $P_0$ and quadrupole $P_2$ are estimated from $k_\perp$-modes and $k_\parallel$-modes as:
\beq
\left\{
\bal
&P(k_\parallel)=P_0+P_2\\
&P(k_\perp)=P_0-\frac{1}{2}P_2\\
\eal
\right.
,\quad\text{i.e.}\quad
\left\{
\bal
&P_0 = \frac{1}{3}\left(2P(k_\perp) + P(k_\parallel)\right)\\
&P_2 = \frac{2}{3}\left(P(k_\parallel) - P(k_\perp)\right)\\
\eal
\right.
\eeq
One would therefore want a similar signal-to-noise in the $k_\perp$-modes and $k_\parallel$-modes.
For instance, if the $k_\perp$-modes are much better measured than the $k_\parallel$-modes, then the uncertainty on $P(k_\parallel)$ dominates and makes the monopole and quadrupole perfectly (anti-)correlated.
In this case, the marginalized uncertainties on the monopole and quadrupole are thus much larger than the unmarginalized ones.
This appears to be the case for SPHEREx, but not for COMAP or CONCERTO, which were designed to cover a smaller area in order to achieve a low detector noise.
This is an important consideration for LIM survey design, given the importance of measuring the supercluster infall effect.

We also note that there is no overlap in the scales probed along and across the LOS by SPHEREx (for $\mathcal{R}=40$; small overlap otherwise), contrary to COMAP and CONCERTO.
This means that the quadrupole, estimated by comparing $P(k_\perp)$ and $P(k_\parallel)$, is always comparing different scales $|\vk|$.
As a result, the quadrupole measurement is entirely degenerate with a variation in the scale-dependence of the power spectrum.
This important effect, not included in our forecast, should be taken into account when designing LIM surveys.

Finally, we have also checked that removing the $k_\text{max}=0.1\,h\,{\rm Mpc}^{-1}$ cut improves the SNR on both the power spectrum monopole and quadrupole. 
This is motivation to improve the modeling of the 1-halo and shot noise terms, in order to extract more information from the 2-halo term, out to smaller scales.

\begin{figure}[h!]
\centering
\includegraphics[width=0.6\textwidth]{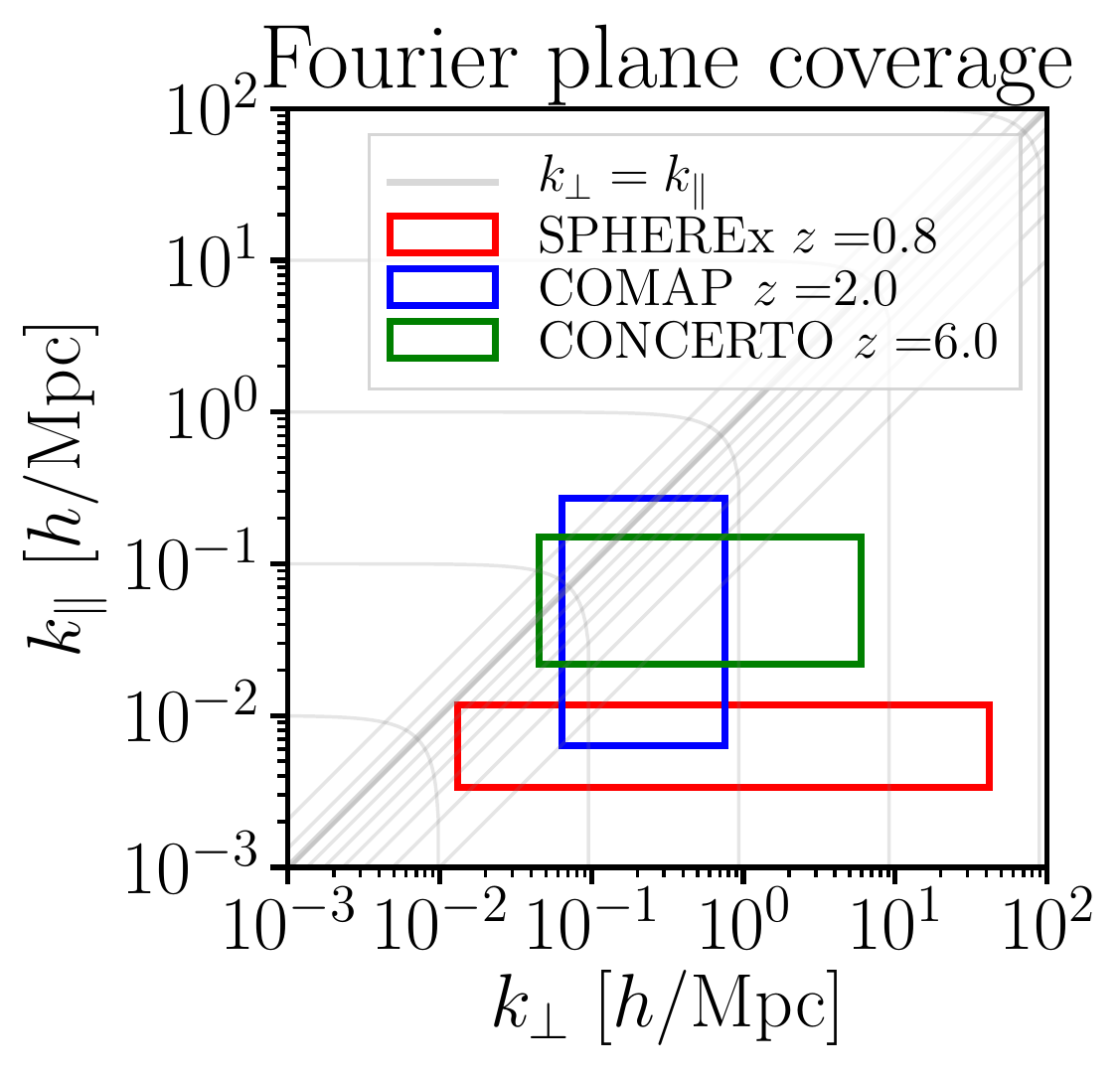}
\caption{
Fourier modes sampled by the various experiments.
Along each axis, the edges of the rectangle represent the fundamental mode and the maximum multipole, respectively.
The iso-$|\vk|$ lines and iso-$\mu$ curves are shown in thin gray.
The RSD power spectrum quadrupole is best measured when the Fourier modes measured sample many $\mu$ values for any given $|\vk|$ scale.
}
\label{fig:fourier_plane_coverage}
\end{figure}

%%%%%%%%%%%%%%%%%%%%%%%%%%%%%%%%%%%%%%%%%%%%%%%%%%%%%%%%%%
\subsubsection{Angular resolution: distinguishing 1-halo and shot noise terms?}
\label{sec:resolution_distinguish_1halo_shot}

An experiment's ability to distinguish the 2-halo, 1-halo and shot noise terms depends on the range of scales it probes across the line of sight.
The largest scale accessible is the fundamental wavevector $k_{\perp f} = \sqrt{\pi / f_\text{sky}} / \chi$, where $f_\text{sky}$ is the sky fraction covered.
The smallest scale accessible is determined by the instrumental point spread function (PSF) or beam, which multiplies the Fourier space density field by the Gaussian kernel $\exp[-k^2(1-\mu^2) \sigma_\text{PSF}^2/2] = \exp[-k^2(1-\mu^2) / (2 k_{\perp\ \text{max}}^2)]$,
where 
$\sigma_\text{PSF} = \text{FWHM} / \sqrt{8 \ln 2}$
and 
$k_{\perp\ \text{max}} = 1/ (\chi\ \sigma_\text{PSF})$.
\begin{figure}[h!]
\centering
\includegraphics[width=0.45\textwidth]{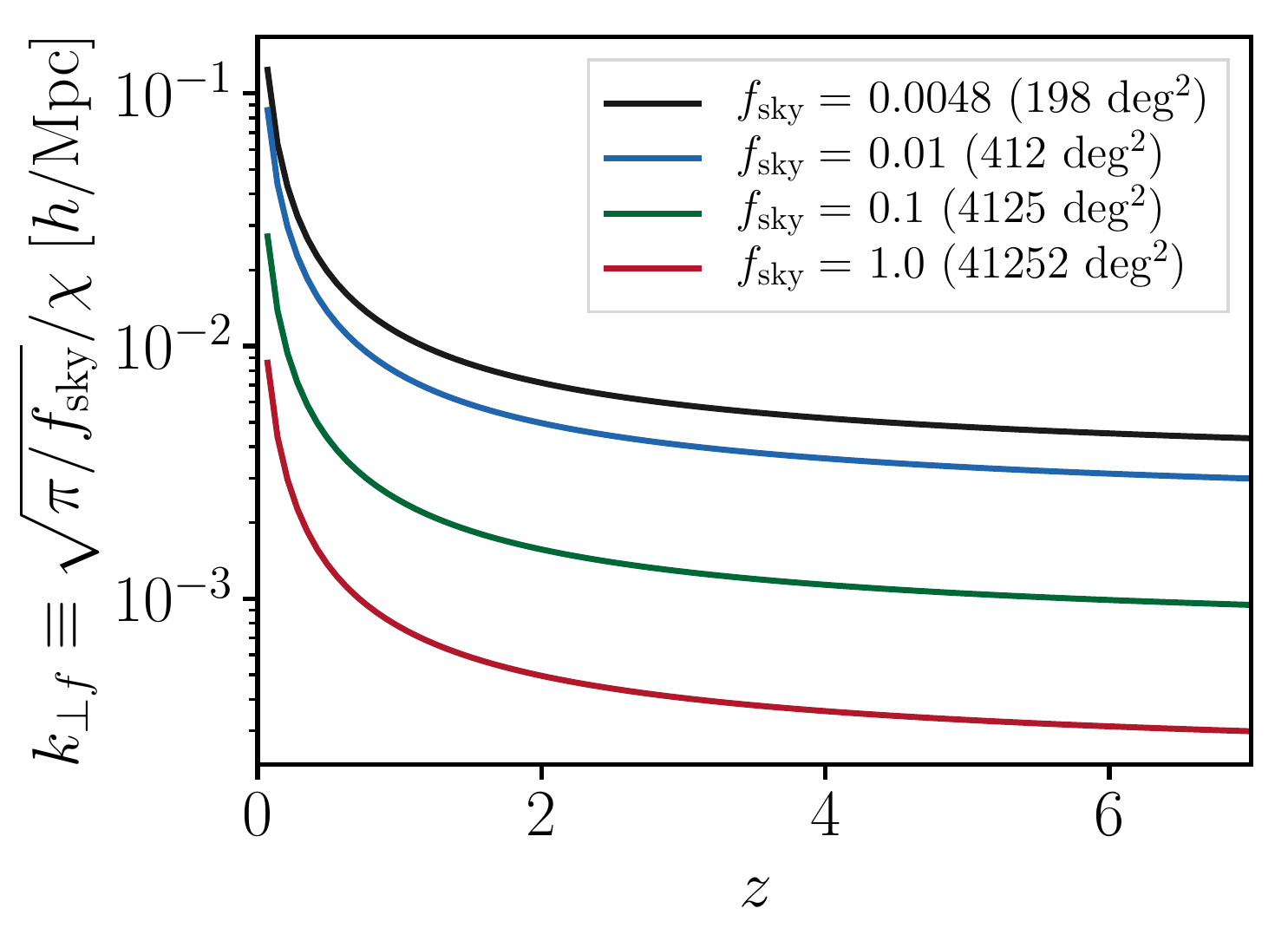}
\includegraphics[width=0.45\textwidth]{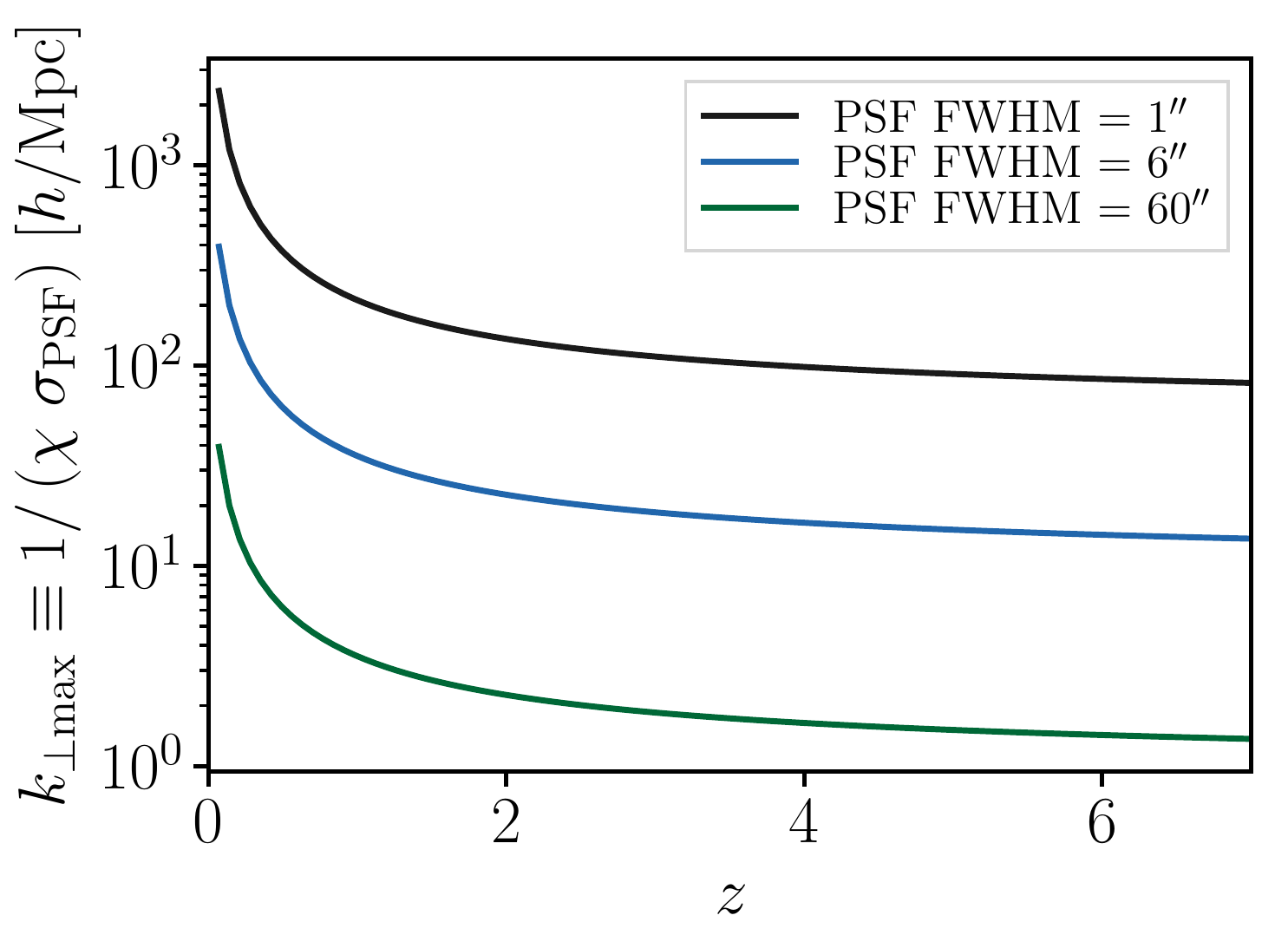}
\caption{
Range of scales accessible across the line of sight for a given survey.
Left: Fundamental wavevector accessible for a survey with a given sky coverage, as a function of redshift. The SPHEREx deep fields, covering $\sim 200$ deg$^2$, correspond to $f_\text{sky}=0.0024$.
Right: Maximum wavevector accessible across the line of sight as a function of redshift, for a given resolution.
A point spread function (PSF) full width at half-maximum (FWHM) of $5-7''$ corresponds to the SPHEREx LIM, and $\sim 1'$ is appropriate for CCAT-prime, TIME and CONCERTO.
}
\label{fig:kperpmax}
\end{figure}
As shown in Fig.~\ref{fig:kperpmax}, an experiment like SPHEREx (200~deg$^2$ of sky coverage, with a PSF FWHM=$5-7'$) will resolve the 2-halo term, and the sum of the 1-halo term and the shot noise on the scales where they are degenerate. 
It may be just sufficient to measure the smallest scales $k\sim 40\, h\,{\rm Mpc}^{-1}$ where the shot noise alone dominates.

%%%%%%%%%%%%%%%%%%%%%%%%%%%%%%%%%%%%%%%%%%%%%%%%%%%%%%%%%%
\subsubsection{Interloper and continuum foregrounds}
\label{sec:interloper_continuum_foregrounds}

Foregrounds are a key concern in all of LIM. 
They may dramatically limit our ability to learn astrophysics and cosmology from LIM.
Unfortunately, our knowledge of foregrounds is typically as uncertain as our knowledge of the LIM signals themselves, making it very difficult to predict their impact.
As such, a complete study of LIM foregrounds and their uncertainties is beyond the scope of this paper. Instead, we simply note that any foreground power spectrum can in principle be included as an effective noise in all our calculations.

As an example, we discuss one case study here: the contamination from low redshift CO emission and from the cosmic infrared background (CIB) to the high redshift [C{\sc ii}] power spectrum.
Ref.~\cite{Cheng16} studies the contamination to [C{\sc ii}] at $z=6$ due to interloper CO-emitting galaxies. Their Fig.~3 shows that the CO emission at lower redshift dominates over the high-redshift [C{\sc ii}] signal of interest, in terms of the observed power spectrum monopole. However, the interloper CO power spectrum has a specific anistropy, which allows it to be effectively separated.
In Ref.~\cite{Pullen18}, the line and continuum contamination to [C{\sc ii}] at $z=2.6$ is quantified. The measured cross-spectrum between BOSS quasars and the \textit{Planck} 545~GHz map probes the sum of the [C{\sc ii}] signal of interest, the CIB, the thermal Sunyaev-Zel'dovich (tSZ) effect and potential interloper lines.
At this low redshift, CO emission is not a concern. 
The low significance [C{\sc ii}] detection, despite the high-significance measurement of the quasar - 545~GHz cross spectrum, indicates that the CIB dominates over [C{\sc ii}] at this redshift. 
Here again, measuring the anisotropy of the 3D power spectrum may allow disentangling the two.

%%%%%%%%%%%%%%%%%%%%%%%%%%%%%%%%%%%%%%%%%%%%%%%%%%%%%%%%%%
%%%%%%%%%%%%%%%%%%%%%%%%%%%%%%%%%%%%%%%%%%%%%%%%%%%%%%%%%%
\section{LIM vs.\ galaxy detection: Astrophysics \& Cosmology}
\label{sec:lim_vs_galaxies}

A line intensity map contains astrophysical information about the population of galaxies that produce it.
It also contains cosmological information, being a tracer of the matter density field.
One approach to try to extract these two sorts of information is to model the LIM mean intensity, 2-halo, 1-halo and shot noise power spectra.
On the other hand, a more traditional approach consists in building the catalog of galaxies bright enough to be detected individually.
One can use this catalog to estimate the galaxy luminosity function, and as a tracer of the matter density.
It is therefore natural to ask whether the catalog of bright, detected galaxies is sufficient, or whether the LIM observables (mean intensity and LIM power spectrum) contain additional information.

In what follows, we explore this question in detail.
Ignoring first the finite experimental sensitivity, we quantify the sparsity of the relevant individual sources -- halos in the 1-halo regime and galaxies in the shot noise regime.
Adding a realistic detector noise, we quantify the luminosity detection threshold for individual galaxies in \S\ref{sec:matched_filter}.
We convert this to a halo mass threshold.
We then ask whether the LIM observables are sensitive to galaxies below the detection threshold in \S\ref{sec:sensitivity_undetected_sources}.
Finally, we compare the catalog of detected galaxies to the intensity field, as tracers of the matter density field, in \S\ref{sec:lim_vs_gal_matter_tracer}.
There, we revisit the study of \cite{Cheng19}, in Fourier space instead of a single pixel, in order to take into account the pixel-to-pixel correlation, and to answer the question as a function of Fourier scale.

We note that we are not directly comparing galaxy surveys to LIM experiments.
This would involve comparing observational efficiencies and comparing at fixed observing time or cost.
Instead, we are simply answering the question of how to best analyze an intensity map, either by building a catalog of bright galaxies or by measuring the mean intensity and the LIM power spectrum.

Although we do not model foreground components, setting them to zero, we include them in our formalism through their power spectrum.
Any foreground thus acts as an additional source of noise, degrading the intensity map and limiting our ability to detect bright galaxies from it.
This effective noise floor from foregrounds can likely be reduced, by relying on their spectral smoothness, masking bright sources, or cross-correlating the intensity map with galaxies at low redshift.
Whenever possible, we present results as a function of the map noise, allowing them to be adjusted in the presence of foregrounds.
The results shown for the fiducial noise levels of the experiments considered will change, perhaps dramatically, in the presence of foregrounds. 
We provide the formalism to compute these changes, by including the foreground power spectrum in the total map power.

%%%%%%%%%%%%%%%%%%%%%%%%%%%%%%%%%%%%%%%%%%%%%%%%%%%%%%%%%%
%%%%%%%%%%%%%%%%%%%%%%%%%%%%%%%%%%%%%%%%%%%%%%%%%%%%%%%%%%
\subsection{Sparsity of the sources -- halos and galaxies}
\label{subsec:interloper_deconfusion}
\label{sec:sparsity_halos_galaxies}

The mantra of LIM is that one does not necessarily need to resolve individual sources, and that the blended emission of a multitude of sources can be a sufficient tracer of the matter on large-scale structure, and a sufficient probe of individual galaxy properties.
One may thus wonder how blended the individual sources are, for a given experimental resolution.
We thus evaluate the effective mean numbers of sources per voxel.
In the 1-halo regime, the relevant sources are halos; in the shot noise regime, they are individual galaxies.

%%%%%%%%%%%%%%%%%%%%%%%%%%%%%%%%%%%%%%%%%%%%%%%%%%%%%%%%%%
\subsubsection{Sparsity of halos}

\begin{figure}[h!]
\centering
\includegraphics[width=0.65\textwidth]{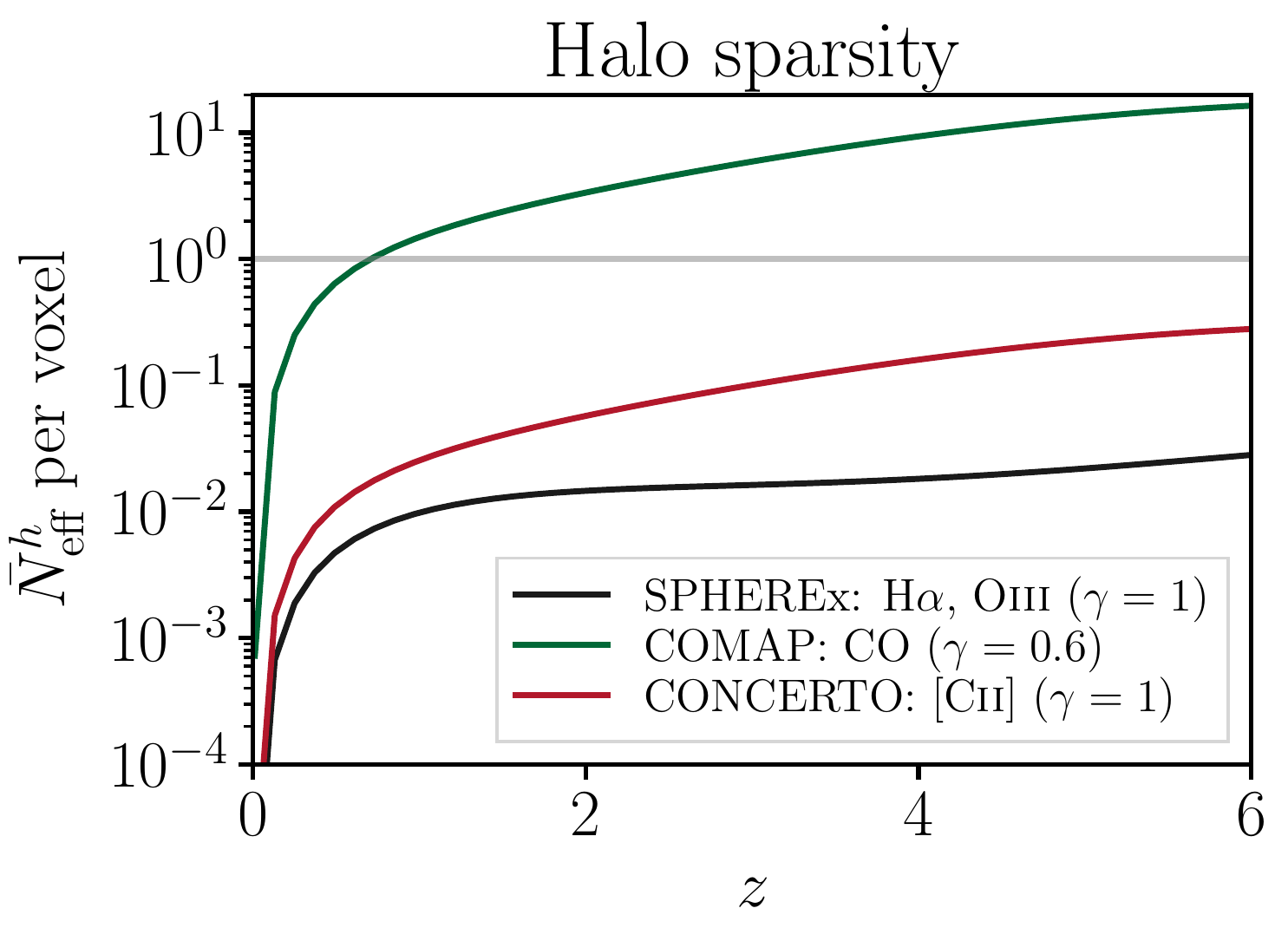}
\caption{
Effective mean number of halos per voxel, for SPHEREx ($6.2''$ pixel and $\mathcal{R}=40$), COMAP ($1.5'$ pixel and $\mathcal{R}=800$) and CONCERTO ($0.12'$ pixel and $\mathcal{R}=300$).
We have assumed that the mean halo luminosity scales with star formation rate:
$L^h \propto \text{SFR}^\gamma(m,z)$,
with $\gamma=1$ for the SPHEREx UV and optical lines, for [C{\sc iii}],
and $\gamma=0.6$ for CO.
The halos are therefore sparse for SPHEREx and CONCERTO at all redshifts but not for COMAP.
}
\label{fig:halo_sparsity}
\end{figure}

Most of the literature assumes that the halo luminosities scale as the star formation rate to some power, $L_j(m) = \text{SFR}^{\gamma_j}(m,z)$.
In ref.~\cite{Fonseca17}, $\gamma_j=1$ for the various SPHEREx lines (Ly$\alpha$, H$\alpha$, H$\beta$ and the [O{\sc iii}] doublet at 500.7\,nm and 495.9\,nm),
and $\gamma_j=0.8-1.1$ for far IR lines ([N{\sc ii}], [N{\sc iii}], [C{\sc ii}] and the CO rotational lines).
Refs.~\cite{Lidz16,Gong17,Gong20} assume a linear scaling ($\gamma_j=1$) for H$\alpha$, [O{\sc iii}], [O{\sc ii}], [C{\sc ii}] and CO. 
Ref.~\cite{Cheng16} assumes $\gamma_\text{CO}=0.6$. 
In what follows, we adopt $\gamma=1$ for all lines except CO for which we assume $\gamma=0.6$.
The effective halo number density is then simply:
\beq
\bar{n}^\text{h eff}_{i,j}
=
\frac{
\left( \int dm \ n(m)\ \text{SFR}^{\gamma_i}(m)  \right)
\left( \int dm \ n(m)\ \text{SFR}(m)^{\gamma_j}  \right)
}
{\int dm \ n(m)\ \text{SFR}^{\gamma_i+\gamma_j}(m)}.
\eeq
Fig.~\ref{fig:halo_sparsity} shows the effective number of halos per pixel, for experiments similar to SPHEREx, COMAP and CONCERTO.
In Fig.~\ref{fig:galaxy_sparsity}, the gray lines show the dependence of this number on the power law index $\gamma$.
As expected, the lower $\gamma$ give a higher number density of halos, since they weight more heavily the more numerous, lower mass halos.
Going from $\gamma=0.6$ to $\gamma=1$ can change the effective halo number density, and thus the amplitude of the 1-halo term, by more than an order of magnitude.
In what follows, we shall use $\gamma=1$ unless specified otherwise, as appropriate for the optical/ultraviolet lines.

%%%%%%%%%%%%%%%%%%%%%%%%%%%%%%%%%%%%%%%%%%%%%%%%%%%%%%%%%%
\subsubsection{Sparsity of galaxies}

\begin{figure}[h!]
\centering
\includegraphics[width=0.32\textwidth]{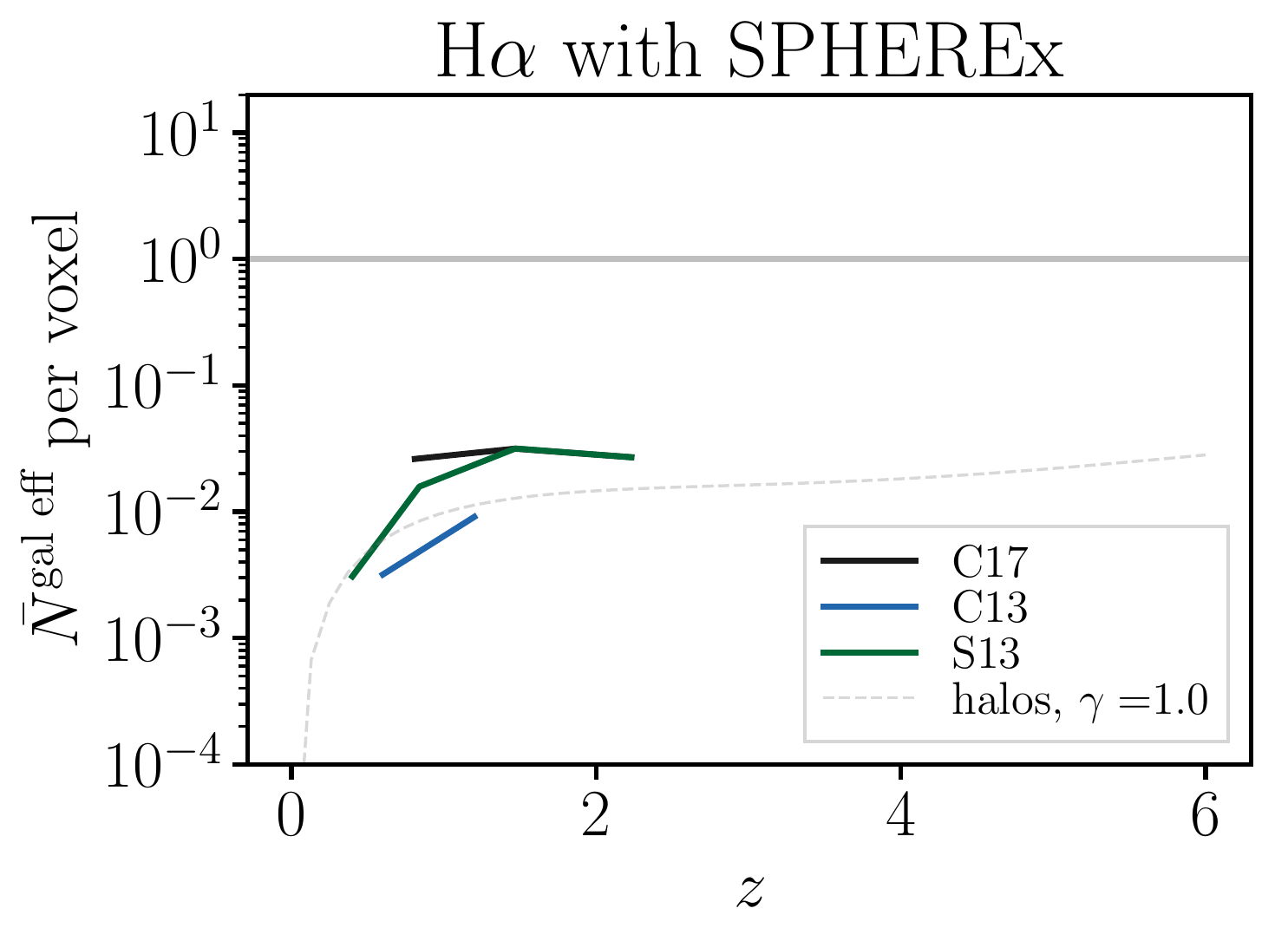}
\includegraphics[width=0.32\textwidth]{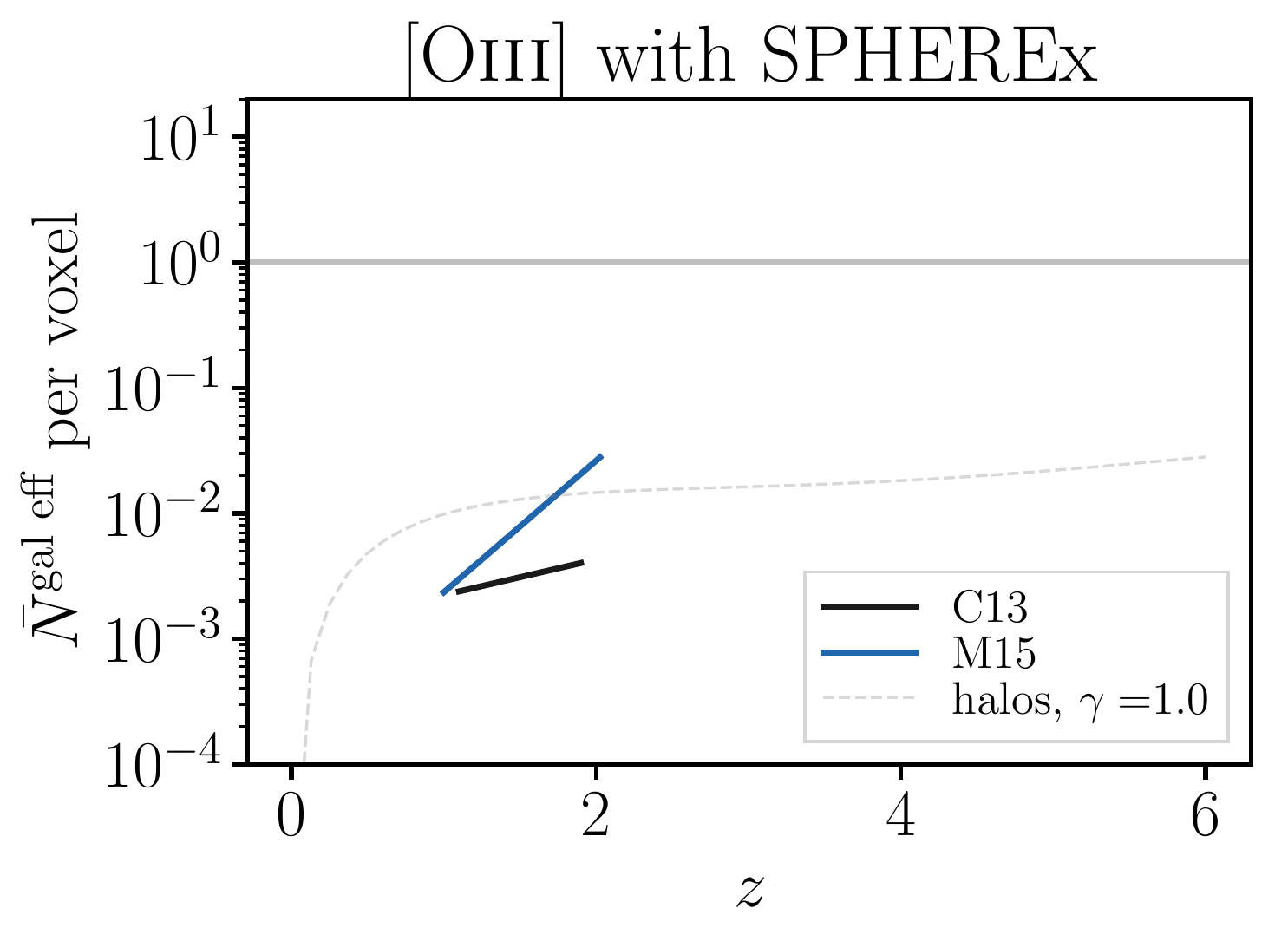}
\includegraphics[width=0.32\textwidth]{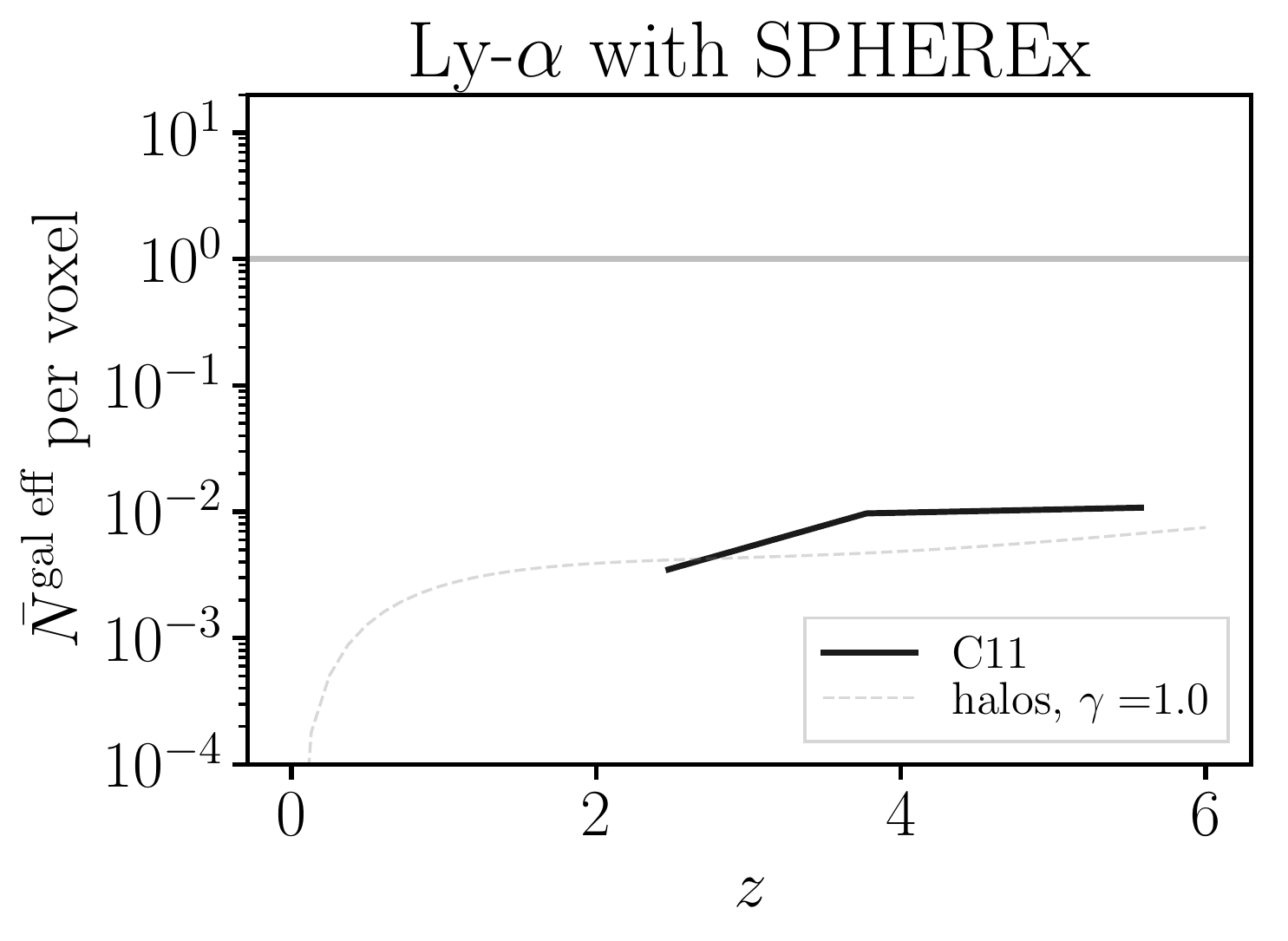}
\includegraphics[width=0.45\textwidth]{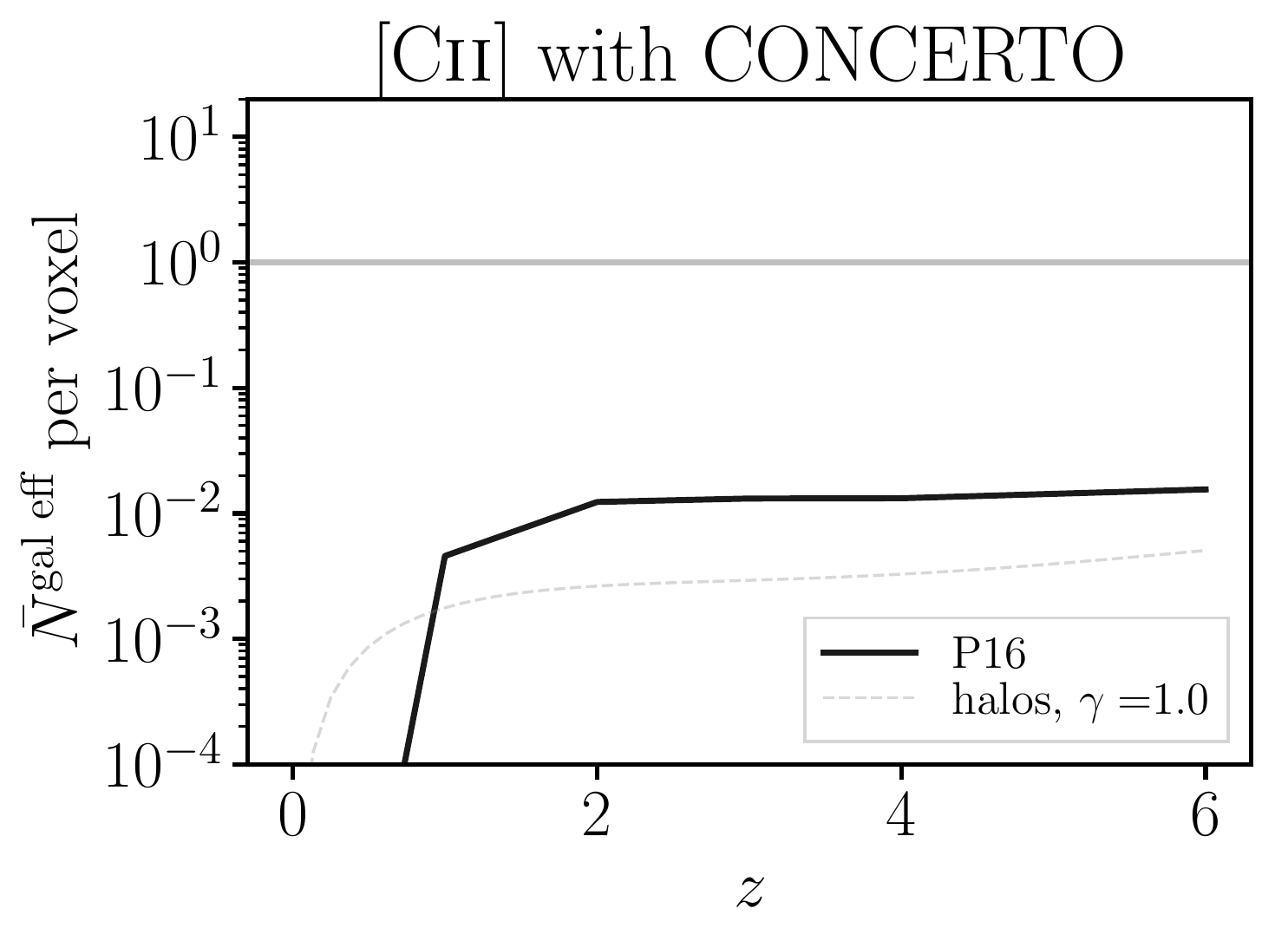}
\includegraphics[width=0.45\textwidth]{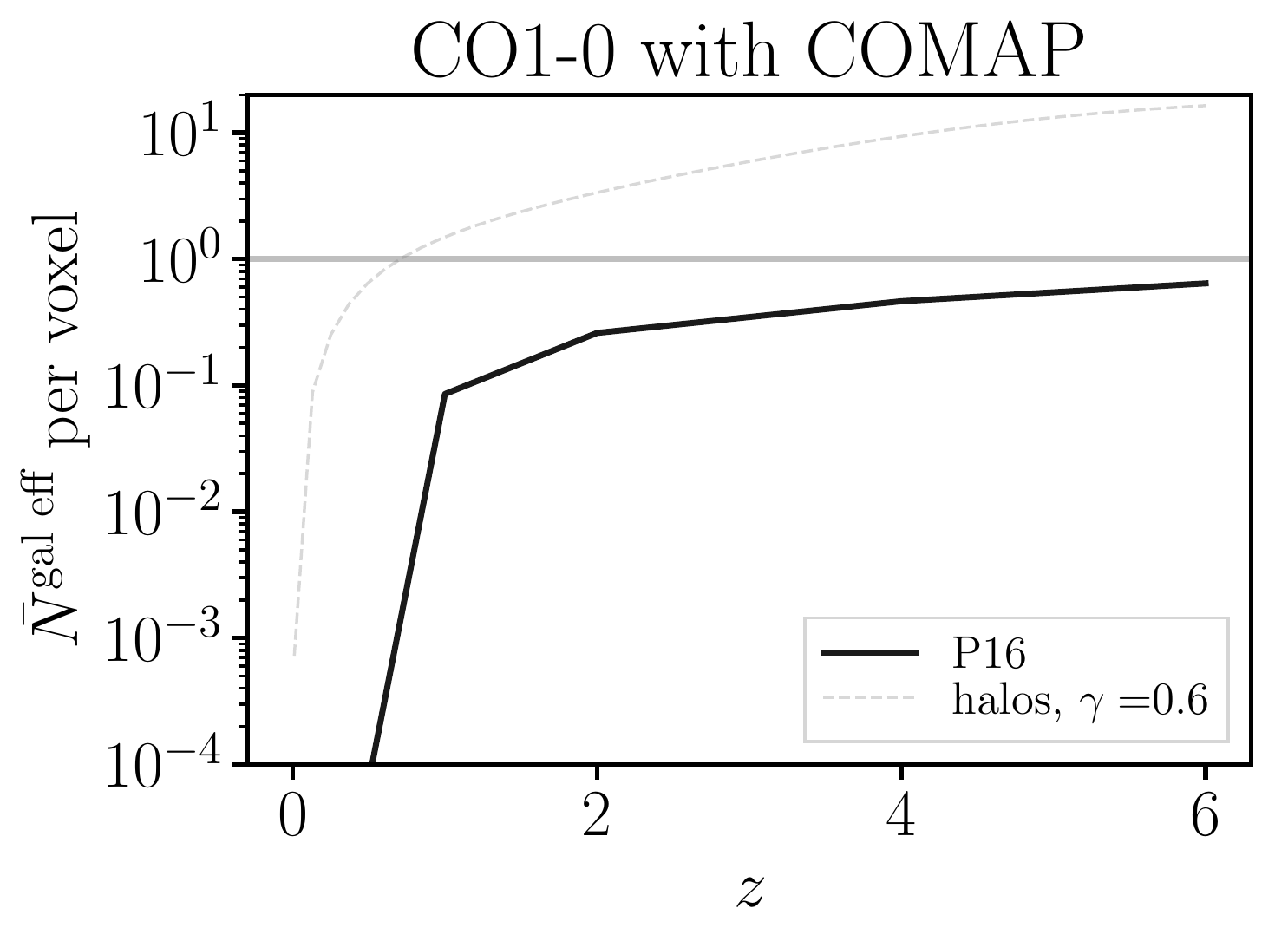}
\caption{
Effective mean number of galaxies per voxel, for SPHEREx (H$\alpha$, [O{\sc iii}] and Ly-$\alpha$, top row) and CONCERTO ([C{\sc ii}], bottom left) and COMAP (CO, bottom right).
Predictions are shown from various observed luminosity functions (S13 \cite{Sobral13}; C13 \cite{Colbert13}; M15 \cite{Mehta15}; C17 \cite{Cochrane17}; C11 \cite{Cassata11} and P16 \cite{Popping16}).
Despite the large theoretical uncertainties, the galaxies appear sparse for all lines and experiments.
For comparison, the thin dashed and dotted lines show the effective number of halos per voxel.
As discussed in \S2.5 of \cite{paper1}, the effective number density of halos does not have to be smaller than that of galaxies.
}
\label{fig:galaxy_sparsity}
\end{figure}

For given line $i$, the effective galaxy number density (Eq.~2.17 in \cite{paper1}),
which determines the amplitude of the shot noise power spectrum, is entirely determined by the line $i$ luminosity functions:
\beq
\frac{1}{\bar{n}^\text{gal eff}_{i,i}}
=
\frac{
\int dL_i\ \Phi(L_i) L_i^2
}
{
\left( \int dL_i \; \Phi(L_i) L_i  \right)^2
}
.
\label{eqn:ngaleff_auto}
\eeq
For the cross-correlation of two lines, this generalizes to:
\beq
\frac{1}{\bar{n}^\text{gal eff}_{i,j}}
=
\frac{
\int dL_i dL_j\ \Phi(L_i, L_j) L_i L_j
}
{
\left( \int dL_i \; \Phi(L_i) L_i  \right)
\left( \int dL_j \; \Phi(L_j) L_j  \right)
}
.
\label{eqn:ngaleff_cross_lf}
\eeq
The effective galaxy number density in cross-correlation can thus be inferred from the auto-correlations and the correlation coefficient $r_{i,j}$ between galaxy luminosities in lines $i$ and $j$:
\beq
\frac{1}{\bar{n}^\text{gal eff}_{i,j}}
=
\frac{r_{i,j}}{
\sqrt{
\bar{n}^\text{gal eff}_{i,i} 
\bar{n}^\text{gal eff}_{j,j}
}
}
\quad .
\label{eqn:ngaleff_cross_r}
\eeq
In Fig.~\ref{fig:galaxy_sparsity}, we show the effective number of galaxies per voxel, for various experiments.
For comparison, the dashed and dot-dashed lines show the sparsity of halos for the corresponding lines.
As expected intuitively, the effective number density of galaxies is higher than that of halos for most lines.
This is the case for H$\alpha$, Ly-$\alpha$ and [C{\sc ii}], at most redshifts, for which $\gamma=1$.
However, as discussed in \S2.5 of \cite{paper1},
this does not have to be the case, in part due to the galaxy line noise.
Indeed, [O{\sc iii}] ($\gamma=1$) at $z=1$ and CO ($\gamma=0.6$) at any redshift have a higher effective number density of halos than galaxies.

From Fig.~\ref{fig:galaxy_sparsity}, it appears that the intensity maps from SPHEREx, COMAP and CONCERTO are sparse, meaning that only a small fraction of voxels contain galaxies at the redshift of interest. 
In particular, voxels with more than one such galaxy are extremely rare.
This sparsity is important in a number of approaches to reject interloper galaxies\cite{Cheng20}.

However, it is worth pointing out that the observed maps still contain a non-zero intensity in most pixels from foregrounds.
Such a foreground is the continuum emission from galaxies within the angular pixel but at any other redshift. These undesired sources are thus much more numerous, hence less sparse, than the sources at the redshift of interest.
Other foregrounds are Galactic emission or solar system emission like the zodiacal light.
They act as a spurious source of noise in the maps, or equivalently as a new source of confusion in the detection of the galaxies at the redshift of interest.

%%%%%%%%%%%%%%%%%%%%%%%%%%%%%%%%%%%%%%%%%%%%%%%%%%%%%%%%%%
\subsubsection{Consequences for interloper line deconfusion}

Quantifying the sparsity of sources is also crucial for interloper line deconfusion.
Indeed, ref.~\cite{Cheng20} shows that an algorithm similar to a multifrequency matched filter is effective at removing interloper line emission, if the sources are sparse.

Emission from an interloper line at a different redshift can be confused with emission from the targeted line at the corresponding redshift.
Various methods of interloper line deconfusion have been proposed.
At the map level, blind masking consists in simply masking the brightest voxel in the map, whose signal presumably comes from foreground sources \cite{Pullen13, Breysse15, Yue15, Silva15, Kogut15}.
Instead, targeted masking consists in masking the map at the positions of known sources, from an external catalog.
This can be further extended by deprojecting a template interloper intensity map, rather than masking pixels.
This amounts to removing from our intensity map of interest any part that is correlated with the template.
The template can be built from a source catalog, or can be an other intensity map, from a different redshift, for continuum foregrounds. This method has the advantage of not only removing the sources themselves, but also any emission correlated with them.

Assuming that the map voxels are small enough, the individual interloper sources contaminating the intensity map become sparse. Using this sparsity assumption, and assuming a fixed ratio of line intensities, ref.~\cite{Cheng20} extracts individual interloper sources from the set of voxels at the same angular position and different wavelengths. 

At the power spectrum level, since the interloper emission comes from a different redshift than assumed, the relative mapping of transverse and longitudinal scales is wrong. This causes the 3D power spectrum of the interlopers to appear anisotropic, similarly to the Alcock-Paczynski test \cite{Pullen13, Lidz16, Cheng16, Gong20, Liu16}. 
Measuring this spurious anisotropy requires a sufficient spectral resolution, and is thus further motivation for pursuing RSD measurements with LIM.
While this interloper effect is typically much larger than the supercluster infall effect \cite{Cheng16}, RSD should still be included when designing survey specifications.

Finally, cross-correlating intensity maps at a pair of observed wavelengths, with a well chosen wavelength ratio, allows cross-correlation of two different lines from the same sources, at the same redshift, and thus extract only these sources, rejecting any interloper line emission \cite{Visbal10, Visbal11, Pullen13, Visbal10, Gong12, Gong14, Switzer17, Switzer19}.
As we show below, two different lines from the same redshift are perfectly correlated in the 2-halo regime only.
Depending on the pair of lines considered, the cross-correlation typically also suppresses the 1-halo and shot noise terms.
This allows us to measure the 2-halo term out to smaller scales, and provides a new linear combination of the 2-halo, 1-halo and shot noise terms, different from the auto-spectrum.

%%%%%%%%%%%%%%%%%%%%%%%%%%%%%%%%%%%%%%%%%%%%%%%%%%%%%%%%%%
\subsection{Detection threshold for individual galaxies \& halos}
\label{sec:matched_filter}

Above, we quantified the effective number density of sources in the intensity map.
In practice, finite sensitivity and resolution only allow us to detect the brightest of them.
In this section, we quantify the detection threshold, in terms of luminosity and halo mass, and use this to compare the intensity map to the catalog of detected sources, as probes of faint galaxies and of the matter density field.

\subsubsection{Minimum galaxy luminosity: Detector VS confusion noise}

Here, we quantify the minimum galaxy luminosity detectable from a given intensity map,
depending on its power spectrum, the detector noise level, and the angular and spectral resolutions.
We assume that individual sources are detected via a matched filter, defined as the minimum-variance, unbiased, linear estimator for a galaxy flux given the intensity map.
This is routinely done in 2D to detect point sources from cosmic microwave or cosmic infrared background intensity maps, and we extend it to 3D, including the effect of RSD\footnote{The matched filter for the case of interferometric, 21-cm intensity mapping was discussed in ref.~\cite{White17}.}.
In 2D, the angular dependence of the point source profile simply follows the PSF.
In 3D, the point source profile also has a radial dependence determined by the spectrograph's point spread function (SPSF).

As we derive in Appendix~\ref{app:matched_filter}, the 3D matched filter can be expressed as:
\beq
\hat{F}
=
\left(\frac{c\chi^2}{aH\nu}\right)
\sigma_{\hat{F}}^2
\int\frac{d^3\vk}{\left( 2\pi \right)^3}
\frac{I(\vk) \ W(\vk)}
{2\left[ W(\vk)^2\ \left(P_\text{LIM}(\vk) + P_\text{fg}(\vk)\right) + N(\vk) \right]},
\eeq
where $\sigma_{\hat{F}}^2$ is the variance of the matched filter in units of flux, given by:
\beq
\sigma_{\hat{F}}^2
=
\left(\frac{aH\nu}{c\chi^2}\right)^2
\left\{
\int\frac{d^3\vk}{\left( 2\pi \right)^3}
\frac{W(\vk)^2}{2\left[ W(\vk)^2\ \left(P_\text{LIM}(\vk) + P_\text{fg}(\vk)\right) + N(\vk) \right]}
\right\}^{-1}
.
\label{eq:variance_matched_filter}
\eeq
In this expression, 
$W(\vk)\equiv \text{PSF}(k_\perp)\ \text{SPSF}(k_\parallel)$
accounts for the convolution of the intensity map by the angular and spectral point-spread functions,
and the total map power spectrum $W^2\ \left(P_\text{LIM} + P_\text{fg}\right) + N$ includes the detector noise and any foreground present in the map.
In Appendix~B of \cite{paper1} and Appendix~\ref{app:2d_matched_filter},
% In Appendices.~\ref{app:angular_clustering} and \ref{app:2d_matched_filter}, 
we derive the analogous expressions for 2D intensity maps instead.
An individual galaxy is detectable if its flux per unit frequency is above a certain number of $\sigma(\hat{F})$.  In what follows, we adopt a $5\,\sigma$ detection threshold, such that the minimum galaxy luminosity detected is:
\beq
L_\text{min} = 
5\ \sigma_{\hat{F}}
\times 4\pi \left[ (1+z)\chi \right]^2.
\eeq
The expression in Eq.~\eqref{eq:variance_matched_filter} can be understood intuitively.
The inverse variance, $\sigma^{-2}_{\hat{F}}$, is a sum over all 3D Fourier modes available in the intensity map.
Indeed, each Fourier mode contributes independent information on the point source flux, such that the overall inverse variance is the sum of the inverse variance from each $\vk$-mode.
For a non-zero detector noise, the angular and spectral point-spread functions act as a cutoff in this integral, such that Fourier modes that are not resolved do not contribute to the overall signal-to-noise. 
As a result, increasing the angular and spectral resolutions can dramatically increase the sensitivity to point sources\footnote{In the absence of detector noise (i.e.\ $N=0$), the point spread functions cancel in the integral, such that resolution no longer matters.
This is also expected, since the Gaussian PSF and SPSF can be deconvolved exactly in the absence of noise.}.

Finally, the variance, $\sigma^{2}_{\hat{F}}$, is a growing function of the total map power spectrum 
$P_\text{total} = W^2\ \left(P_\text{LIM}+P_\text{fg}\right)+N$.
If the detector noise level is dominant, i.e.\ 
$P_\text{total} \sim N$,
the matched filter is detector-noise limited, 
and the detectability of individual sources is independent of the LIM power spectrum.
On the other hand, when the LIM power dominates, i.e.
$P_\text{total} \sim W^2\ P_\text{LIM}$,
the matched filter is confusion-noise limited and
the detectability of point sources becomes independent of the detector noise level.  
In this limit, our ability to detect a point source is limited by the LIM fluctuations due to other, potentially undetected, sources present in the map.
In the final key regime, the map power spectrum is dominated by foregrounds:
$P_\text{total} \sim W^2\ P_\text{fg}$.
This is analogous to the confusion-limited regime above.
Here, what limits our ability to detect a point source at the redshift of interest is no longer the other sources at the same target redshift, but instead other sources at different redshifts.
One example is continuum emission: the continuum emission from galaxies at many different redshifts contaminates the LIM, and these galaxy continua can represent a large source of confusion.

\begin{figure}[h!]
\centering
\includegraphics[width=0.45\textwidth]{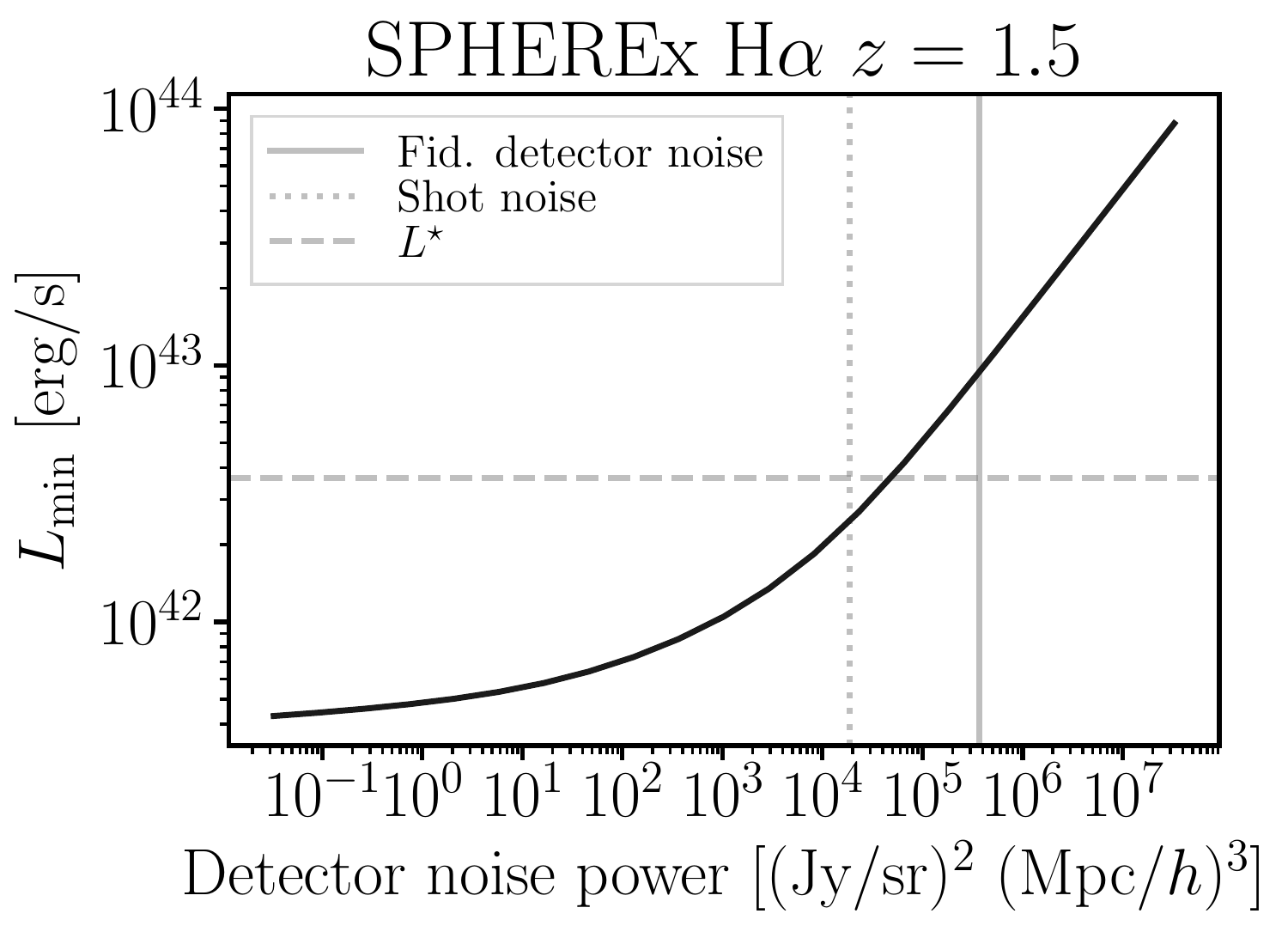}
\includegraphics[width=0.45\textwidth]{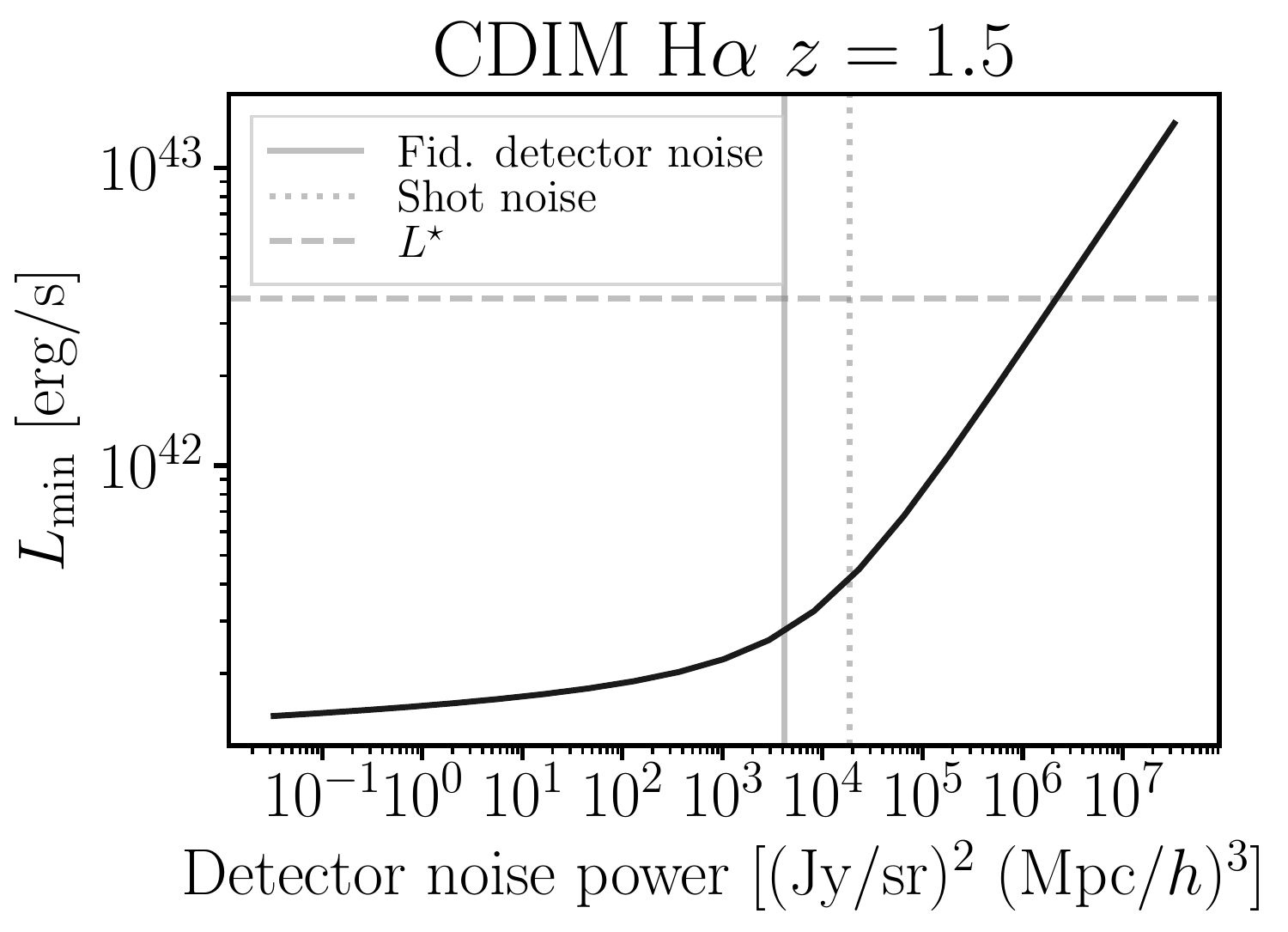}
\caption{
Galaxy luminosity detection threshold in H$\alpha$ for SPHEREx (left) and CDIM (right), as a function of the detector noise.
The fiducial noise level for each experiment is indicated as a vertical gray line, and the shot noise as a vertical dotted line.
Any galaxy with luminosity higher than $L_\text{min}$ is detectable at $5~\sigma$, assuming the detector noise level along the x-axis.
This includes the effect of confusion by other sources in the LIM, but no foregrounds such as continuum emission from other galaxies.
These foregrounds would act as a noise floor.
Many galaxies will be detectable if the luminosity threshold $L_\text{min}$ is smaller or comparable to the Schechter luminosity scale $L^\star$ (horizontal gray line). 
This is the case for CDIM much more so than SPHEREx, mostly due to the different angular resolutions and noise levels.
When the detector noise is below the shot noise level, we are confusion noise limited: the limiting source of noise in the detection of point sources is from the shot noise of other sources.
The minimum luminosity detectable $L_\text{min}$ becomes roughly independent of the detector noise.
}
\label{fig:matched_filter_lmin_mmin_spherex}
\end{figure}

Although Eq.~\eqref{eq:variance_matched_filter} includes the effect of foregrounds, we do not attempt to model their power spectrum, and set them to zero in the figures.
In the absence of foregrounds, the matched filter is therefore confusion-noise (detector-noise) limited for a given Fourier mode if and only if this Fourier mode is signal-dominated (noise-dominated) in the observed intensity map.
We illustrate the detector noise-dominated and confusion noise-dominated regimes in the case of H$\alpha$ maps from SPHEREx in Fig.~\ref{fig:matched_filter_lmin_mmin_spherex}.

\begin{figure}[h!]
\centering
\includegraphics[width=0.45\textwidth]{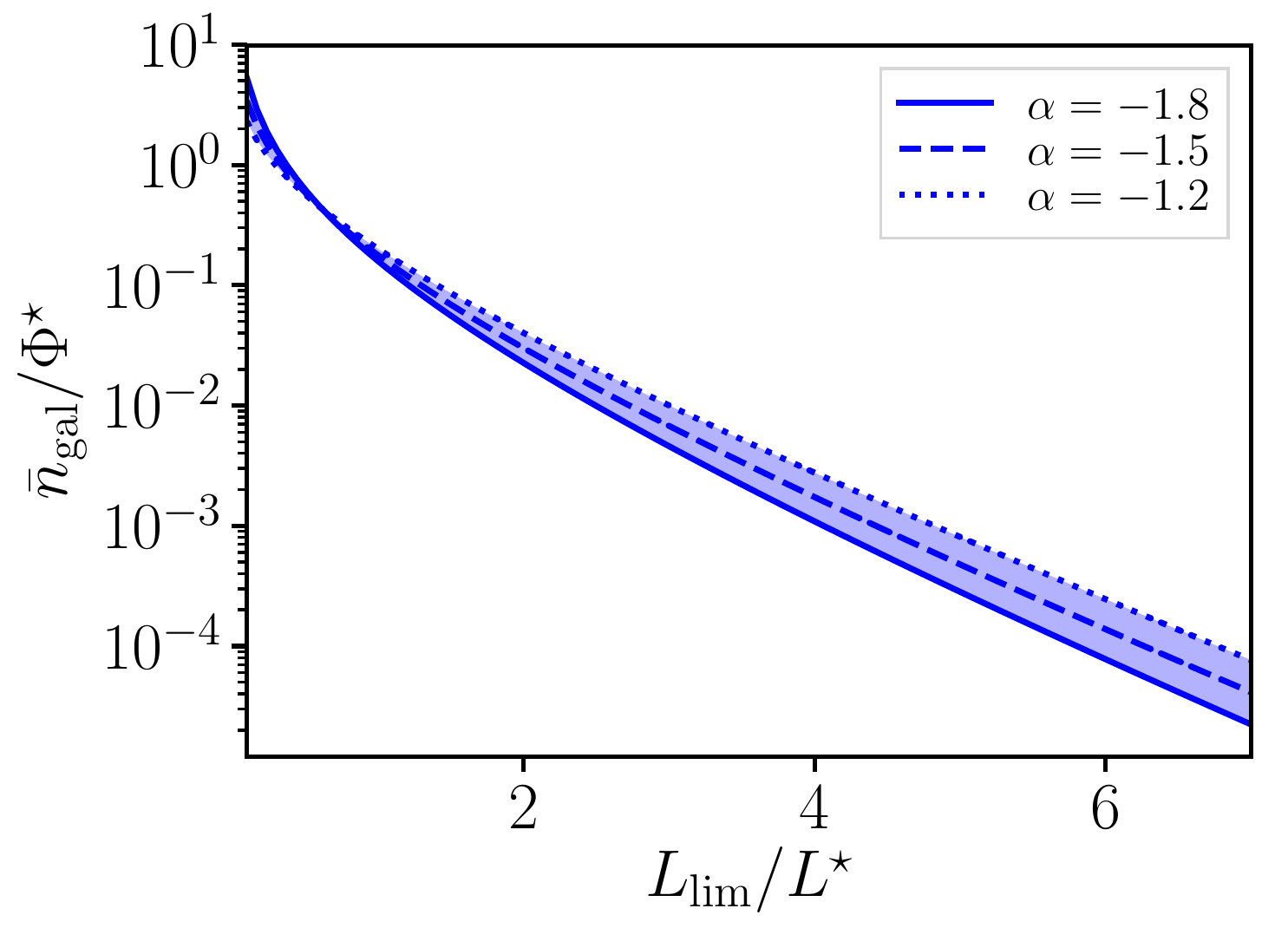}
\includegraphics[width=0.45\textwidth]{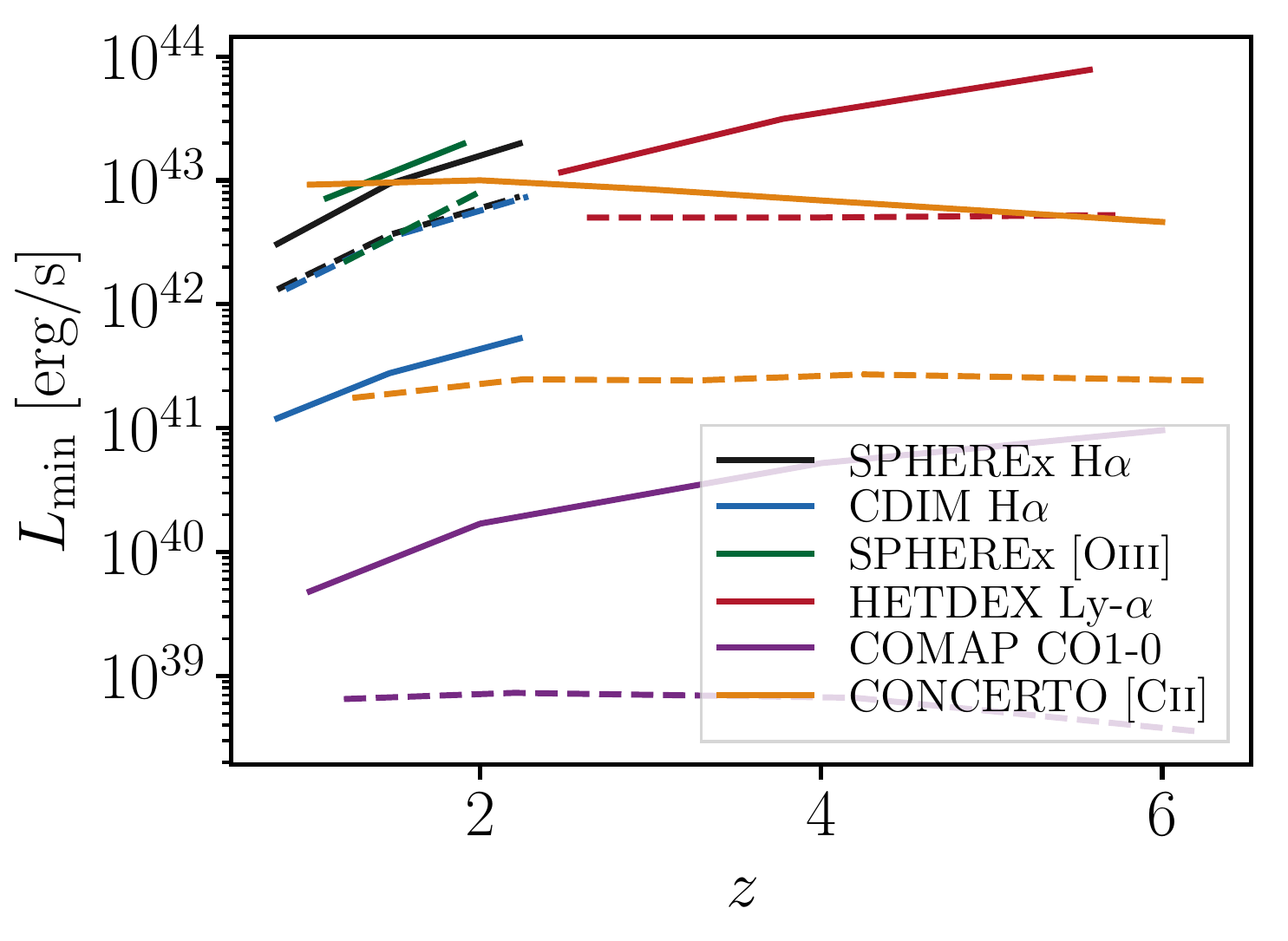}
\caption{
\textbf{Left:}
For a Schechter LF, the number density of galaxies brighter than the luminosity threshold $L_\text{min}$ is a function of $L_\text{min}/L^\star$ and $\alpha$.
It varies slowly with the power law index $\alpha$, for the values encountered here.
However, it decreases exponentially with $L_\text{min}/L^\star$.
Thus the question of the detectability of galaxies is approximately answered by comparing the detection threshold $L_\text{min}$ to the Schechter scale luminosity $L^\star$.
\textbf{Right:}
We thus compare the luminosity detection threshold $L_\text{min}$ for each experiment (solid lines) to the Schechter luminosity scales $L^\star$ (dashed lines), as a function of redshift.
HETDEX, CDIM and SPHEREx thus have the ability to detect a non-negligible number of point sources, whereas COMAP and CONCERTO won't.
}
\label{fig:schechter_ngal}
\end{figure}

Are there any galaxies bright enough to detect, given the luminosity threshold $L_\text{min}$?
For Schechter LFs, the answer can be expressed analytically in terms of $L_\text{min}/L^\star$ and the Schechter index $\alpha$.
If
\beq
\Phi(L) = \frac{\Phi^\star}{L^\star}
\left( \frac{L}{L^\star} \right)^\alpha
e^{-L/L^\star},
\eeq
then the mean number density of galaxies bright enough to be detected is simply:
\beq
\bar{n}_\text{gal} (L\geq L_\text{min})
=
\Phi^\star\
\Gamma_u(\alpha+1, L\text{min}/L^\star),
\eeq
where the upper incomplete gamma function is 
$\Gamma_u(a, x) \equiv \int_x^\infty dt\ t^{a-1} e^{-t}$.
As shown in Fig.~\ref{fig:schechter_ngal} (left panel), the answer is basically obtained by comparing $L_\text{min}$ and $L^\star$.
For $L_\text{min}$ greater than a few $L^\star$, the number density of galaxies detected drops dramatically.
We compare $L_\text{min}$ (solid lines) to $L^\star$ (dashed lines) for each experiment in Fig.~\ref{fig:schechter_ngal} (right panel).
From this comparison, we do not expect COMAP and CONCERTO to detect a significant number of point sources at the redshift of interest.
For them, the LIM observables are thus trivially the best approach to extract astrophysical and cosmological information.
On the other hand, we find that HETDEX, CDIM and SPHEREx should detect a significant number of point sources.
We thus investigate these experiments further and compare the properties of the LIM observables and the galaxy catalog.
Again, foregrounds may change this conclusion, by increasing the effective map noise. This can be fixed by including their power spectrum in Eq.~\eqref{eq:variance_matched_filter}.

\subsubsection{Minimum halo mass: Kennicutt-Schmidt relation}

The minimum luminosity derived above applies to any non-resolved source in the map.
As we show in Figs.~\ref{fig:tradeoff_a_co_comap}-\ref{fig:tradeoff_a_cii_concerto}, LIM experiments do not necessarily have the angular resolution to see the turn over of the 1-halo term.
In that case, halos can thus be considered point-like.
The luminosity detection threshold is thus also a halo luminosity threshold, which we can convert into halo mass.
This conversion is only approximate, however.
First, the scatter between halo mass and halo luminosity means that a strict luminosity threshold is not a sharp mass threshold.
Second, some LIM experiments (e.g. SPHEREx) actually do resolve halos, such that they are not strictly point-like. In that case, the halo detection threshold estimated here is typically underestimated.
Nonetheless, this approach is a useful starting point.

\begin{figure}[h!]
\centering
\includegraphics[width=0.32\textwidth]{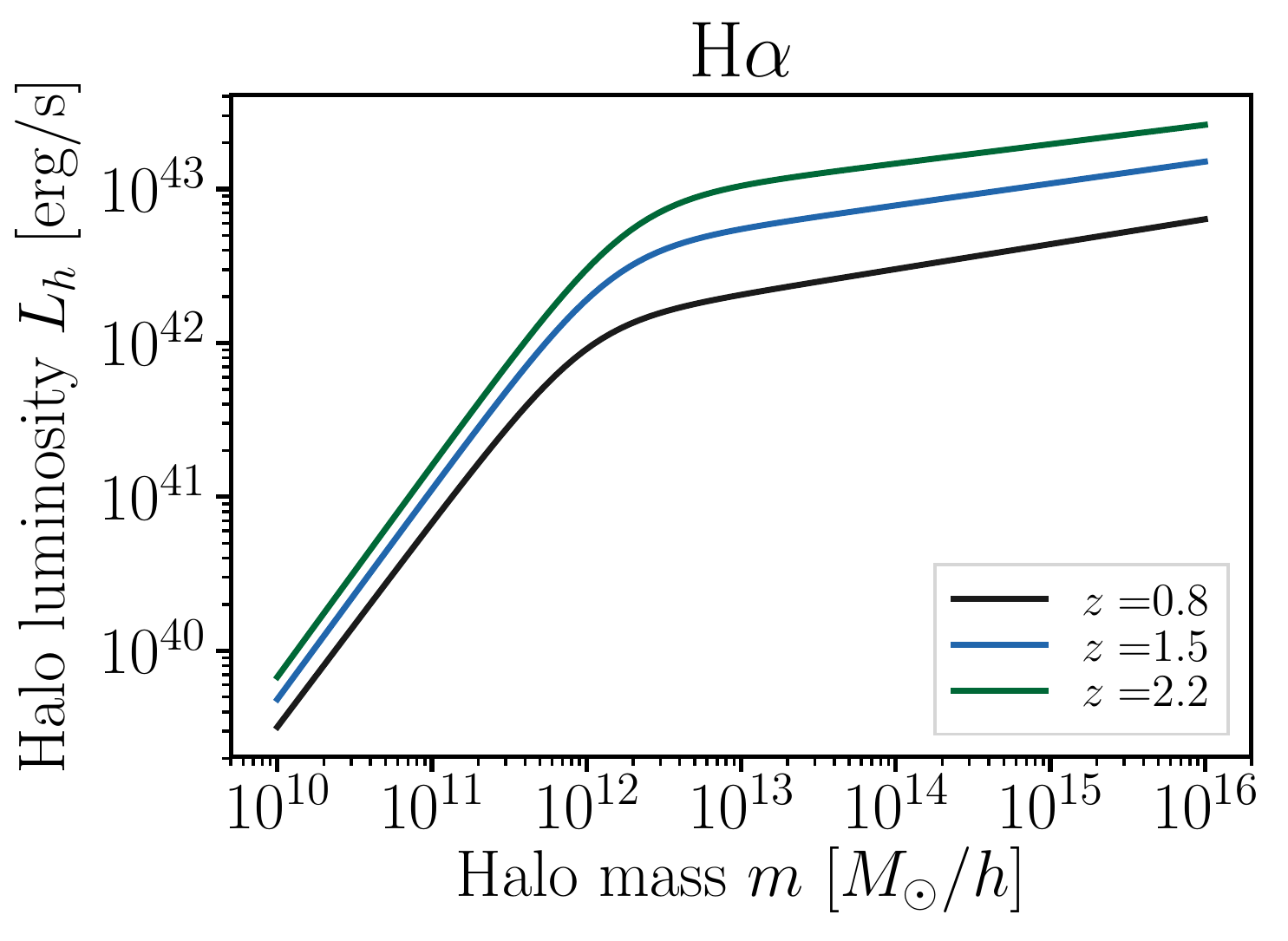}
\includegraphics[width=0.32\textwidth]{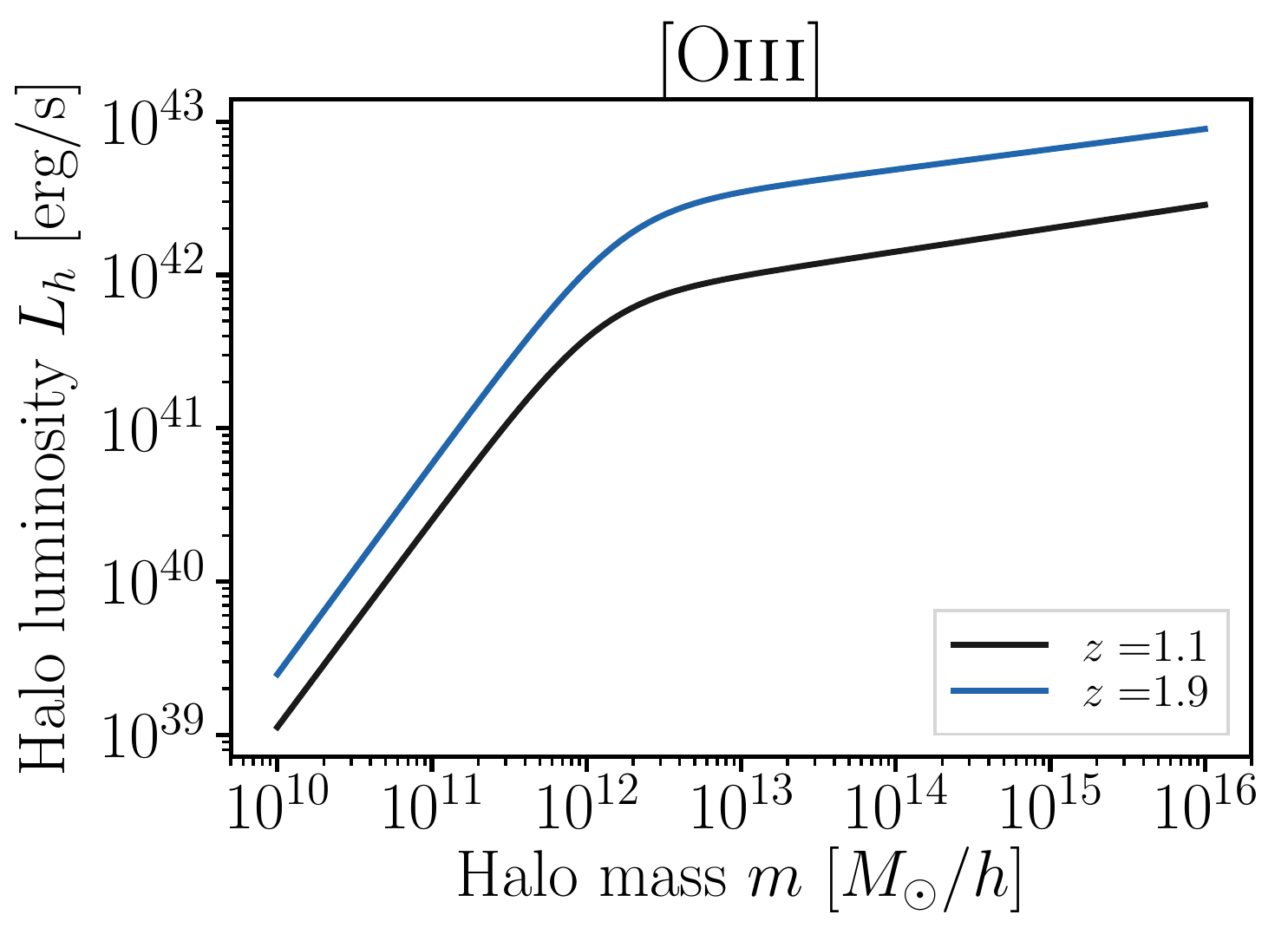}
\includegraphics[width=0.32\textwidth]{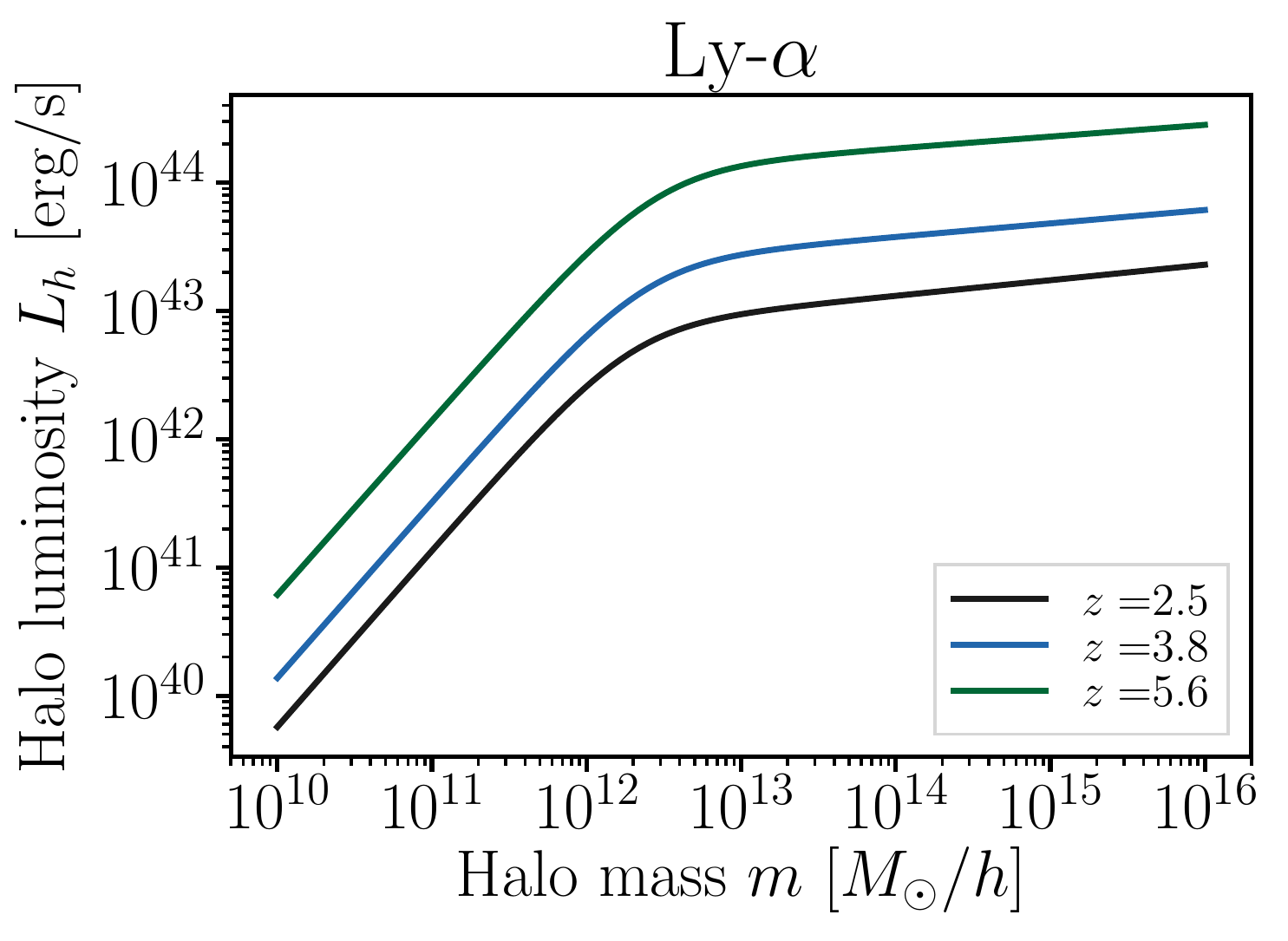}
\includegraphics[width=0.45\textwidth]{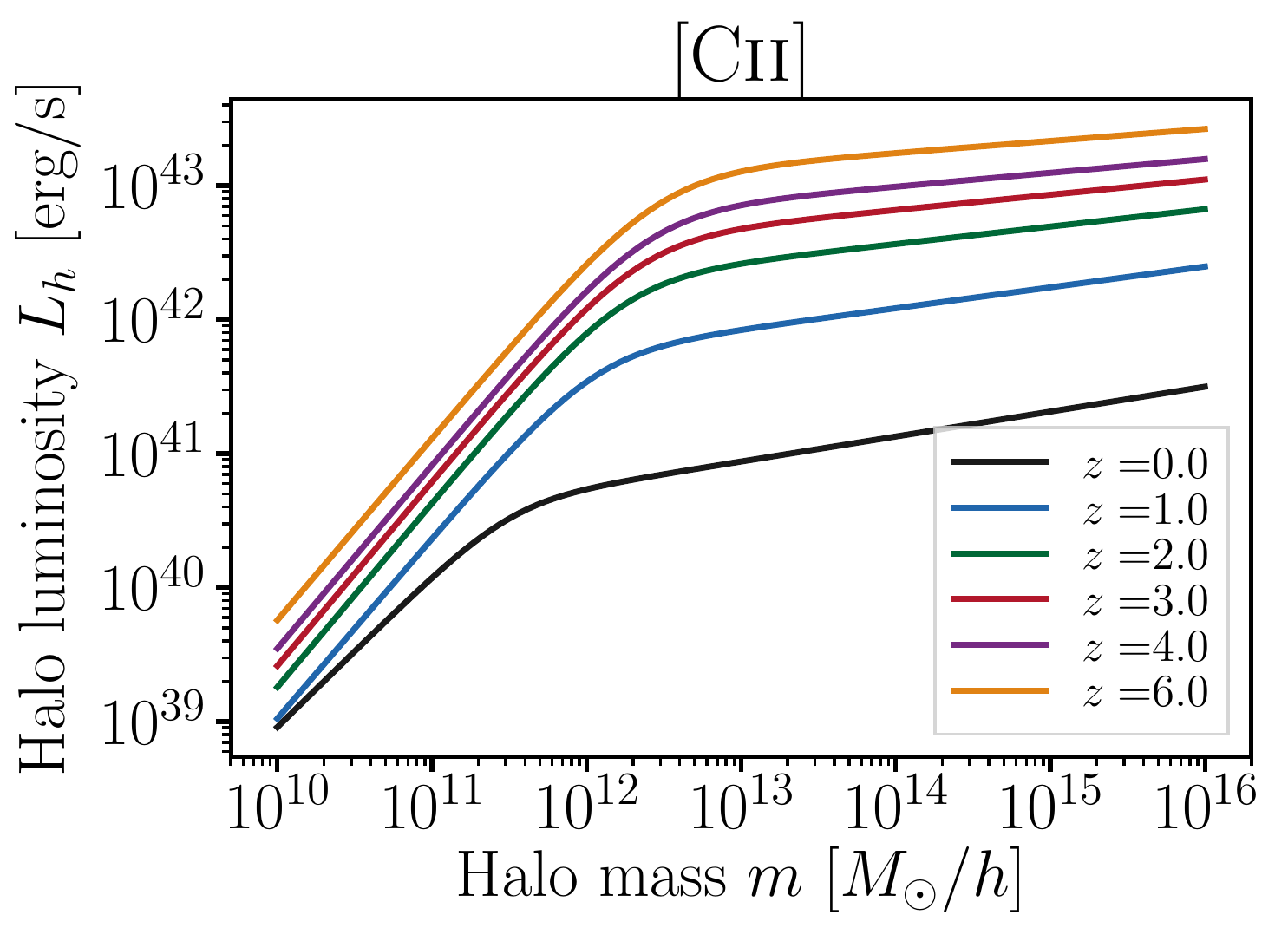}
\includegraphics[width=0.45\textwidth]{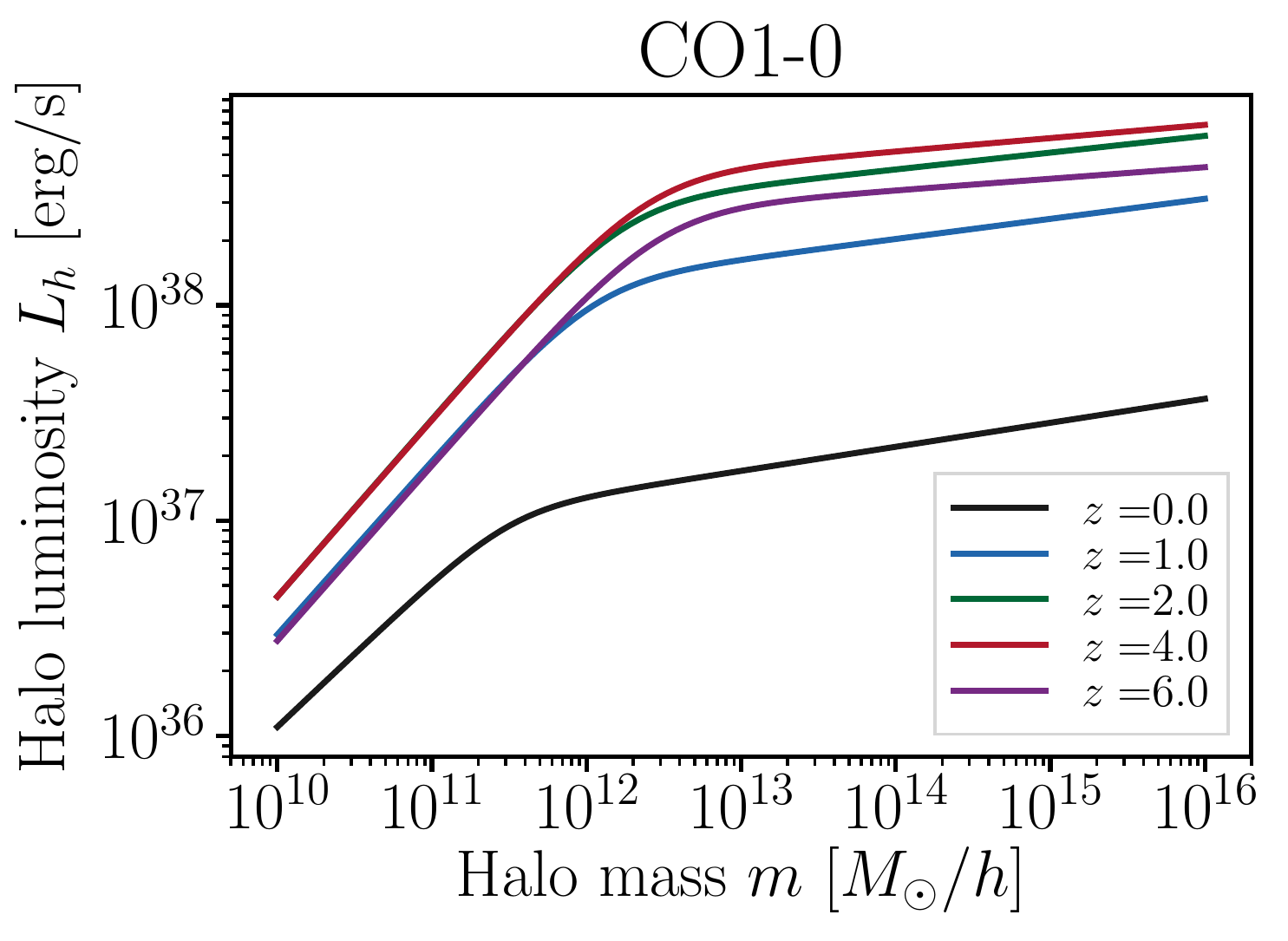}
\caption{
Relationship between halo mass and halo luminosity, from the Kennicutt-Schmidt relation $L_h(m) = K\times \text{SFR}^\gamma(m)$.
Here, $\gamma=1$ for H$\alpha$, [O{\sc iii}], Ly-$\alpha$ and [C{\sc ii}], and $\gamma=0.6$ for CO.
The constant $K$ is inferred for each line and redshift by matching the mean intensity to that inferred from the observed galaxy LF.
}
\label{fig:halo_luminosity}
\end{figure}

In practice, we convert halo luminosity to halo mass using the mean relation
$L_h(m) = K\times \text{SFR}^\gamma(m)$,
where again $\gamma=1$ for H$\alpha$, [O{\sc iii}], Ly-$\alpha$ and [C{\sc ii}], and $\gamma=0.6$ for CO.
This is consistent with the CLF ansatz throughout this paper.
We estimate the Kennicutt-Schmidt constant $K$ by matching the mean intensity inferred from the galaxy luminosity function:
\beq
\int dm\ n(m)\ K\ \text{SFR}^\gamma(m)
=
\int dL\ \Phi(L)\ L
\quad\text{i.e.}\quad
K=
\frac{\int dL\ \Phi(L)\ L}{\int dm\ n(m)\ \text{SFR}^\gamma(m)}.
\eeq
The resulting relation between halo mass and luminosity is shown in Fig.~\ref{fig:halo_luminosity} for the various lines considered.

\begin{figure}[h!]
\centering
\includegraphics[width=0.45\textwidth]{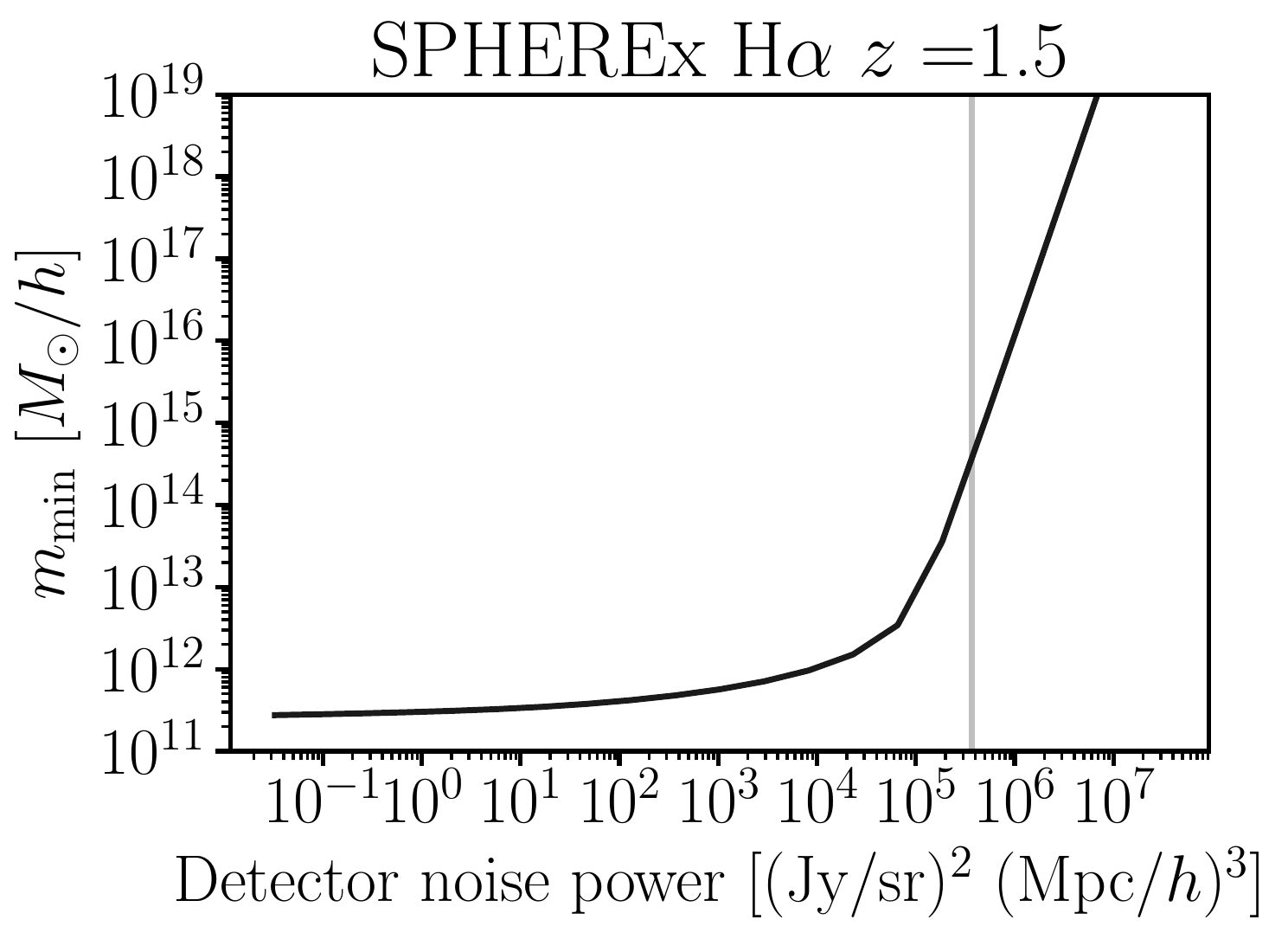}
\includegraphics[width=0.45\textwidth]{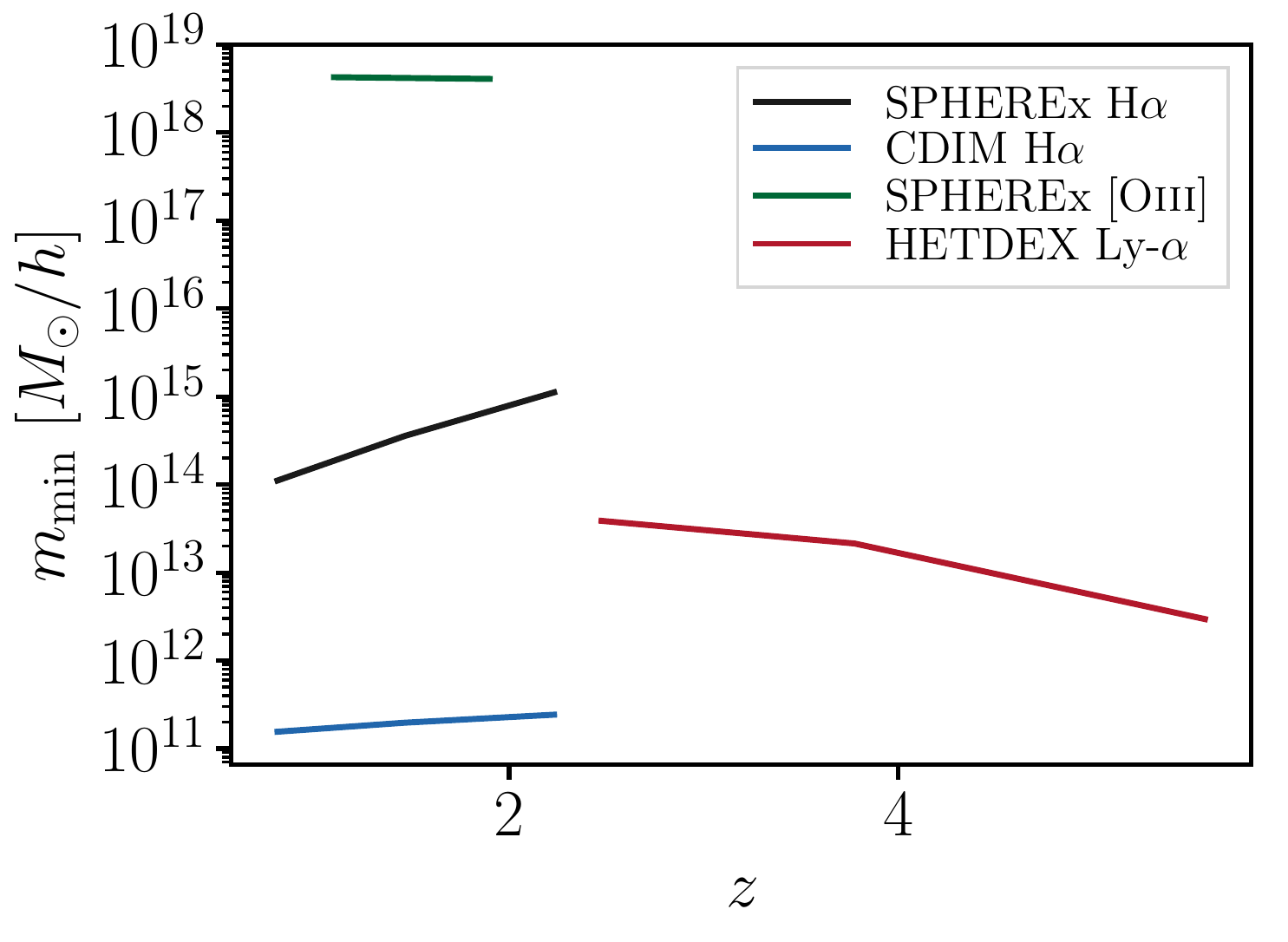}
\caption{
\textbf{Left:} Minimum halo mass detectable in H$\alpha$ with SPHEREx as a function of the detector noise level.
The fiducial noise level is indicated by the vertical gray line.
This curve shows again the confusion-dominated regime, for low detector noise values, and the detector noise-dominated regime, when the latter is large.
\textbf{Right:} Minimum halo mass detectable as a function of redshift for each experiment, assuming their fiducial noise levels.
The mass thresholds for COMAP and CONCERTO, extremely high, are not shown: essentially no halo is bright enough to be detected.
Again, any foreground present in the map will increase the effective map noise over the fiducial value.
}
\label{fig:matched_filter_lmin_mmin_all}
\end{figure}

Finally, the minimum halo mass detectable is inferred by solving:
\beq
K\times \text{SFR}^\gamma(m_\text{min}) = L_\text{min}.
\eeq
The result is shown as a function of detector noise for H$\alpha$ maps from SPHEREx in Fig.~\ref{fig:matched_filter_lmin_mmin_all} (left panel). This highlights again the confusion and detector noise-dominated regimes for low and high noise values.
We show the mass detection threshold as a function of redshift for SPHEREx, CDIM and HETDEX in Fig.~\ref{fig:matched_filter_lmin_mmin_all} (right panel), assuming their fiducial detector noise levels.
We do keep in mind that potential foregrounds will act to enhance the effective total map noise, over the fiducial values assumed here.
The mass detection thresholds are not shown for COMAP and CONCERTO, as they would be unphysically high ($M_\text{min} \geq 10^{17} M_\odot$): essentially no halo at the redshift of interest is bright enough to be detected.

In summary, the minimum galaxy luminosity detectable depends on the detector noise and the confusion noise, from sources in the LIM itself or from foregrounds.
The corresponding halo mass threshold is then determined by the minimum luminosity threshold, and the power law index in the relation between SFR and halo luminosity.
Combined with our halo model, the mass and luminosity thresholds for detection are all we need to compare the properties of the catalog of detected galaxies to the intensity map.

%%%%%%%%%%%%%%%%%%%%%%%%%%%%%%%%%%%%%%%%%%%%%%%%%%%%%%%%%%
\subsection{Sensitivity to galaxies below the detection threshold}
\label{sec:sensitivity_undetected_sources}

What fraction of the LIM observables is produced by galaxies too faint to detect individually?
If undetected sources produce a large fraction of the LIM observables, then we conclude that LIM does contain additional information, over the catalog of bright galaxies.
Ref.~\cite{Silva17} addresses this question for the mean H$\alpha$ intensity, and we extend this work to other lines and to the LIM power spectrum.
We thus answer this question for the mean intensity and galaxy shot noise, using the luminosity threshold $L_\text{min}$, and for the 2-halo and 1-halo terms using the mass threshold $m_\text{min}$.

\subsubsection{Mean intensity and shot noise: $L_\text{min}$ VS $L^\star$}

\begin{figure}[h!]
\centering
\includegraphics[width=0.45\textwidth]{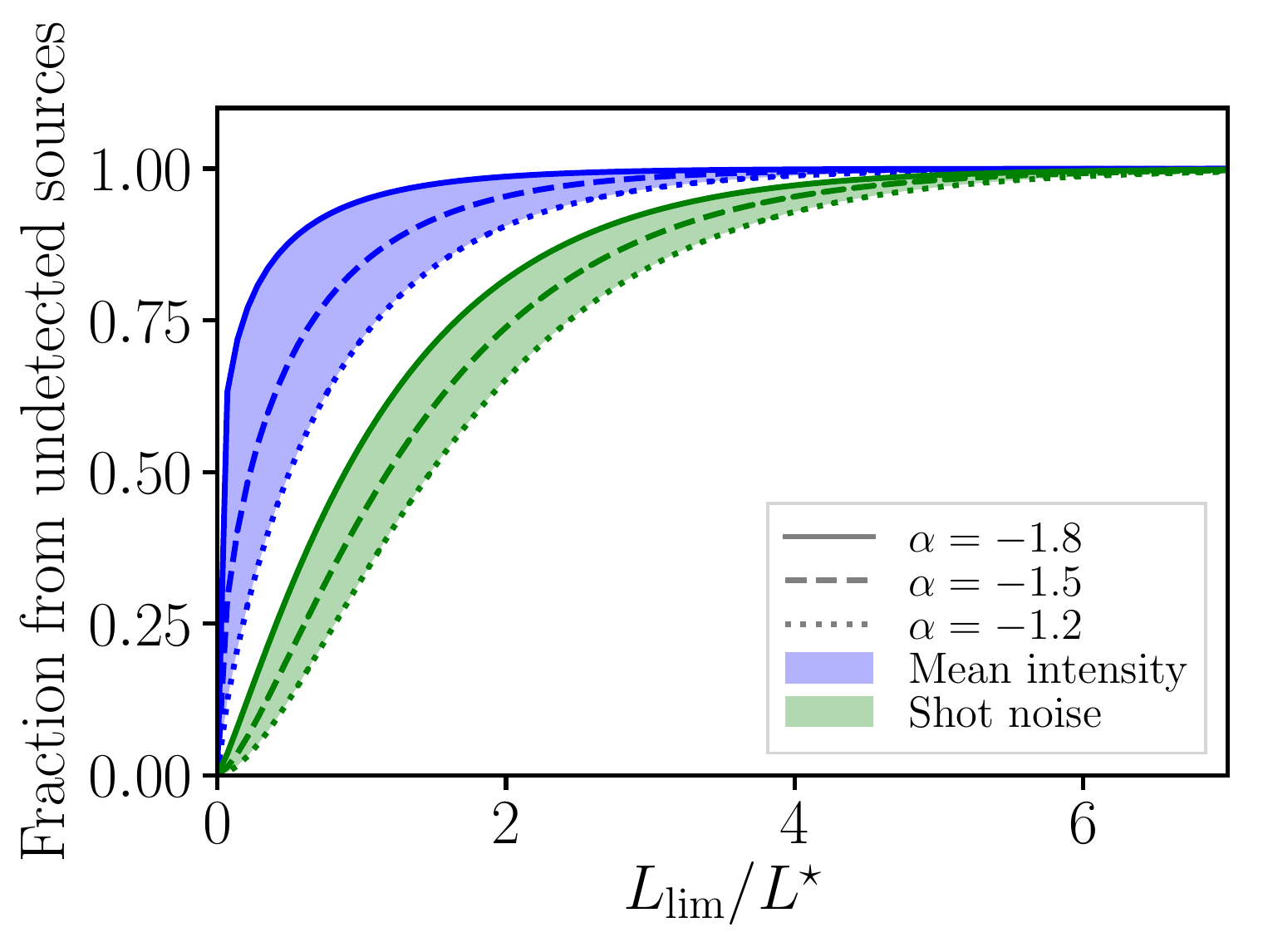}
\caption{
Fraction of the luminosity moments produced by sources below the detection threshold $L_\text{min}$, for Schechter LFs with luminosity scale $L^\star$ and power index $\alpha$.
The result varies fast with $L_\text{min}/L^\star$, such that it is mostly determined by comparing $L_\text{min}$ to $L^\star$.
}
\label{fig:schechter_frac_undetected}
\end{figure}

Given the luminosity threshold $L_\text{min}$, we compute the contribution of undetected galaxies to the mean intensity and shot noise power spectrum.
For Schechter LFs, the answer can be expressed again in terms of $L_\text{min}/L^\star$ and the Schechter index $\alpha$:
\beq
\text{Fraction of $n$-th moment from undetected sources}
=
\frac{\Gamma_l(\alpha+n+1, L_\text{min}/L^\star)}{\Gamma(\alpha+n+1)},
\eeq
where the Euler gamma function is
$\Gamma(a) \equiv \int_0^\infty dt\ t^{a-1}e^{-t}$
and the lower incomplete gamma function is
$\Gamma_l(a, x) \equiv \int_0^x dt\ t^{a-1}e^{-t}$.
As shown in Fig.~\ref{fig:schechter_frac_undetected}, these functions vary slowly with $\alpha$ for the range of values relevant to our LFs, and fast with $L_\text{min}/L^\star$. 
Thus the fraction of the luminosity moment from undetected sources is mostly determined by comparing $L_\text{min}$ to $L^\star$.

\subsubsection{2-halo and 1-halo terms}

\begin{figure}[h!]
\centering
\includegraphics[width=0.45\textwidth]{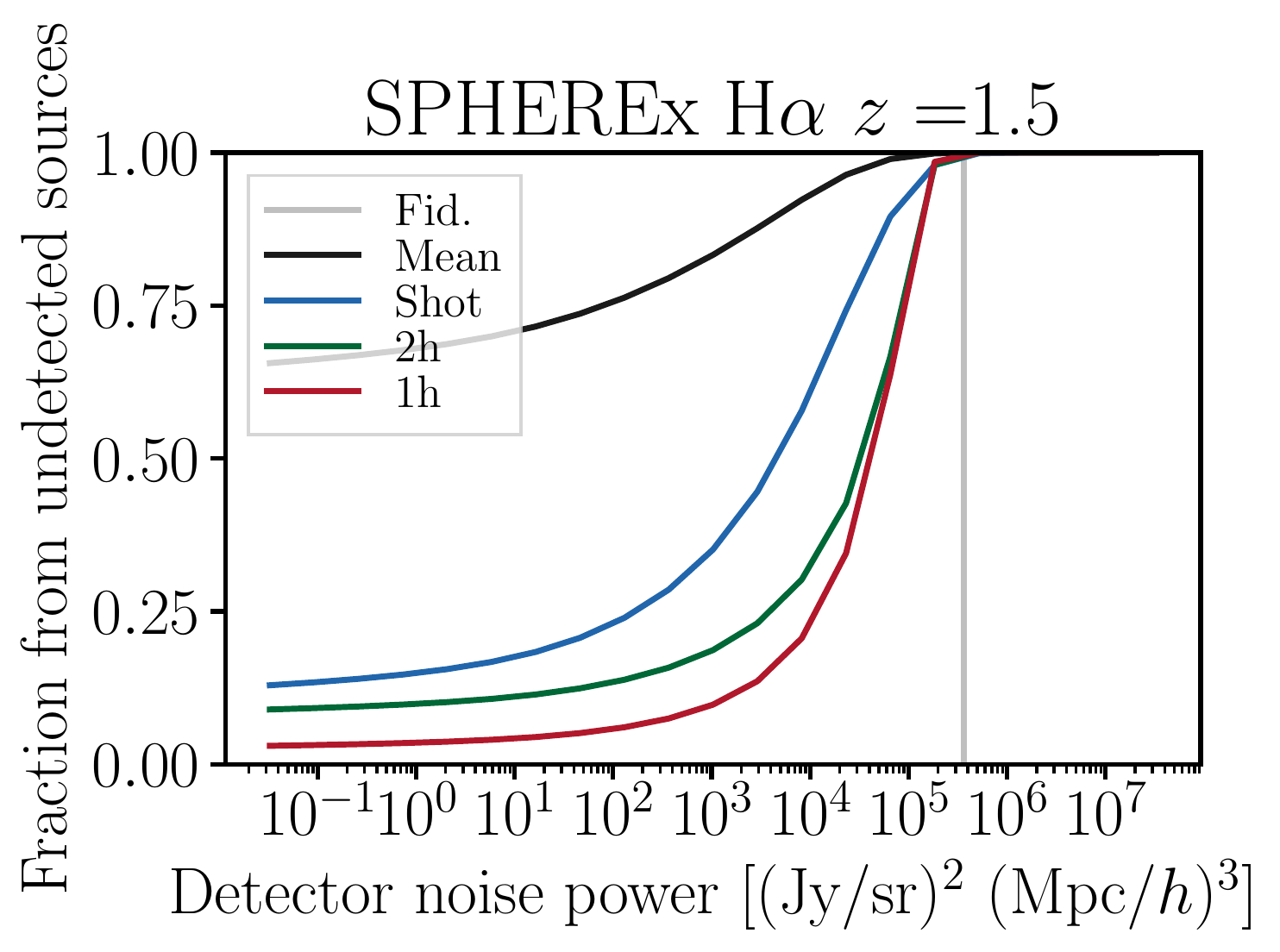}
\includegraphics[width=0.45\textwidth]{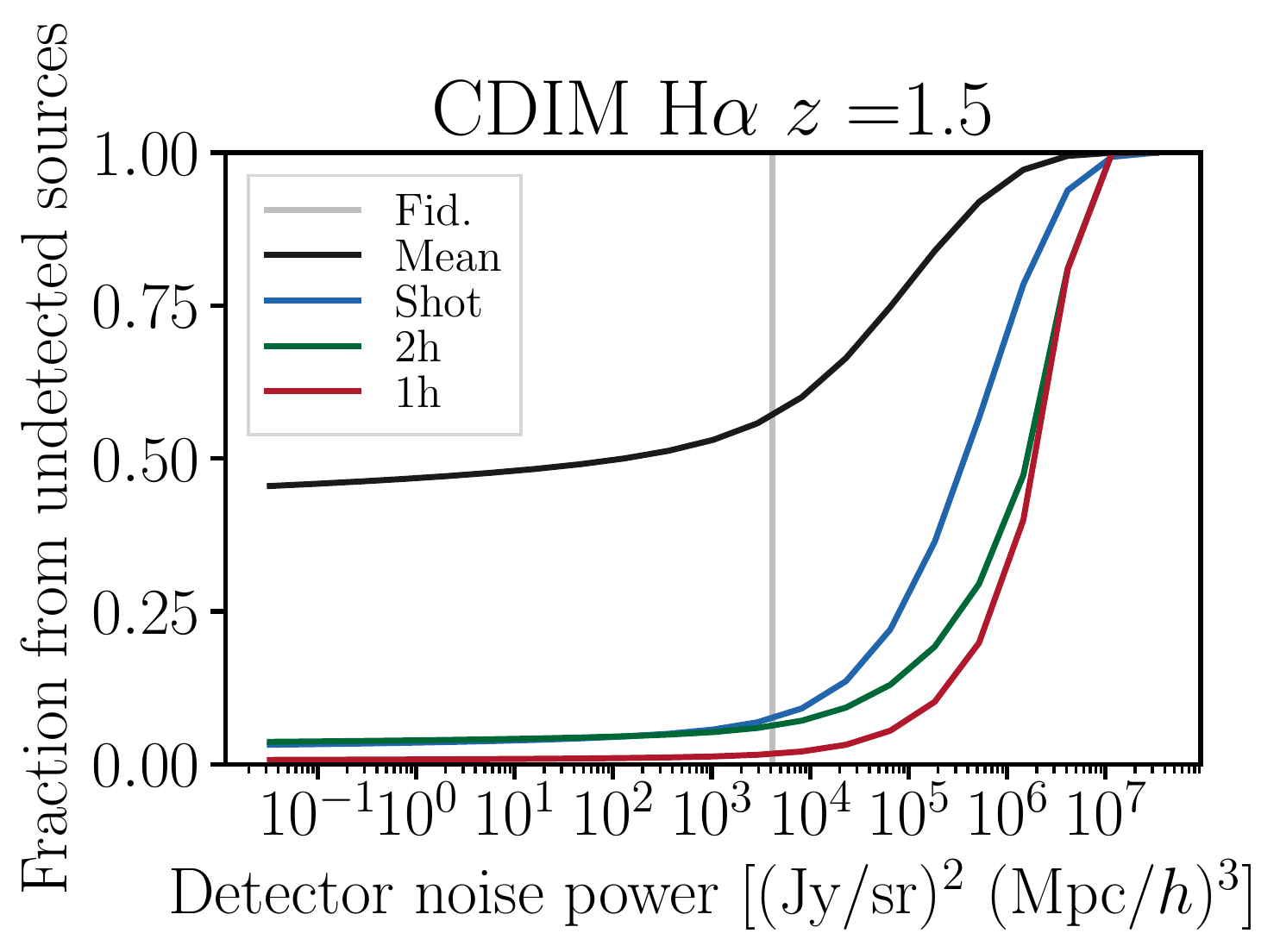}
\caption{
What fraction of the LIM observables comes from undetected sources?
The fraction of the mean intensity, shot noise, 1-halo and 2-halo terms coming from sources too faint to detect, are shown and contrasted for SPHEREx (left) and CDIM (right).
No foregrounds (e.g., continuum emission from other galaxies) were included: they would enhance the effective map noise.
As expected, as the detector noise increases, fewer sources are individually detected, and most of the LIM signal comes from faint, undetected galaxies.
The fiducial detector noise levels for the two experiments are shown as vertical gray lines. 
While most of SPHEREx observables comes from undetected sources, this is not the case for CDIM, due to the different angular resolution and detector noise.
}
\label{fig:frac_undetected_spherex}
\end{figure}

Given the mass threshold $m_\text{min}$, we ask the same question for the 2-halo and 1-halo terms: what fraction comes from undetected halos?
Comparing their values with and without the mass cutoff $m_\text{min}$,
we show the result for H$\alpha$ maps from SPHEREx in Fig.~\ref{fig:frac_undetected_spherex}, as a function of the detector noise.
The fraction of the power spectrum coming from undetected sources may vary as a function of scale.
We focused on $k=0.01 h/$Mpc for the 2-halo term and $k=0.1 h/$Mpc for the 1-halo term, but the calculation can be reproduced for any desired wavevector $k$.

As expected, the fraction from undetected sources is higher for the mean intensity than the shot noise, since the shot noise upweighs bright galaxies.
The 2-halo term differs from the mean intensity in two ways. Both are determined by a similar mass integral (with an additional halo bias for the 2-halo term), however this integral is squared for the 2-halo term. As a result, the fraction of the 2-halo term from undetected sources is much smaller than for the mean intensity.
Because the 1-halo term upweighs massive halos compared to the 2-halo term, its fraction from undetected sources is smaller, as expected.

\begin{figure}[h!]
\centering
\includegraphics[width=0.45\textwidth]{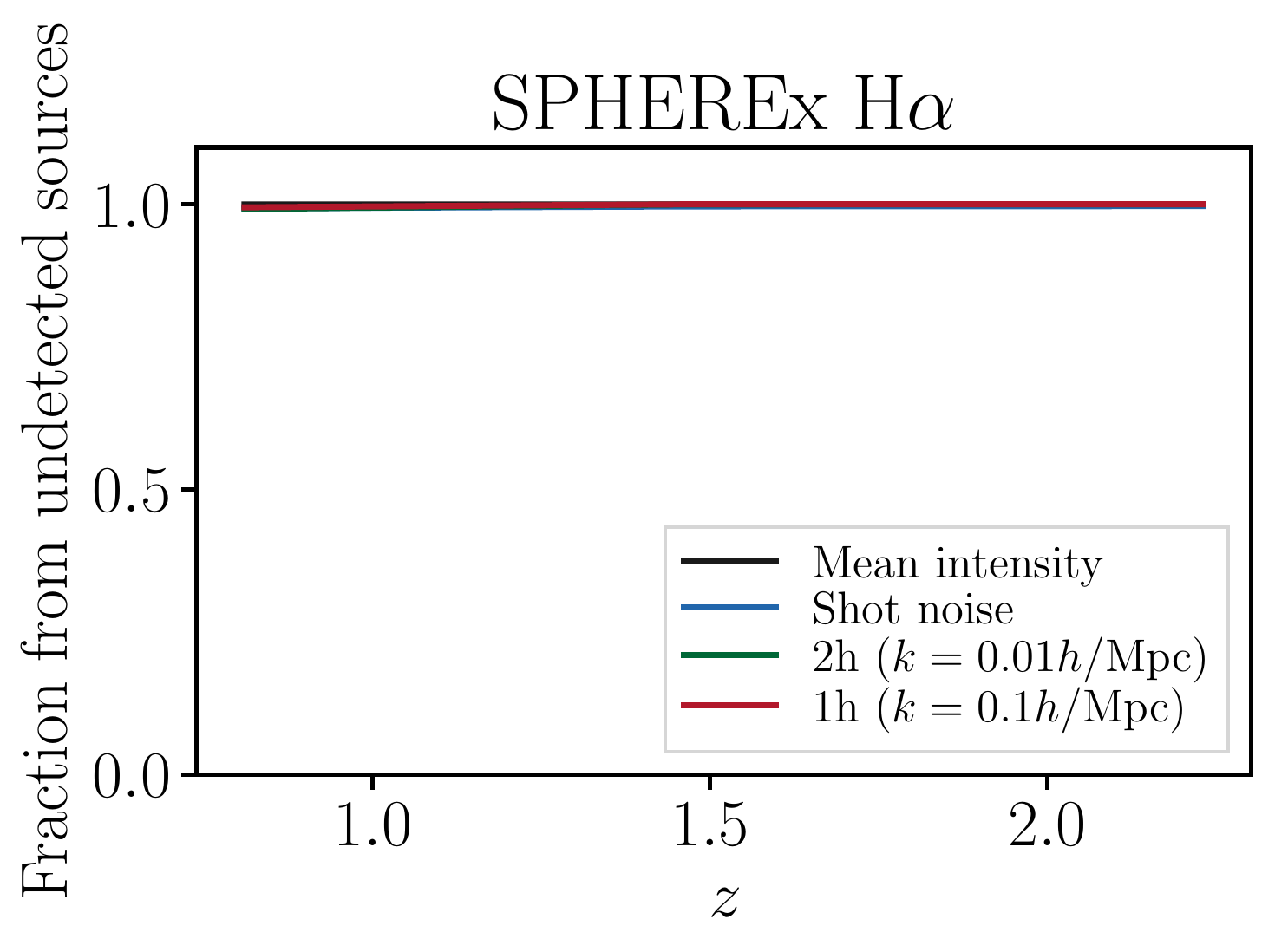}
\includegraphics[width=0.45\textwidth]{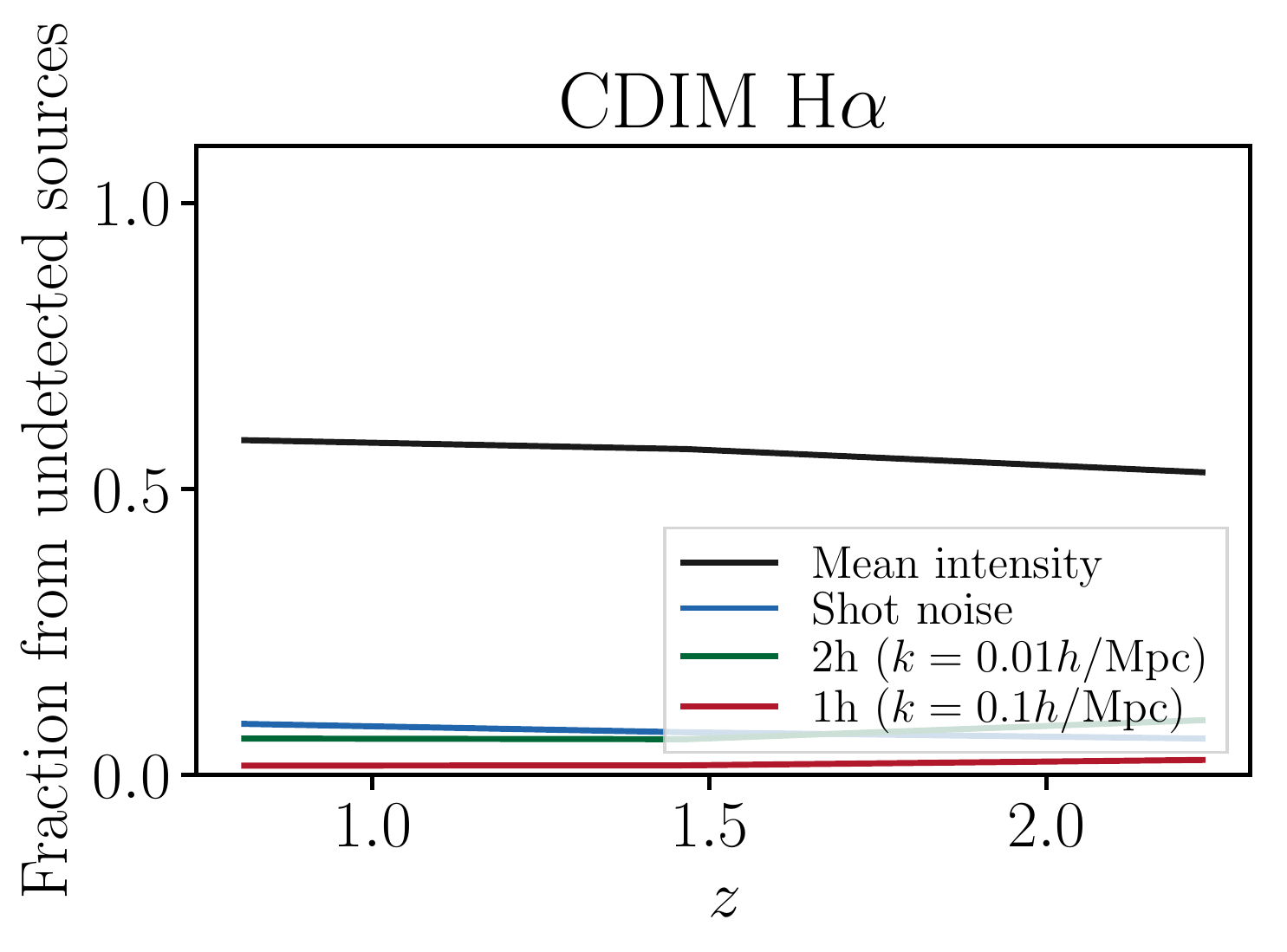}
\includegraphics[width=0.45\textwidth]{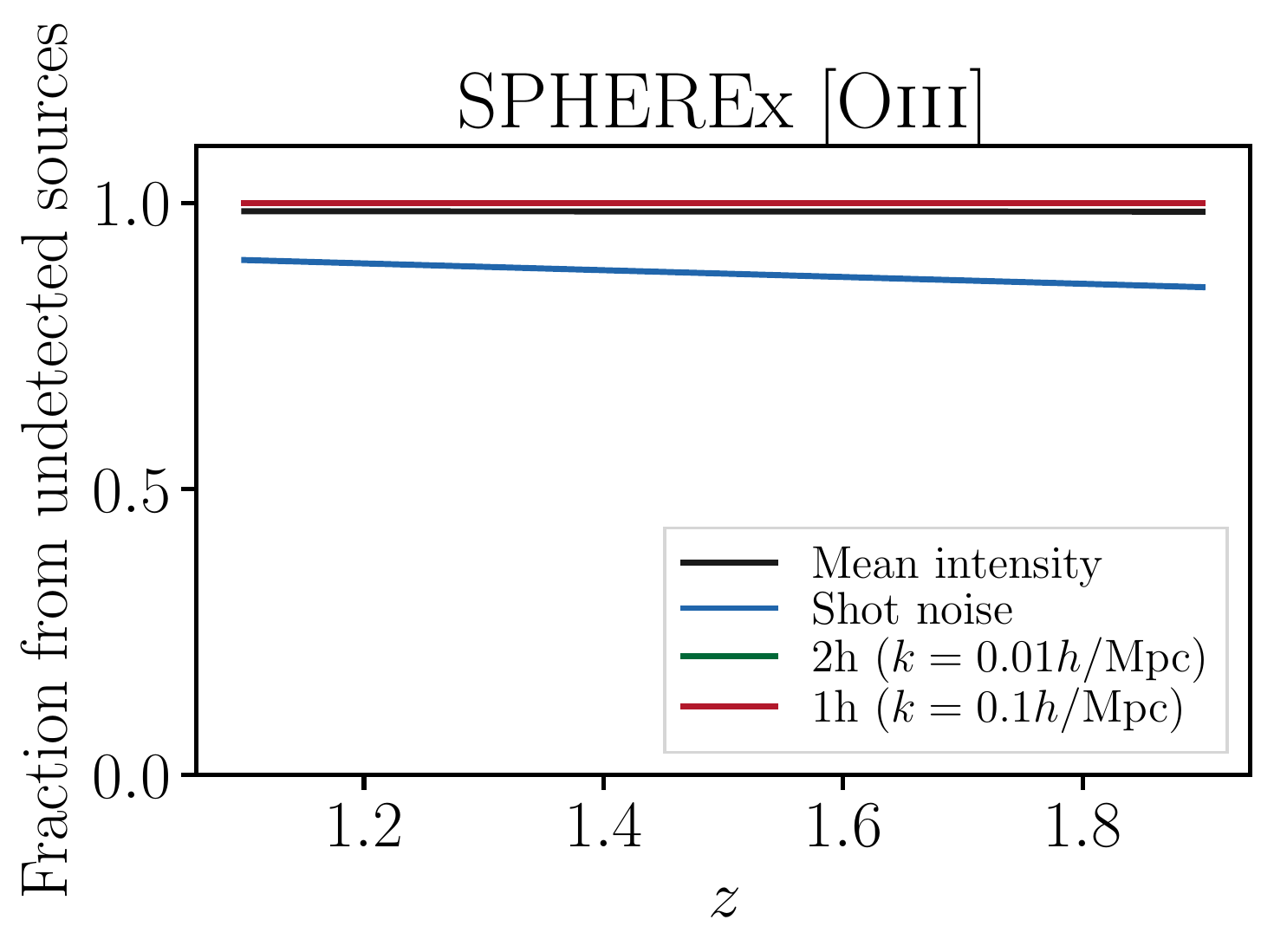}
\includegraphics[width=0.45\textwidth]{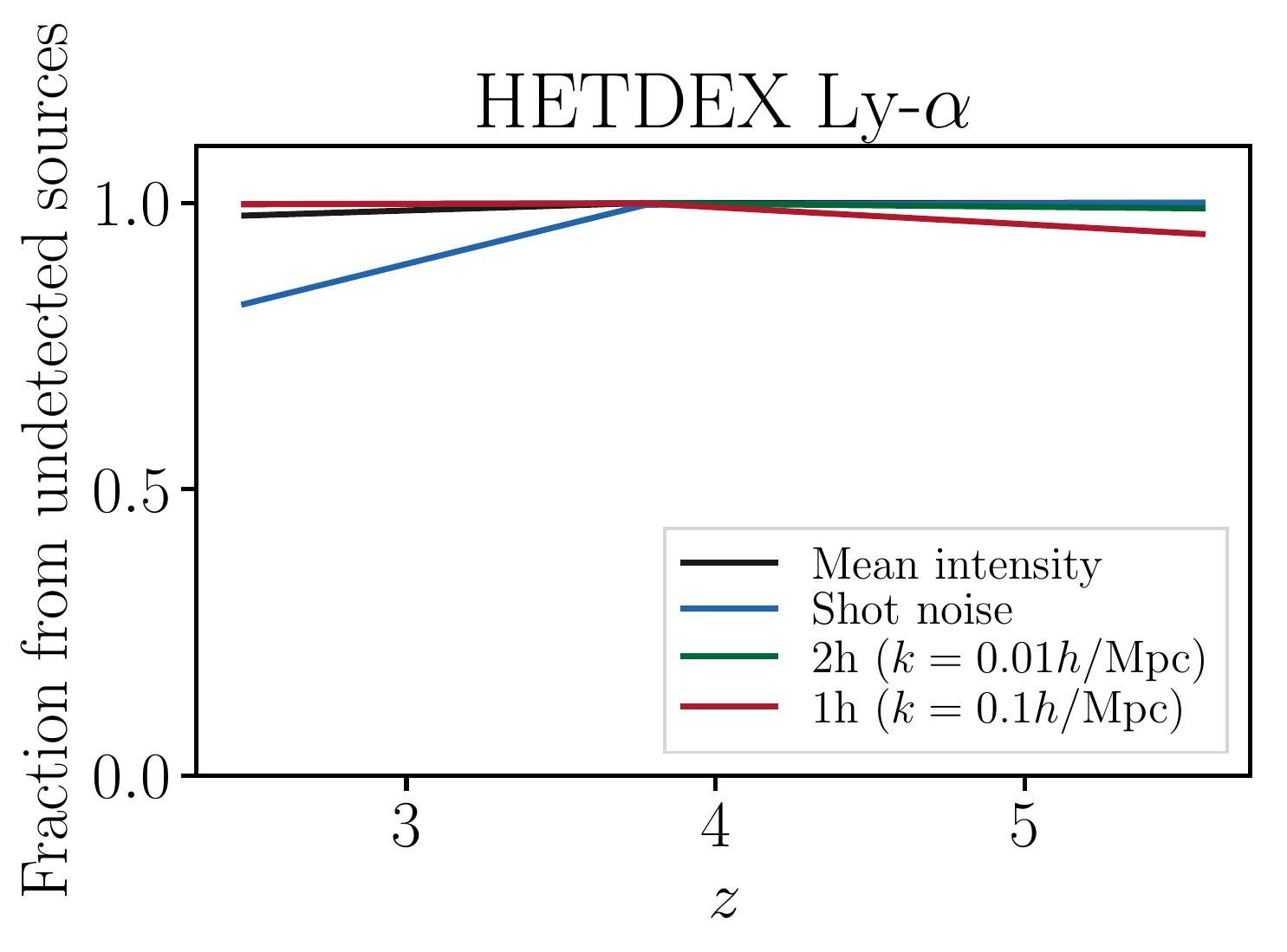}
\caption{
What fraction of the LIM observables comes from undetected sources?
We answer this question as a function of redshift for H$\alpha$
from SPHEREx (top left) and CDIM (top right),
as well as SPHEREx [O{\sc iii}] (bottom left)
and HETDEX Ly-$\alpha$ (bottom right).
For the SPHEREx-like and HETDEX-like experiments, most of the LIM observables are due to undetected sources. This implies that the LIM observables carry independent astrophysical information from the catalog of detected sources.
On the other hand, for the CDIM-like experiment, almost none of the power spectrum (2-halo, 1-halo and shot noise) comes from undetected sources: the LIM power spectrum thus contains no additional astrophysical information compared to the catalog of detected sources.
}
\label{fig:frac_undetected_all}
\end{figure}

Keeping only the fiducial detector noise values for each experiments, we show the same results for each experiment in Fig.~\ref{fig:frac_undetected_all}
We focus on SPHEREx, CDIM and HETDEX, since they detect a significant number of point sources at the target redshifts.
Again, the assumed specifications follow Table~\ref{tab:exp_specs} and the detector noise levels are described in Appendix.~\ref{app:experimental_detector_noise}.
Interestingly, the different experiments are in different regimes.
For SPHEREx, the H$\alpha$ LIM observables are exclusively produced by undetected sources,
and this is also the case for most [O{\sc iii}] observables.
This implies that SPHEREx LIM observables do contain valuable information about faint galaxies, beyond the catalog of detected sources.
The Ly-$\alpha$ LIM observables from HETDEX are in the same regime.
On the other hand, the CDIM H$\alpha$ power spectrum terms are mostly due to resolved sources, and about half of the H$\alpha$ mean intensity is.
Thus, H$\alpha$ LIM from CDIM only contains limited new information about faint galaxies, compared to the catalog of detected sources.
We only interpret fractions close to zero or to unity, as the precise value of any intermediate fraction is sensitive to uncertainties in the LIM modeling, and to small changes in the detector noise.

In summary, whether the LIM observables indeed probe galaxies too faint to detect individually depend on the experiment (resolution and detector noise), the line considered (through $L^\star$) and redshift.
Our analysis provides a general way to answer this question, for any set of experimental specifications.
Although foregrounds may change our conclusions, they can be straightforwardly included in the analysis if their power spectra are known.

%%%%%%%%%%%%%%%%%%%%%%%%%%%%%%%%%%%%%%%%%%%%%%%%%%%%%%%%%%
\subsection{Comparing tracers of the matter density}
\label{sec:lim_vs_gal_matter_tracer}

The intensity map and the catalog of bright sources detected in it constitute two probes of the large-scale structure in the Universe.
Which of the two is the better tracer of the matter density field?
This question was explored in detail at the field level in ref.~\cite{Cheng19}, for a given real-space pixel of an intensity map.
They found LIM to be a better estimator of the matter density field in the regimes of high detector noise or high confusion.
Here, we wish to compare LIM and individual galaxy detection for a given Fourier wavevector, rather than real-space pixel.
This is useful because the comparison between the two probes can be scale-dependent.
Furthermore, in Fourier space, the answer is simply to compare the bias and shot noise of the two fields: a good tracer of the mass has high bias and low noise, as we formalise below.

\begin{figure}[h!]
\centering
\includegraphics[width=0.45\textwidth]{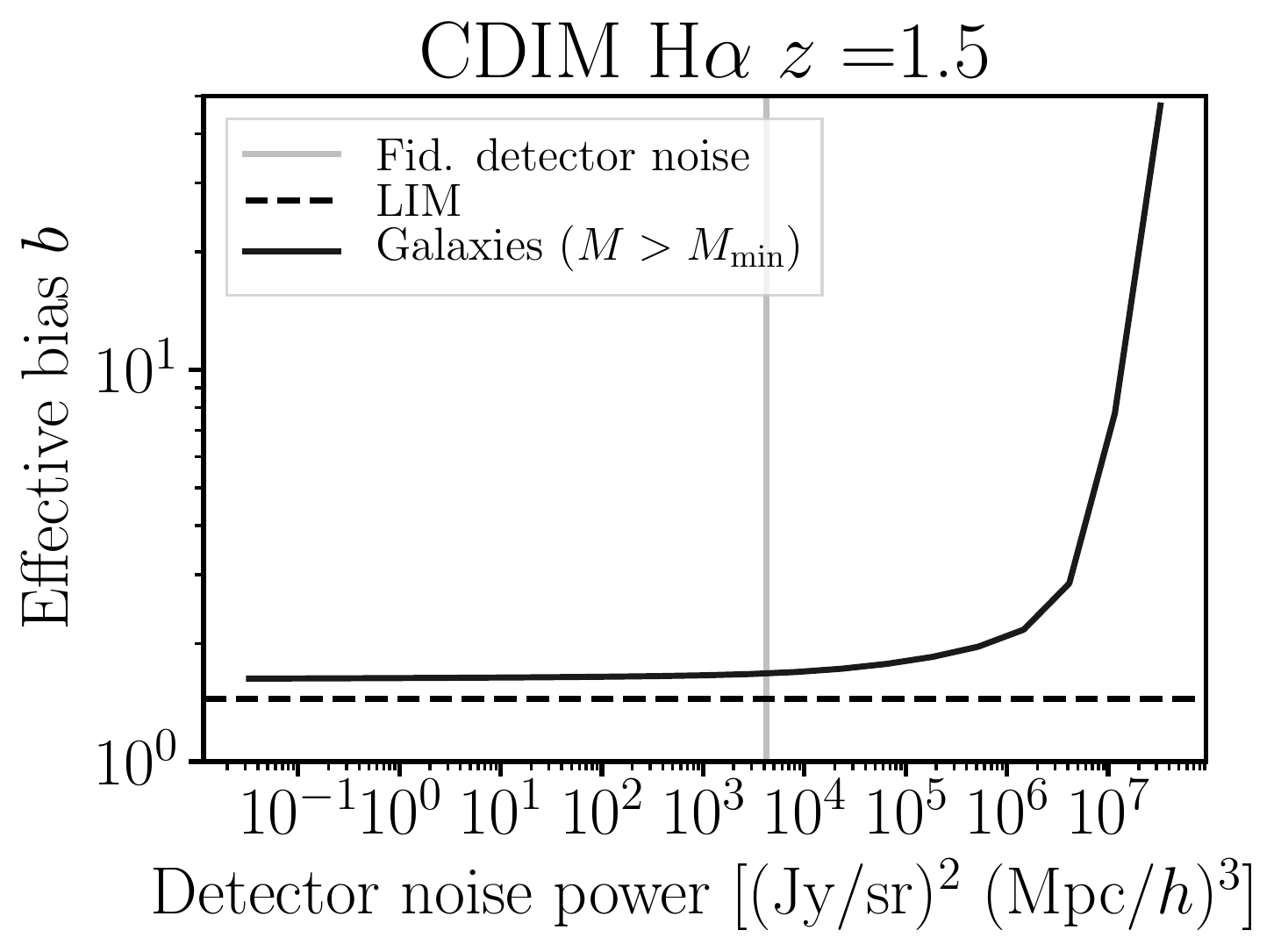}
\includegraphics[width=0.45\textwidth]{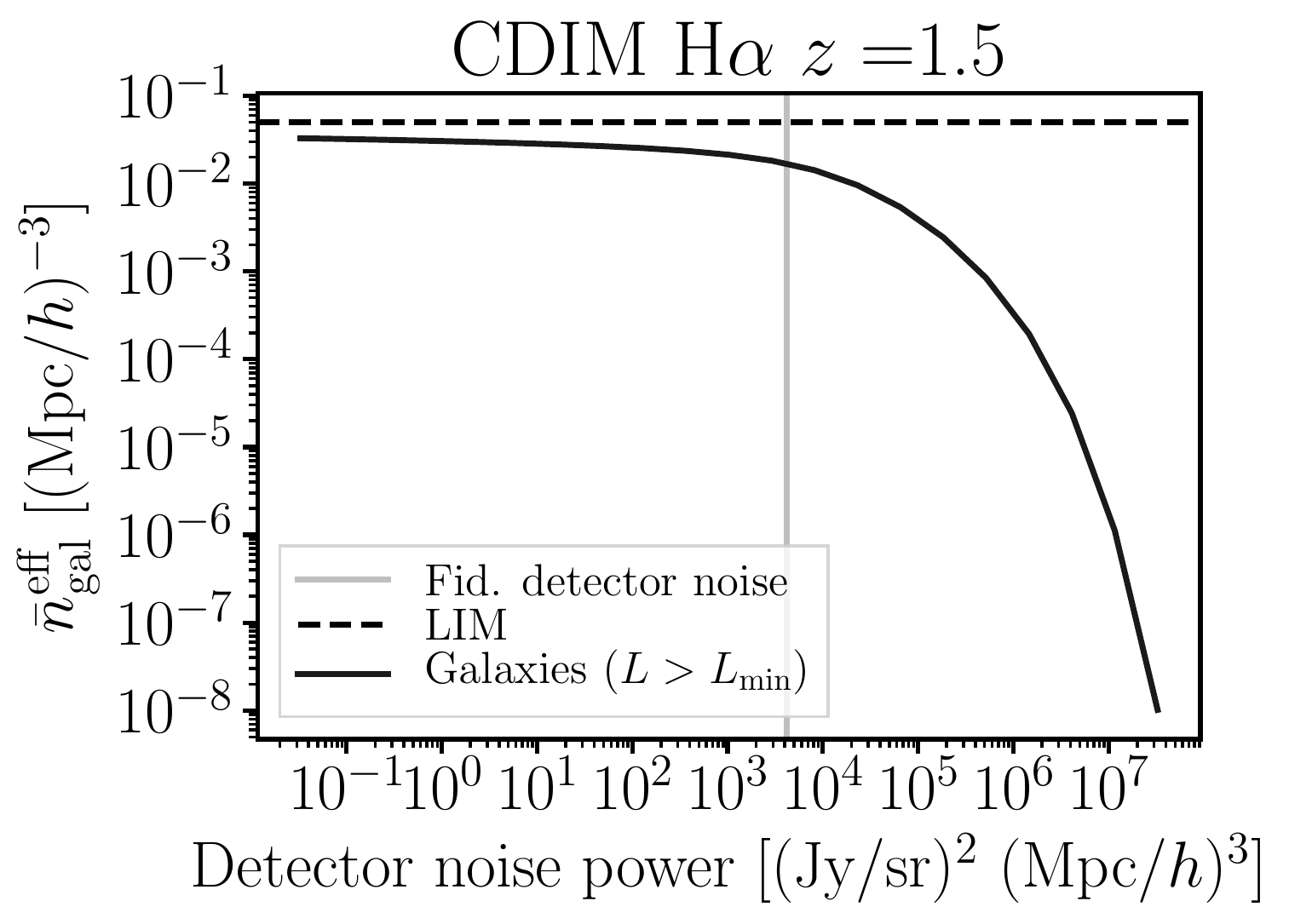}
\caption{
Increasing the detector noise level can dramatically reduce the number density of detected sources (right panel), and selects for the brightest, most highly biased galaxies (right panel).
The performance of the galaxy catalog as a tracer of the matter density field is determined by the product $\bar{n}^\text{gal eff} b^2$.
}
\label{fig:matched_filter_bias_ngaleff}
\end{figure}

The halo model formalism, based on the CLF, predicts both the power spectrum of the line intensity field, and of the number density of detected galaxies, as reviewed in Appendix~A.4 of \cite{paper1}.
% \ref{app:tracer_auto_correlation}.
In particular, the bias and shot noise of the detected galaxies are given by:
\beq
\left\{
\bal
&b_g(k, \mu, z)
\equiv
\frac{1}{\bar{n}_g}
\int
dm\ n(m)\;
b(m) 
\;
N_\text{gal}
u(k,m,z)
e^{-k^2 \mu^2\sigma_d^2(m) / 2}
\\
&P^\text{shot}_g
=
\frac{1}{\bar{n}_g}
=
\left[\int dm\ n(m) N_\text{gal}(m) \right]^{-1},\\
\eal
\right.
\eeq
where $N_\text{gal} = \int_{L_\text{min}}^\infty dL\ \Phi(L|m,z)$.
Thus the LIM and clustering power spectra are simply given by:
\beq
\left\{
\bal
&P_\text{LIM} = I^2 b_\text{LIM}^2 P_\text{lin} + N_\text{LIM}
&\quad\text{with}\quad
&N_\text{LIM}
=
P^{1h}_\text{LIM}
+ \frac{I^2}{n_\text{gal eff}}
+ W^{-2}\ N_\text{det}
\\
&P_\text{g} = b_\text{g}^2 P_\text{lin} + N_\text{g}
&\quad\text{with}\quad
&N_\text{g}
=
P^{1h}_\text{g}
+ \frac{1}{n_\text{gal}}
\\
\eal
\right.
\eeq
Ignoring the 1-halo term, one naturally favors a tracer of the matter density with a high bias and a low shot noise, i.e.\ a high effective number density of sources.
As shown in Fig.~\ref{fig:matched_filter_bias_ngaleff} (left panel),
the bias of the bright, individually detected galaxies can be much higher than that of the LIM. This is expected since these brighter sources occupy more massive host halos.
However, the effective number density of sources can be dramatically lower for the individually detected galaxies (Fig.~\ref{fig:matched_filter_bias_ngaleff}, right panel).
As the detector noise increases, both effects intensify, as only the rarest, brightest galaxies, occupying the most massive halos, are detected.

However, this comparison of bias and number density of sources ignores an important noise contribution to the LIM power.
The detector noise, which determines the galaxy luminosity detection threshold, also adds noise to the measured LIM power spectrum, and not to the galaxy clustering power spectrum.
This effect is most important in the high detector noise regime, where LIM would otherwise be vastly superior to galaxy detection, due to the lack of detectable sources.
To examine this more carefully, we compute the signal-to-noise on the linear power spectrum per Fourier mode, from both LIM and the bright galaxy clustering. 
Including the cosmic variance, we get:
\beq
\left\{
\bal
&\text{SNR}_{P_\text{lin}}^\text{LIM}
=
\frac{I^2 b_\text{LIM}^2 P_\text{lin}}{I^2 b_\text{LIM}^2 P_\text{lin} + N_\text{LIM}}
&&=
\frac{n_\text{gal eff}\ b_\text{LIM}^2 P_\text{lin}}
{n_\text{gal eff}\ b_\text{LIM}^2 P_\text{lin} + \left( 1 + W^{-2}I^{-2}\ n_\text{gal eff} N_\text{det} \right)}\\
&\text{SNR}_{P_\text{lin}}^\text{g}
=
\frac{b_\text{g}^2 P_\text{lin}}{b_\text{g}^2 P_\text{lin} + N_\text{g}}
&&=
\frac{n_\text{g} b_\text{g}^2 P_\text{lin}}
{n_\text{g} b_\text{g}^2 P_\text{lin} + 1}\\
\eal
\right.
\eeq
where we ignored the 1-halo term as a source of noise for both tracers in the last equalities.

Several comments are in order. 
The second line in the equation above is the usual function of $n_\text{g} b_\text{g}^2 P_\text{lin}$ found in galaxy clustering, which grows fast until $n_\text{g} b_\text{g}^2 P_\text{lin}\sim 1$.
The first line is the LIM analog.
In the absence of detector noise, it is the same function of 
$n_\text{gal eff} b_\text{LIM}^2 P_\text{lin}$, i.e. where the galaxy bias is replaced with the LIM bias, and the total number of galaxies is replaced by the effective number density of luminosity-weighted galaxies.
For a non-zero detector noise, the signal-to-noise from LIM is degraded.
In this SNR, we included the cosmic variance of the power spectrum. 
As a result, two tracers that are cosmic variance limited will be considered equivalent, even though one may have a lower noise than the other.
One important case where this approach breaks is cosmic variance cancellation techniques \cite{Seljak09, McDonald09} aiming at measuring primordial non-Gaussianity \cite{Moradinezhad19, Moradinezhad20}.
We do not explore this further here.
Again, no foregrounds were included here.
They would have the effect of increasing the luminosity threshold for detection, and increase the LIM noise power spectrum.
Both effects are thus similar to enhancing the detector noise, compared to the fiducial detector noise.
Finally, since the SNR computed is per Fourier mode, a given power spectrum bandpass can still be detected despite a low SNR per mode, as long as enough Fourier modes are included in the bandpass.

\begin{figure}[h!]
\centering
\includegraphics[width=0.32\textwidth]{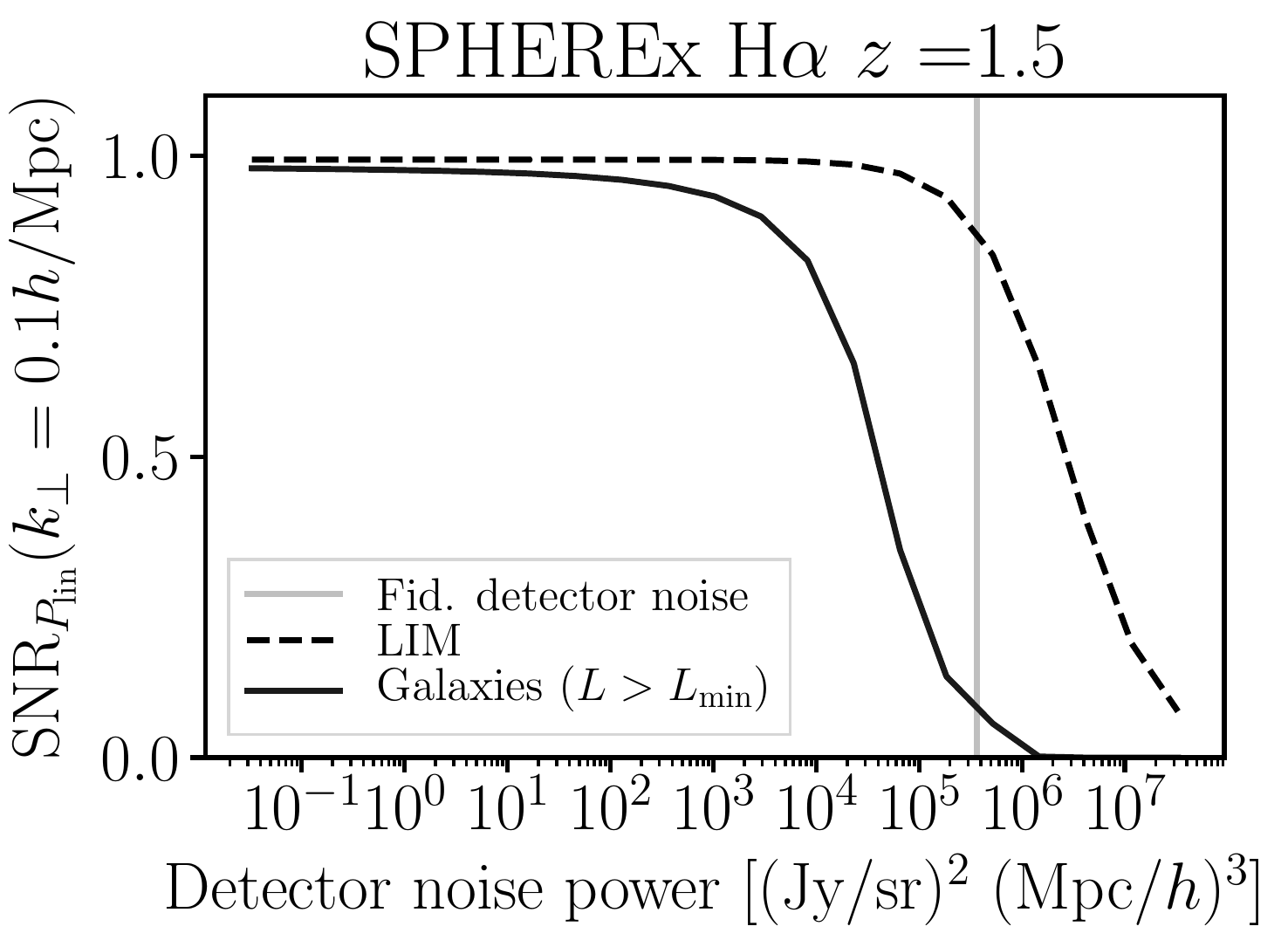}
\includegraphics[width=0.32\textwidth]{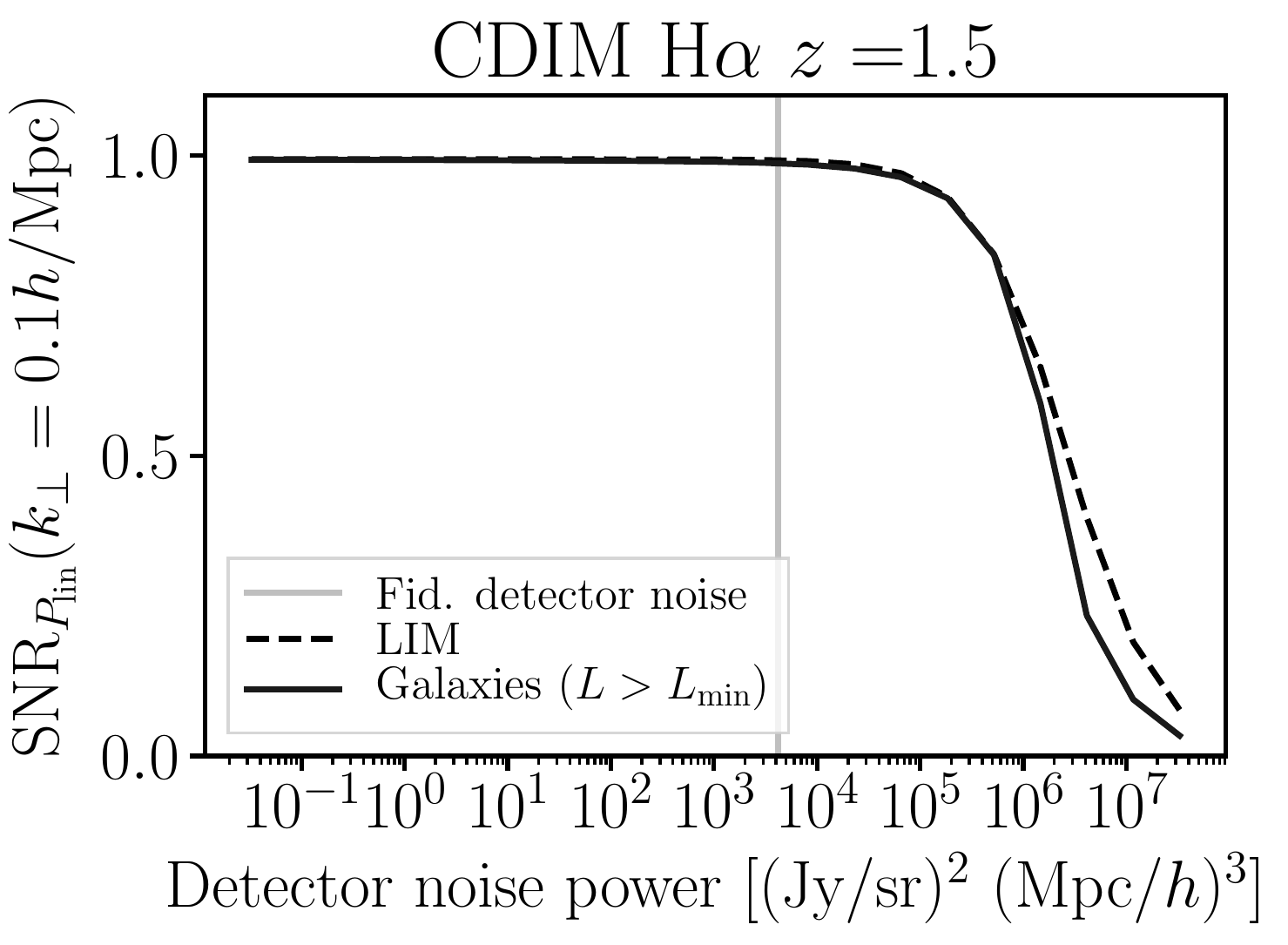}
\includegraphics[width=0.32\textwidth]{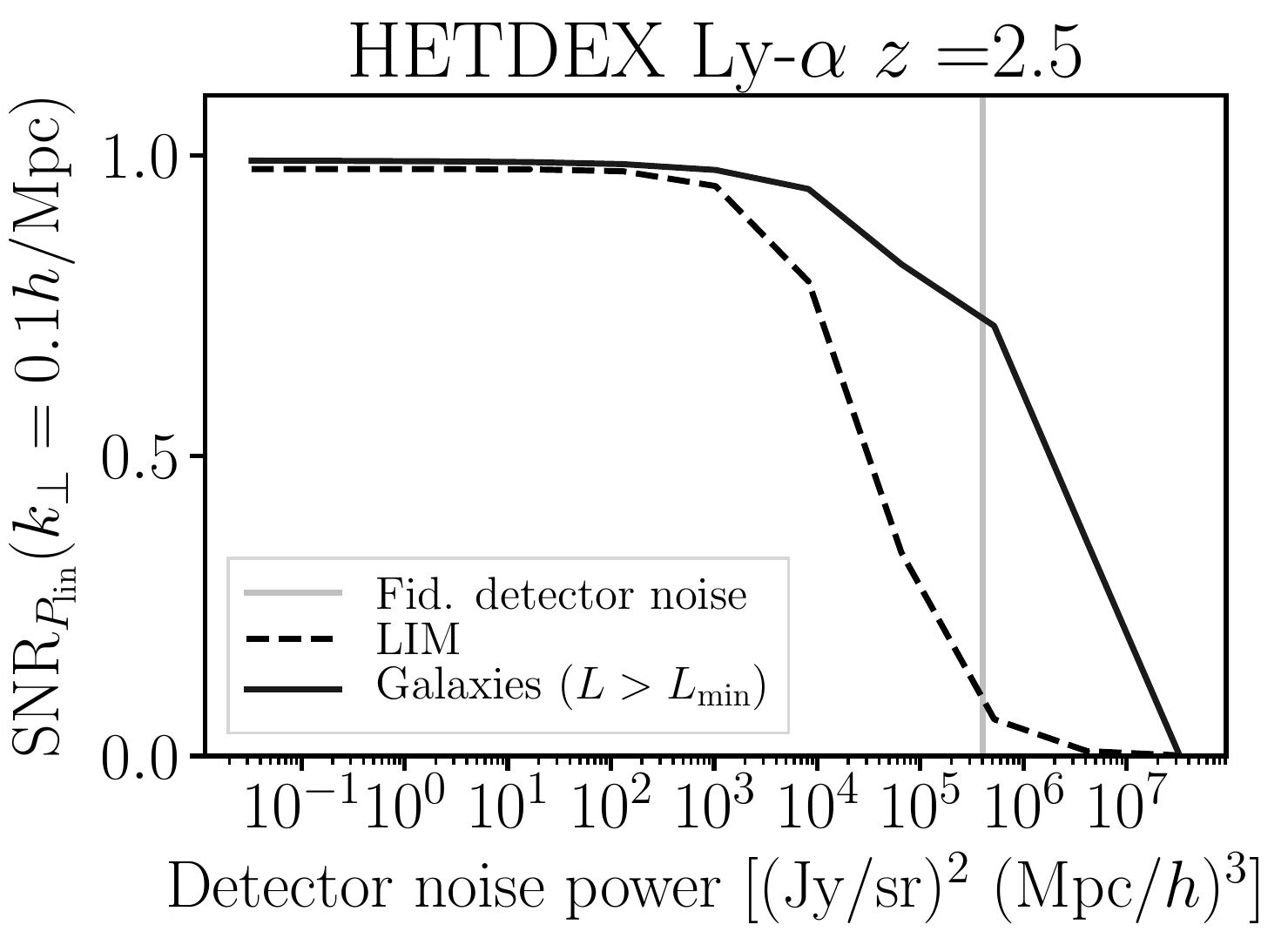}
\includegraphics[width=0.32\textwidth]{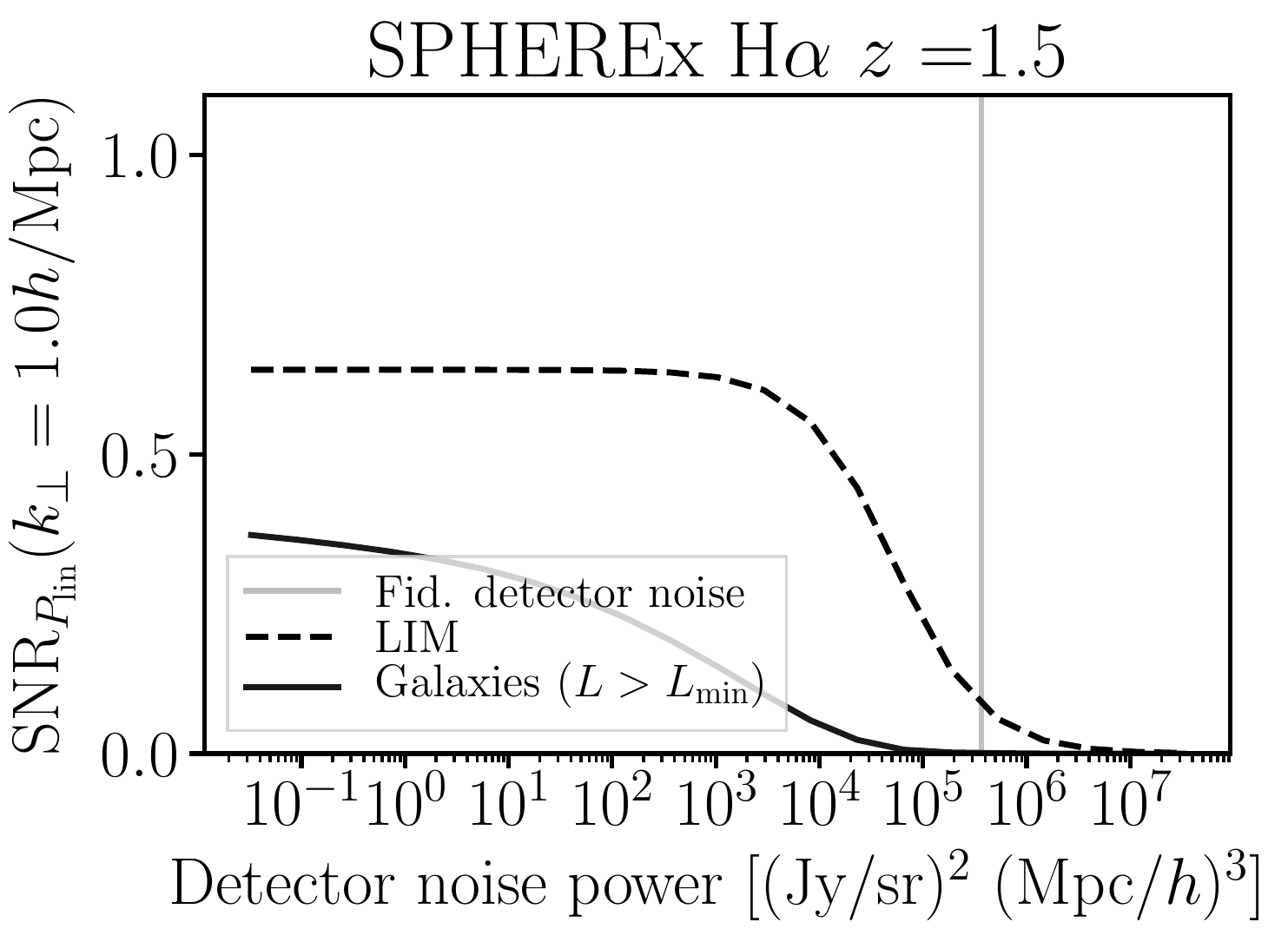}
\includegraphics[width=0.32\textwidth]{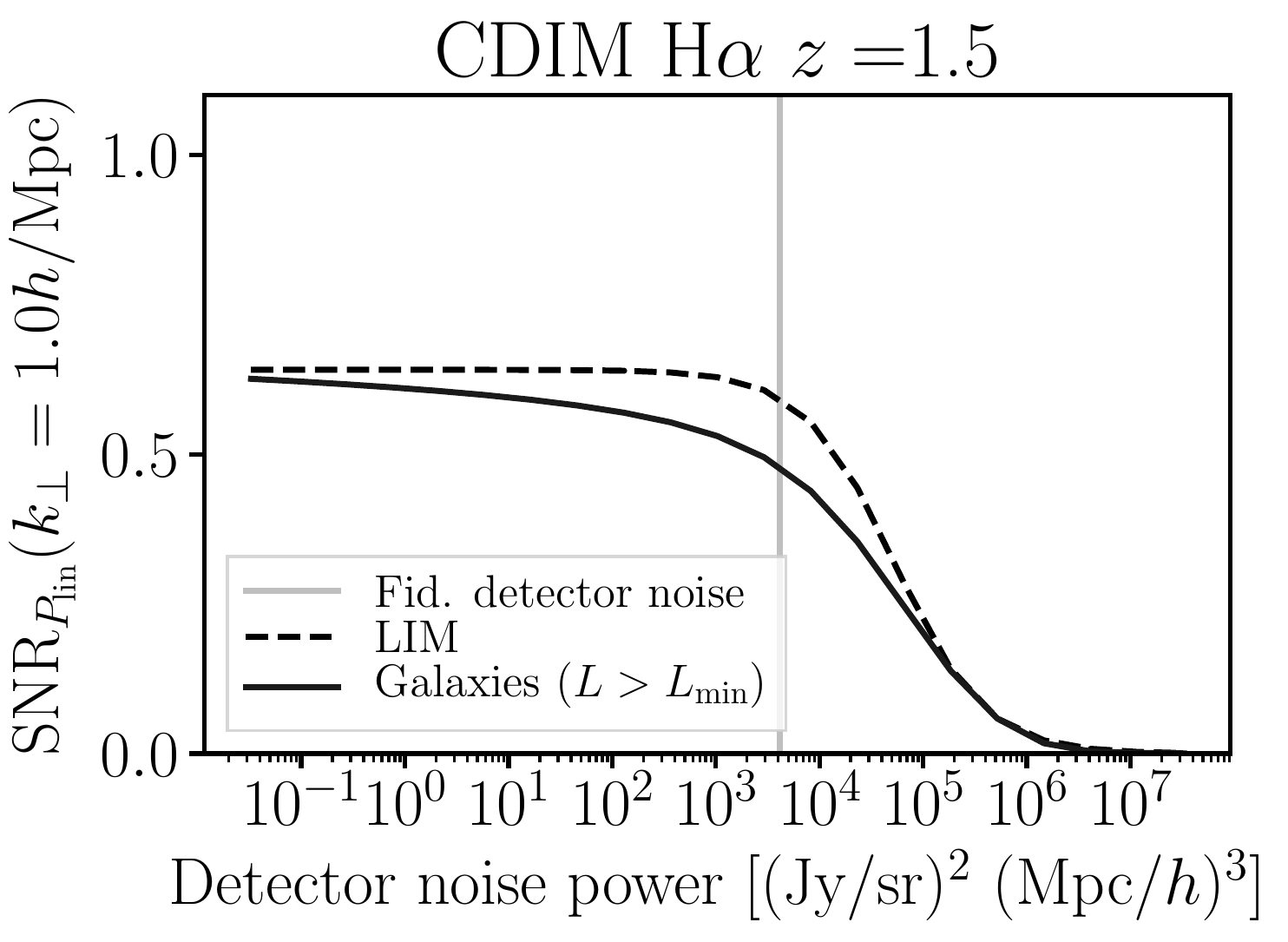}
\includegraphics[width=0.32\textwidth]{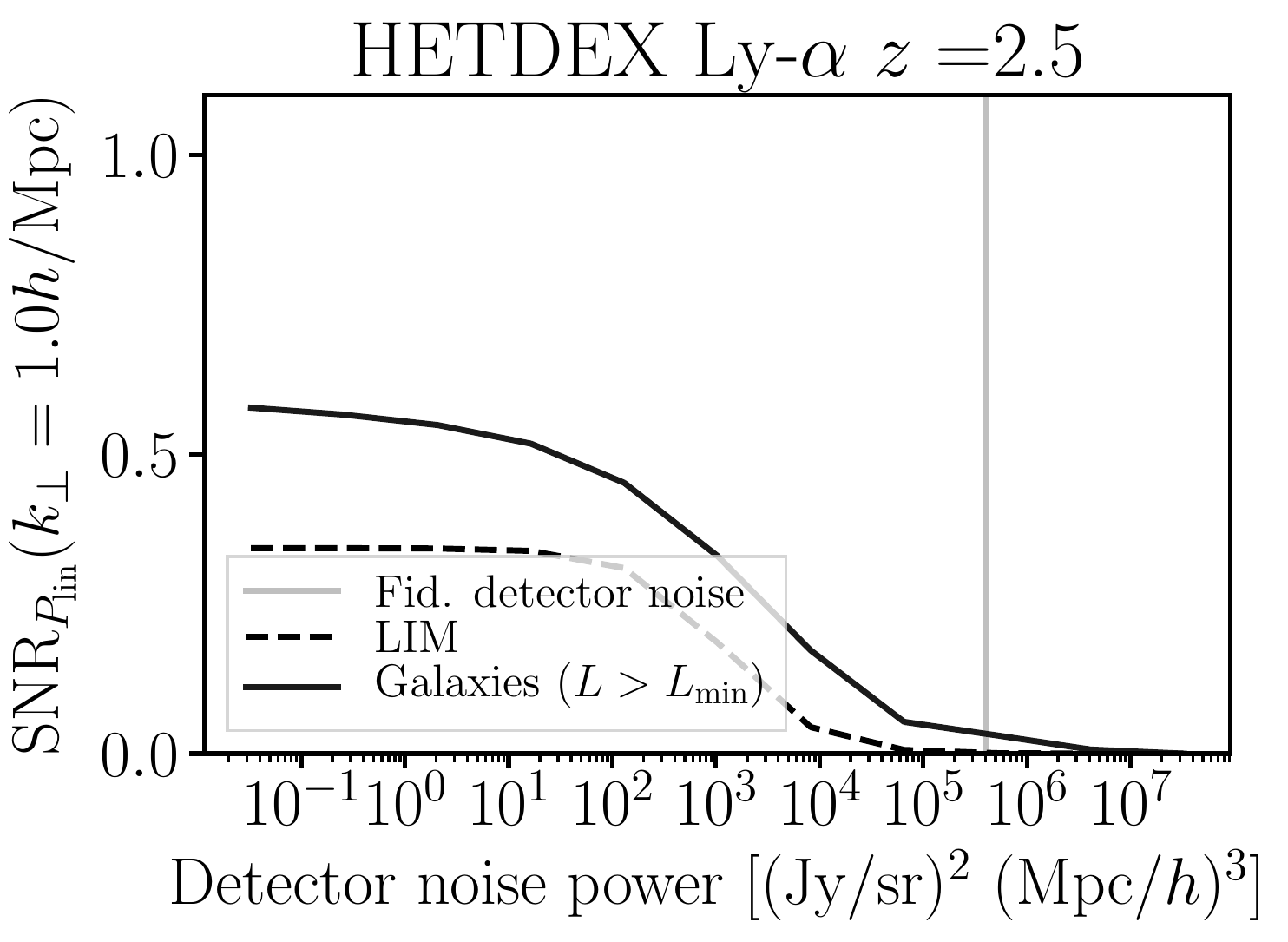}
\caption{
How does the intensity map compare to the catalog of detected sources, as a tracer of the matter density field?
We show this comparison for three experiments, SPHEREx H$\alpha$ (left column), CDIM H$\alpha$ (center column) and HETDEX Ly-$\alpha$ (right column).
Focusing on $k=0.1h/$Mpc (top row),
in all cases, LIM and galaxy detection are equivalent for low detector noise, with unit SNR per mode, because they are all cosmic variance-limited.
There, the detector noise in LIM and the shot noise in LIM and galaxy detection are negligible; the differences in tracer bias are irrelevant.
At higher noise, the LIM SNR drops once the detector noise becomes larger than the cosmic variance.
For galaxy detection, the SNR drops when the shot noise becomes larger than the cosmic variance, due to the reduction in the number of detectable sources in the map.
These transitions occur in different orders for the three experiments: LIM has higher SNR for SPHEREx (left), galaxy detection is best for HETDEX (right), and they are equivalent for CDIM (center).
The answer can change depending on the Fourier scale considered.
Switching to $k=1h/$Mpc, LIM and galaxy detection are not cosmic variance-limited for low detector noise.
They are limited by their respective shot noises, which are not identical: LIM and galaxy detection are no longer equivalent.
}
\label{fig:matched_filter_snr}
\end{figure}

We compare the SNR per Fourier mode for LIM and galaxy detection for three experiments, illustrating three different regimes, in Fig.~\ref{fig:matched_filter_snr}.
On large scales, $k=0.1h/$Mpc, as the detector noise tends to zero, LIM and galaxy detection reach SNR per Fourier mode of unity.
This is shown in the top row of Fig.~\ref{fig:matched_filter_snr}.
Indeed, when the noise is low, the dominant source of noise at $k=0.1 h/$Mpc is cosmic variance. In this case, the shot noise is negligible, such that the slightly different bias and source density of LIM and galaxy detection are irrelevant.
Thus, LIM and galaxy detection are equally good tracers of the matter density in the low noise regime.
In practice, a noise floor from foregrounds may prevent us from reaching this cosmic variance-limited regime.

For galaxy detection, as the detector noise increases, the dramatic reduction of the detected galaxy density eventually wins over the increase in galaxy bias. In other words, the galaxy catalog becomes more and more shot noise-dominated, causing the SNR per mode to drop.
For LIM, the source bias and effective number density are fixed. 
However, when the detector noise is large enough, LIM becomes detector noise-dominated, as opposed to cosmic variance-limited.
When this happens, the SNR per mode starts diminishing.
Both for LIM and galaxy detection, it is thus clear when the SNR per mode starts dropping.
However, whether this occurs first for LIM or galaxy detection is a quantitative question, as illustrated in Fig.~\ref{fig:matched_filter_snr}.
For SPHEREx H$\alpha$, galaxy detection loses SNR before LIM, so LIM is preferred.
For HETDEX Ly-$\alpha$, the opposite occurs, so galaxy detection is preferred.
For CDIM H$\alpha$, LIM and galaxy detection lose SNR at the same time, such that they remain equally good tracers of the matter density field.

On smaller scales, $k=1h/$Mpc, the situation changes.
For low detector noise, the shot noise, instead of the cosmic variance, dominates.
As a result, LIM and galaxy detection are no longer equivalent: the tracer with the higher $\bar{n}^\text{gal eff} b^2$ is the better tracer of the matter density.

In all cases, Fig.~\ref{fig:matched_filter_snr} shows a similar dependence on the detector noise. 
In the low noise regime, the SNR per Fourier mode asymptotes to a non-negligible value of order unity, indicating a useful tracer of the matter density field.
In the high noise regime, the SNR per Fourier mode becomes negligible, such that LIM and galaxy detection are no longer useful tracers.
In the transition region where the SNR per mode drops fast, one should be careful, as a slight change in detector noise or a small uncertainty in the LIM modeling may dramatically the usefulness of the tracers of the matter density.

\begin{figure}[h!]
\centering
\includegraphics[width=0.45\textwidth]{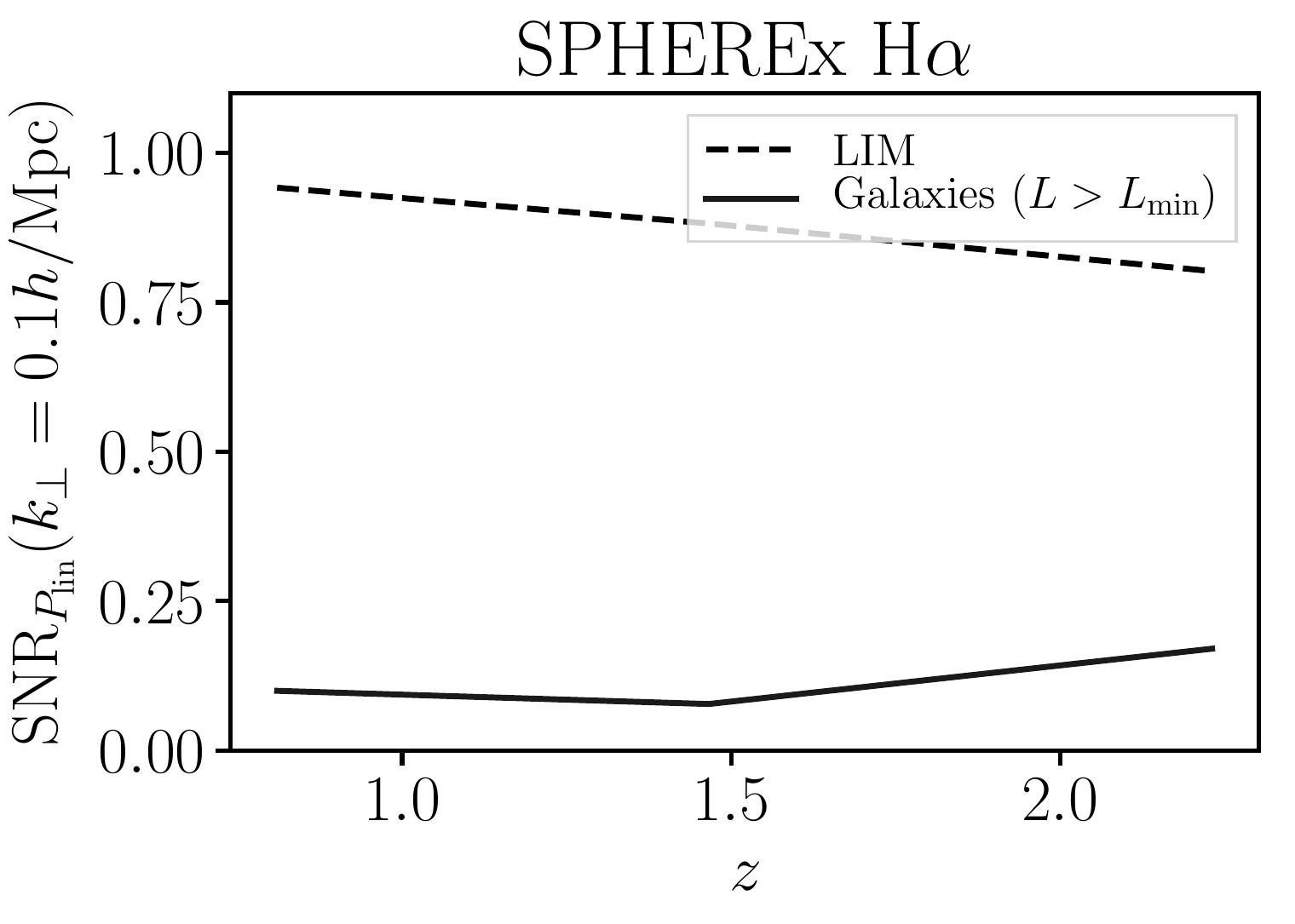}
\includegraphics[width=0.45\textwidth]{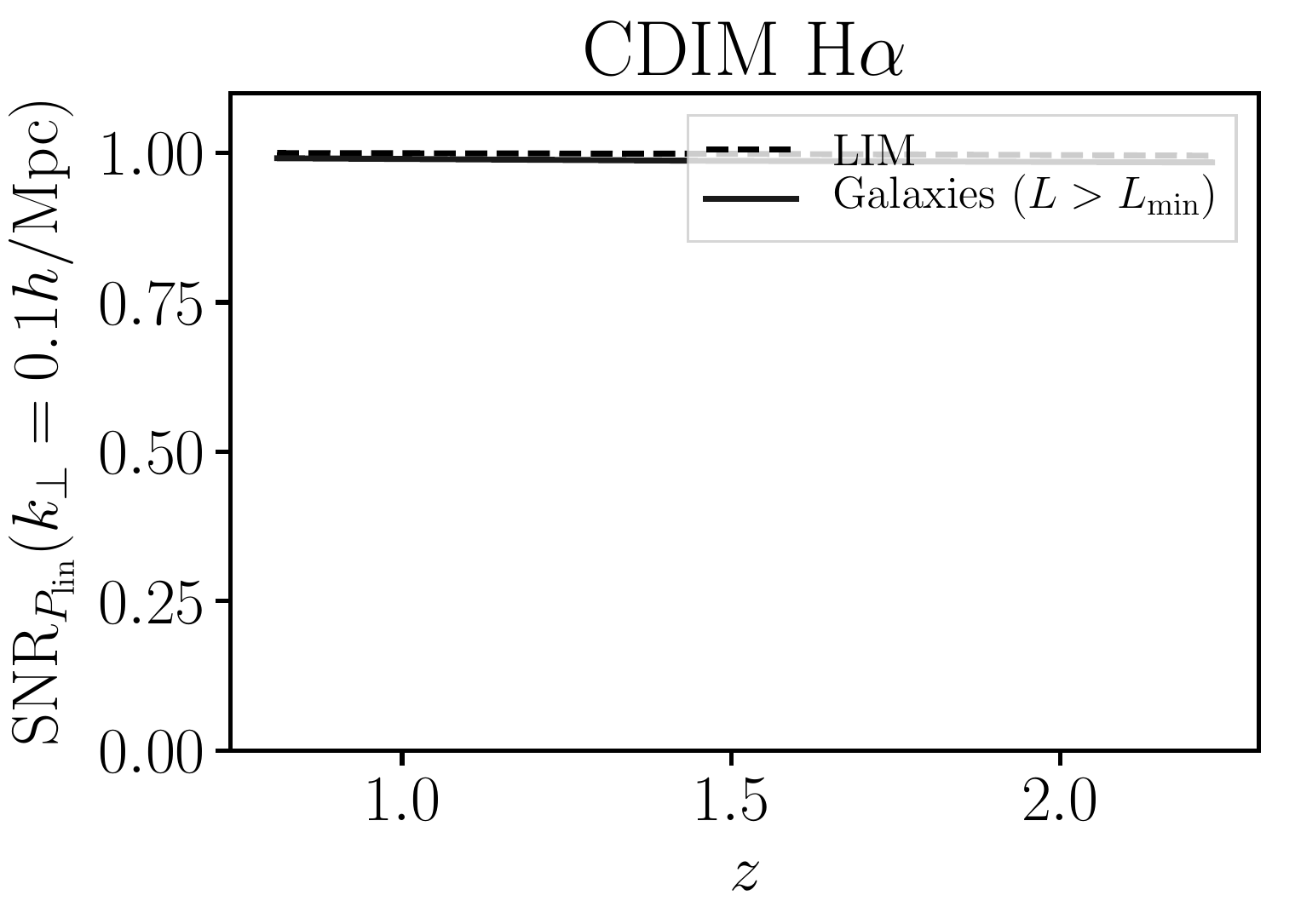}
\includegraphics[width=0.45\textwidth]{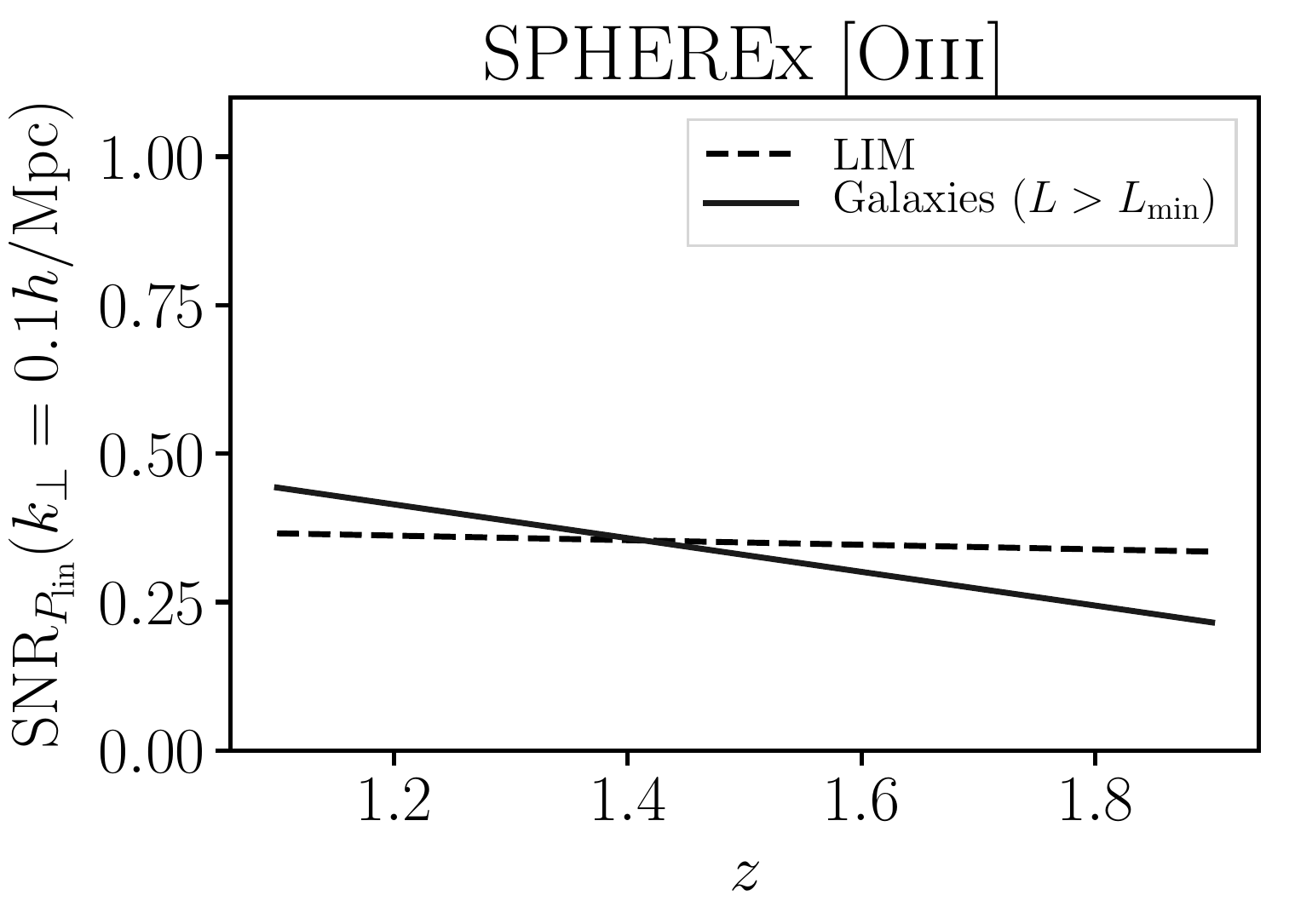}
\includegraphics[width=0.45\textwidth]{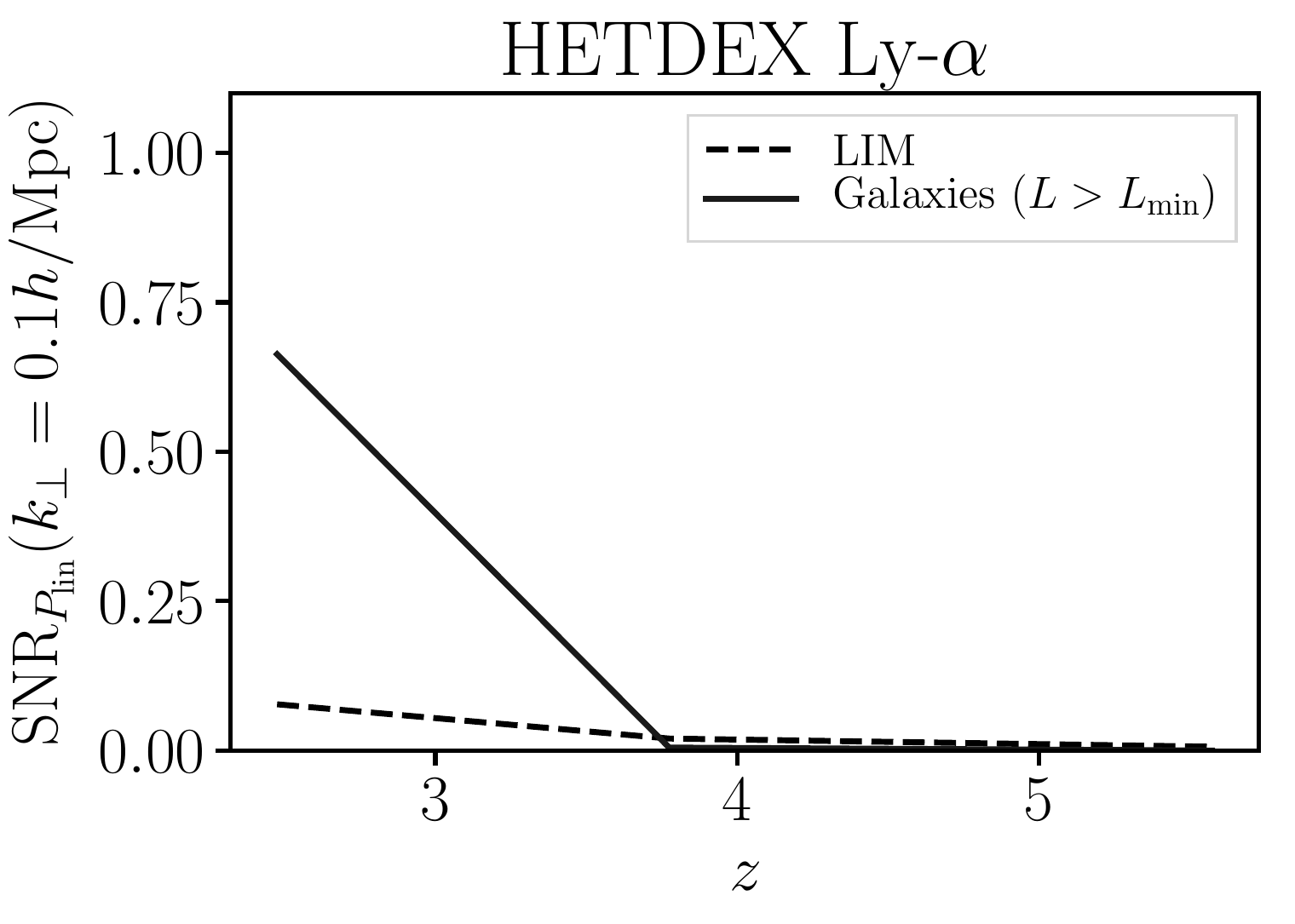}
\caption{
As in Fig.~\ref{fig:matched_filter_snr}, 
we compare LIM and galaxy detection as tracers of the matter density field, via their SNR per Fourier mode.
This time, we focus on
H$\alpha$ from SPHEREx (top left) and CDIM (top right),
[O{\sc iii}] from SPHEREx (bottom left) 
and Ly-$\alpha$ from HETDEX (bottom right).
We set the detector noise level to the fiducial value for each experiment, choose a Fourier scale of interest (here $k=0.1h/$Mpc) and vary redshift.
This illustrates how the comparison between LIM and galaxy detection depends on the line, the experiment and redshift.
It also depends on the Fourier scale considered.
}
\label{fig:np_lim_vs_galaxies}
\end{figure}

Finally, we reproduce this analysis as a function of redshift for various experiments in Fig.~\ref{fig:np_lim_vs_galaxies}, assuming the fiducial experimental noises.
This illustrates how our formalism identifies the better probe of the matter density field, and how the answer can depend on the line, the experimental configuration, the Fourier scale and the redshift.

%%%%%%%%%%%%%%%%%%%%%%%%%%%%%%%%%%%%%%%%%%%%%%%%%%%%%%%%%%
\subsection{Summary}
\label{sec:summary_lim_vs_gd}

In this section, we have given a formal answer to the comparison of LIM and galaxy detection, both as astrophysical probes of faint galaxies, and as cosmological probes of the matter density field.
We have extended the results of \cite{Silva17} and \cite{Cheng19}
(see App.~\ref{app:comparison_cheng19} for a detailed comparison with \cite{Cheng19}).
We have included shot noise, detector noise, confusion noise and cosmic variance.
We have explored various experimental configurations, lines, redshifts and Fourier scales.
The answers can be understood intuitively, as we summarize below.

Focusing on probing faint sources, the minimum galaxy luminosity detectable $L_\text{min}$ is determined by the angular resolution, which determines the number of Fourier modes detectable.
It is limited by either detector noise or confusion noise (typically shot noise or 1-halo term), whichever is larger.
Given the luminosity detection threshold $L_\text{min}$,
the number of galaxies detectable is large if $L_\text{min} \gtrsim L^\star$, where $L^\star$ is the Schechter luminosity scale.
In this case, only a small fraction of the LIM mean intensity and shot noise are sourced by undetected galaxies.
In other words, the galaxy catalog contains most of the information in this case.
In the opposite case $L_\text{min} \lesssim L^\star$, few galaxies will be detected and the LIM mean intensity and shot noise will be mostly sourced by undetected sources.
The fraction of the LIM 2-halo and 1-halo terms coming from undetected halos is a bit more subtle, involving the luminosity dependence of the bias and the halo mass function. We have provided a formalism to compute it.

Let us now focus on probing the matter density field, for cosmology.
If $L_\text{min} \gtrsim L^\star$, the detectable bright galaxies are too rare to be a good tracer of the matter: their shot noise is too high.
In the opposite case, we need to compare several power spectra.
For galaxy detection, we compare the signal power spectrum (or cosmic variance) to the galaxy shot noise. 
As the detector noise increases, fewer and fewer sources (i.e. with higher shot noise), with higher and higher bias, are detected.
The increase in shot noise overwhelms the increase in bias, such that the SNR from the galaxy catalog decreases.
For LIM, we compare the signal power spectrum (or cosmic variance) to the galaxy shot noise, the halo shot noise (or 1-halo term) and the detector noise.
For low detector noises, cosmic variance generally dominates for both LIM and galaxy detection, making them equivalent.
As the detector noise is increased, the LIM SNR drops as soon as the detector noise is similar to the cosmic variance.
Instead, the galaxy detection starts dropping as soon as the enhanced shot noise becomes similar to the cosmic variance.
Which of these happens first is a quantitative problem, which we have solved.

Although we have not sought to model foregrounds, we have shown that they can be directly included in our analysis, as long as their power spectra are known.

%%%%%%%%%%%%%%%%%%%%%%%%%%%%%%%%%%%%%%%%%%%%%%%%%%%%%%%%%%
%%%%%%%%%%%%%%%%%%%%%%%%%%%%%%%%%%%%%%%%%%%%%%%%%%%%%%%%%%
\section{Conclusions}
\label{sec:conclusions}

% intro
In this paper we have investigated what can be learned about astrophysics and cosmology from studying the mean and 2-point function of line intensity maps.
We applied the halo model formalism of the companion paper \cite{paper1}, based on the multi-line conditional luminosity function (CLF).

% What can be learned from LIM: astro and cosmo
% cosmology: 2-halo term
Much of the cosmological information in LIM comes from the 2-halo term, which dominates on large scales.
% importance of RSD
We emphasize the important role that redshift-space distortions will play in disentangling the different contributions to the observed power spectrum in this limit (\S\ref{sec:degeneracies_rsd}, see also refs.\ \cite{Visbal10,Visbal11,Pullen13,Lidz16,Gong20} for related discussion).
% 1-halo and shot noise
On intermediate scales, the 1-halo and/or shot noise terms dominate.
% importance of RSD
While the 1-halo term vanishes on extremely small scales, leaving only the shot noise, measuring and modeling these scales is challenging.
By modulating the 1-halo term and not the shot noise, FOGs offer an easier way to break the degeneracy between the two
(\S\ref{sec:resolution_distinguish_1halo_shot}).
% astrophysics from 1-halo and shot noise
Once separated, the 1-halo and shot noise terms contain independent astrophysical information.
The amplitude of the 1-halo term is an integral constraint on the halo mass--luminosity relation, and typically scales as the mean squared star-formation rate in halos.
The mean intensity and shot noise provide the first two moments of the galaxy luminosity distribution. 
The shot noise thus probes the stochasticity of the luminosity generation mechanism.
In cross-correlation, it encodes the correlation between the luminosity of a given galaxy in two distinct lines.
This therefore contains useful information on the correlation between the intrinsic galaxy properties which produce these lines.
Finally, cross-correlations with photometric or spectroscopic galaxy catalogs, as well as galaxy or CMB lensing, allow to further break degeneracies between astrophysics and cosmology.
All the degeneracies are summarized in \S\ref{sec:summary_degeneracies}.

% LIM VS galaxy detection: matched filter formalism
Two claimed strengths of LIM are its ability to probe galaxies too faint to be detected individually (Astrophysics), and to trace the matter density field better than the catalog of individually detected galaxies (Cosmology).
To test these, we predicted the luminosity detection threshold for individual galaxies,
using a matched filter formalism for 3D redshift-space and 2D projected maps (Appendix.~\ref{app:matched_filter}).
% Astrophysics: probing fainter sources
% Cosmology: tracers of the matter density
Using our halo model, we then computed the fraction of the mean intensity and LIM power spectrum sourced by galaxies too faint to detect (\S\ref{sec:sensitivity_undetected_sources}), extending the results of ref.~\cite{Silva17}.
Finally, we compared the signal-to-noise on the linear power spectrum from LIM and from the catalog of bright sources (\S\ref{sec:lim_vs_gal_matter_tracer}).
In both cases, we have found that the LIM observables indeed outperform galaxy detection from the same intensity map, in the regime of high noise and low resolution.
A more detailed summary is shown in \S\ref{sec:summary_lim_vs_gd}.

Our Fourier space approach agrees with the study of ref.~\cite{Cheng19},
and extends it in a number of ways, discussed in detail in App.~\ref{app:comparison_cheng19}.
We compare LIM and galaxy detection not only as comological probes, but also as astrophysical probes.
Switching from a single-pixel to a Fourier analysis allows us to answer the question as a function of scale, and to include pixel-to-pixel correlations.
We also take into account the luminosity dependence of the clustering bias and redshift-space distortions.
Finally, we include cosmic variance, which makes LIM and galaxy detection equivalent in the low noise regime.
However, \cite{Cheng19} derives an optimal estimator of the matter density field, beyond LIM and galaxy detection, which we did not attempt here.

% theoretical uncertainty: galaxy luminosity functions
One main source of uncertainty in our analysis comes from our limited knowledge of the galaxy luminosity functions at high redshift. This paper did not attempt to reduce this uncertainty, but simply to provide a method to convert a given luminosity function into astrophysical and cosmological forecasts. 
We refer the reader to \S3.2 in\cite{paper1} for a more detailed discussion of this uncertainty.

% foregrounds
The second major source of uncertainty in our analysis is the level of continuum and interloper foregrounds present in the intensity maps.
The formalism we presented is built to include the effect of foregrounds, through their power spectrum.
While the foregrounds can likely be reduced in a number of ways (frequency cleaning, masking, cross-correlations, etc), residual foregrounds will act as a noise floor in the maps.

% code publicly available
We have made our code publicly available at \url{https://github.com/EmmanuelSchaan/HaloGen/tree/LIM}.
This modular code was used to produce the RSD forecasts and the comparison between LIM and galaxy detection for any line and experimental configuration.

%%%%%%%%%%%%%%%%%%%%%%%%%%%%%%%%%%%%%%%%%%%%%%%%%%%%%%%%%%
%%%%%%%%%%%%%%%%%%%%%%%%%%%%%%%%%%%%%%%%%%%%%%%%%%%%%%%%%%
\acknowledgments
We would like to thank Yun-Ting Cheng, Tzu-Ching Chang and Olivier Dor\'e for their kind hospitality and for all our useful discussions during the preparation of this manuscript.
We thank Corentin Schreiber for sharing with us details of the line luminosity implementation in his code \texttt{EGG}.
We thank the organizers and participants of the L2S2: Lines in the Large-scale Structure conference\footnote{\url{https://l2s2.sciencesconf.org/}} for many useful discussions.
We thank Patrick Breysse, Simone Ferraro, Adam Lidz, Abhishek Maniyar, Anthony Pullen, Uros Seljak and David Spergel for their helpful feedback on this paper.
We thank the anonymous referee for their useful comments.
E.S.~is supported by the Chamberlain fellowship at Lawrence Berkeley National Laboratory. 
M.W.~is supported by the DOE and the NSF.
This research has made use of NASA's Astrophysics Data System and the arXiv preprint server.
This research used resources of the National Energy Research Scientific Computing Center (NERSC), a U.S. Department of Energy Office of Science User Facility operated under Contract No. DE-AC02-05CH11231.

%%%%%%%%%%%%%%%%%%%%%%%%%%%%%%%%%%%%%%%%%%%%%%%%%%%%%%%%%%
%%%%%%%%%%%%%%%%%%%%%%%%%%%%%%%%%%%%%%%%%%%%%%%%%%%%%%%%%%
\appendix

%%%%%%%%%%%%%%%%%%%%%%%%%%%%%%%%%%%%%%%%%%%%%%%%%%%%%%%%%%
\section{Detecting individual sources: matched-filtering}
\label{app:matched_filter}

%%%%%%%%%%%%%%%%%%%%%%%%%%%%%%%%%%%%%%%%%%%%%%%%%%%%%%%%%%
\subsection{In 3D redshift space}

In order to derive a matched filter for point sources, we need to model how these point sources appear in the observed LIM.
This is determined by the source flux $F$, the angular point-spread function (PSF) and the spectral point-spread function (SPSF):
\beq
\underbrace{I(\vx)}
_{[\text{erg/s/m}^2\text{/sr/Hz}]}
=
\underbrace{F}
_{\text{Flux [erg/s/m}^2]}\
\underbrace{\text{PSF}}
_{[1/\text{sr}]}
\left(\bm{\theta} = \frac{\vx_\perp}{\chi}\right)\
\underbrace{\text{SPSF}}
_{[1/\text{Hz}]}
\left(d\nu = \frac{d\nu}{d\chi} dx_\parallel\right)
.
\eeq
We assume the PSF and SPSF are Gaussian
\beq
\left\{
\bal
&\text{PSF}(\bm{\theta})
=
\frac{e^{-\theta^2 / 2\sigma_\theta^2}}{2\pi\sigma_\theta^2}
&\quad\text{with}\quad
&\sigma_\theta = \frac{\text{FWHM}}{\sqrt{8\ln 2}}\\
&\text{SPSF}(\nu)
=
\frac{e^{-\nu^2 / 2\sigma_\nu^2}}{\sqrt{2\pi\sigma_\nu^2}}
&\quad\text{with}\quad
&\sigma_\nu = \frac{\nu}{\mathcal{R}}\\
\eal
\right.
,
\eeq
where FWHM is the PSF full width at half-maximum, $\mathcal{R}$ is the spectral resolving power, and $\nu$ is the observed frequency (not the rest-frame line frequency).

The matched filter is most easily derived in Fourier space. 
There, the point source model is:
\beq
\underbrace{I(\vk)}
_{[\text{m.erg/s/sr/Hz}]}
=
\underbrace{F}_{[\text{erg/s/m}^2]}\
\underbrace{\chi^2}
_{[\text{m}^2\text{/sr}]}
\underbrace{\frac{c}{a H \nu}}
_{[\text{m/Hz}]}\
W(\vk)
\quad\text{with}\quad
W(\vk) \equiv
\text{PSF}\left( \vk_\perp \right)\
\text{SPSF}\left(k_\parallel\right)
,
\eeq
where the Fourier-space PSF and SPSF are now given by
\beq
\left\{
\bal
&\text{PSF}(\vk_\perp)
=
e^{-k_\perp^2 / 2 \sigma_{k_\perp}^2}
&\quad\text{with}\quad
&&\sigma_{k_\perp}
=
\frac{1}{\chi \sigma_\theta}\\
&\text{SPSF}(k_{\parallel})
=
e^{-k_\parallel^2 / 2 \sigma_{k_\parallel}^2}
&\quad\text{with}\quad
&&\sigma_{k_\parallel}
=
\frac{d\nu}{d\chi} \frac{1}{\sigma_\nu}
=
\frac{aH\mathcal{R}}{c}\\
\eal
\right.
.
\eeq

The matched filter $\hat{F}$ is defined as the minimum variance, unbiased estimator for the source flux $F$ that is linear in the data $I$.
Given the point source model, deriving it is straightforward. 
For any given Fourier mode $\vk=(\vk_\perp, k_\parallel)$,
The quantity 
$\hat{F}(\vk) 
\equiv 
\left( aH\nu/ c\chi^2 \right)
I(\vk)\ /\ W(\vk)$
is an unbiased estimator of the source flux, $F$.
The variance of this estimator is 
$\sigma^2(\vk) 
=
\left( aH\nu/ c\chi^2 \right)^2
\left[ W^2(\vk)\ P_\text{LIM}(\vk) + N(\vk) \right] /\ W^2(\vk)$,
where $N(\vk)$ is the detector noise power spectrum.
These estimators, for two distinct Fourier modes $\vk$ and $\vk^\prime$, are uncorrelated due to statistical homogeneity:
$\langle I(\vk) I(\vk^\prime) \rangle \propto \delta^D(\vk+\vk^\prime)$.
The minimum variance, unbiased linear combination $\hat{F}$ of these estimators is therefore simply their inverse-variance weighted average:
\beq
\hat{F}
=
\int_{\vk}%\frac{d^3\vk}{\left( 2\pi \right)^3}\
\hat{F}(\vk) / \sigma^2(\vk)
\ \Bigg/
\ \int_{\vk}%\frac{d^3\vk}{\left( 2\pi \right)^3}\
1 / \sigma^2(\vk)
.
\eeq
In this expression, $\int_{\vk}$ is the sum over all independent Fourier modes.
Since the field $I(\vx)$ is real-valued in configuration space, $I(-\vk) = I^\star(\vk)$ and only half of the Fourier modes are independent.
Hence $\int_{\vk} = \frac{1}{2} \int \frac{d^3\vk}{\left( 2\pi \right)^3}$, and the matched filter becomes:
\beq
\hat{F}
=
\left(\frac{c\chi^2}{aH\nu}\right)
\sigma_{\hat{F}}^2
\int\frac{d^3\vk}{\left( 2\pi \right)^3}
\ \frac{I(\vk) \ W(\vk)}
{2\left[ W^2(\vk)\ P_\text{LIM}(\vk) + N(\vk) \right]},
\label{eq:3d_matched_filter_1}
\eeq
where $\sigma_{\hat{F}}^2$ is indeed the variance of the matched filter in units of flux, given by:
\beq
\sigma_{\hat{F}}^2
=
\left(\frac{aH\nu}{c\chi^2}\right)^2
\left\{
\int\frac{d^3\vk}{\left( 2\pi \right)^3}
\ \frac{W^2(\vk)}{2\left[ W^2(\vk)\ P_\text{LIM}(\vk) + N(\vk) \right]}
\right\}^{-1}
.
\label{eq:3d_matched_filter_2}
\eeq
Here again, $\nu$ represents the observed frequency, related to the rest-frame line frequency $\nu_0$ via $\nu = a\nu_0$.
The flux variance can finally be converted to luminosity units via:
\beq
\sigma_{L} = 4\pi \left[ (1+z)\chi \right]^2\ \sigma_{\hat{F}}
\quad
[\text{erg/s}].
\eeq

%%%%%%%%%%%%%%%%%%%%%%%%%%%%%%%%%%%%%%%%%%%%%%%%%%%%%%%%%%
\subsection{In 2D projection}
\label{app:2d_matched_filter}

Consider a the 2D intensity field $I_{2D}(\vtheta)$,
a projection of the 3D intensity field $I(\vx)$ over a thin shell centered around $\chi$ with comoving width $\Delta \chi$:
\beq
I_{2D}(\vtheta) 
\equiv 
\int d\chi' \ W(\chi) I(\vx_\perp=\chi'\vtheta, x_\parallel=\chi')
\quad\text{with}\quad
W(\chi') = \mathbb{1}_{|\chi'-\chi|\leq\Delta \chi} / \Delta\chi.
\eeq
In the 2D projected map, a point source appears as:
\beq
\underbrace{I_{2D}(\vtheta)}
_{[\text{erg/s/m}^2\text{/sr/Hz}]}
=
\underbrace{F}
_{\text{Flux [erg/s/m}^2]}\
\underbrace{\text{PSF}}
_{[1/\text{sr}]}
\left(\vtheta\right)\
\underbrace{\frac{1}{\Delta \nu}}
_{[1/\text{Hz}]}
\quad\text{with}\quad
\Delta\nu = \frac{aH\nu}{c} \Delta \chi.
\eeq
In Fourier space, this simply becomes:
\beq
I(\vl)
=
F\
\text{PSF}(\vl)
\quad\text{with}\quad
\text{PSF}(\vl)
\equiv
e^{-\frac{\vl^2}{2}\sigma_\theta^2}
.
\eeq
Following the same derivation as in 3D, we obtain the 2D matched filter by inverse-variance weighting all the independent $\vl$-modes into the minimum-variance, unbiased, linear estimator for F:
\beq
\hat{F}
=
\sigma_{\hat{F}}^2
\int\frac{d^2\vl}{\left( 2\pi \right)^2}
\ \frac{I_{2D}(\vl) \ \left(\text{PSF}(\vl) / \Delta\nu\right)}
{2\left[ \text{PSF}^2(\vl)\ C_\ell + N_\ell \right]},
\eeq
where $\sigma_{\hat{F}}^2$ is indeed the variance of the matched filter in units of flux, given by:
\beq
\sigma_{\hat{F}}^2
=
\left\{
\int\frac{d^2\vl}{\left( 2\pi \right)^2}
\frac{ \left(\text{PSF}(\vl) / \Delta\nu\right)^2}
{2\left[ \text{PSF}^2(\vl)\ C_\ell + N_\ell \right]}
\right\}^{-1}
.
\eeq

Alternatively, this 2D matched filter can also be derived from the 3D expression Eq.~\eqref{eq:3d_matched_filter_1}-\eqref{eq:3d_matched_filter_2}, in the limit where
\beq
\text{SPSF}(k_\parallel) 
\longrightarrow 
\mathbb{1}_{|k_\parallel| \leq \frac{\Delta k_\parallel}{2} = \frac{\pi}{\Delta\chi}}
,
\eeq
using the Limber approximation and noticing that:
\beq
\left\{
\bal
&\frac{aH\nu}{c\chi^2} = \frac{\Delta\nu}{\mathcal{V}}
\quad\text{with}\quad
\mathcal{V} \equiv \chi^2\Delta \chi\\
&P_{3D}(k=\frac{\ell}{\chi}, \mu=0, z) = \mathcal{V}\ C_\ell\\
&\int\frac{d^3\vk}{(2\pi)^3} =
\underbrace{\int\frac{d^2\vk_\perp}{(2\pi)^2}}
_{\frac{1}{\chi^2}\! \int \! \frac{d^2\vl}{(2\pi)^2}}\;
\underbrace{\int\frac{dk_\parallel}{(2\pi)}}
_{\frac{1}{\Delta\chi}}
=
% \frac{1}{\chi^2\Delta\chi}
\mathcal{V}^{-1}
\int\frac{d^2\vl}{(2\pi)^2}\\
&I_{3D}(\vk=\ell/\chi)
\longrightarrow
\chi^2\Delta\chi \ I_{2D}(\vl)\\
\eal
\right
.
\eeq

%%%%%%%%%%%%%%%%%%%%%%%%%%%%%%%%%%%%%%%%%%%%%%%%%%%%%%%%%%
\section{Experimental configurations: detector noise}
\label{app:experimental_detector_noise}

\subsection{SPHEREx}

Following ref.~\cite{Dore14}, we assume that the minimum flux density detectable at 5~$\sigma$ in one voxel is
$m_{AB}^{5\sigma} = 22$.
This corresponds to a 1~$\sigma$ flux limit of 
$\sigma_{f, \text{source}} = (1/5)\ 10^{\frac{8.9 - m_{AB}^{5\sigma}}{2.5}}$~Jy,
i.e.
$\sigma_{f, \text{source}} = 1.15\times 10^6$~Jy.

To convert this point source flux noise to the voxel flux noise, we account for the fact that the PSF of SPHEREx covers an effective number 
$N_\text{pix eff} = 2-5$ of pixels \cite{Dore14}.
We set $N_\text{pix eff} = 3$ here.
The voxel flux noise is therefore:
$\sigma_{f, \text{voxel}} 
= \sigma_{f, \text{source}} / \sqrt{N_\text{pix eff}}
=
6.6\times 10^{-7}$~Jy.

We then convert the flux uncertainty to intensity uncertainty via
$\sigma_{I, \text{voxel}} 
=
\sigma_{f, \text{voxel}} / \Omega_\text{pixel}
$,
with 
$\Omega_\text{pixel} = 8.46\times 10^{-10}$~sr,
i.e.
$\sigma_{I, \text{voxel}} 
=
780$~Jy/sr.

Finally, the noise on the voxel intensity is converted to white noise power spectrum as
$N(\vk) = \sigma_{I, \text{voxel}}^2 \mathcal{V}_\text{voxel}$~[(Jy/sr)$^2$(Mpc/$h$)$^3$],
where the voxel comoving volume is given by:
$\mathcal{V}_\text{voxel}
=
\chi^2 \Omega_\text{pixel} \Delta\chi
= \Omega_\text{pixel} c/\left( a H \mathcal{R} \right)$.

\subsection{COMAP}

Following \cite{Li16}, the voxel noise for COMAP is given in Rayleigh-Jeans temperature units as
$\sigma_{T, \text{voxel}}
=T_\text{system}
/\sqrt{N_\text{feeds}\ \Delta\nu\ \tau_\text{pixel}}$~[$\mu$K],
where the observing time per pixel is
$\tau_\text{pixel}
=\tau_\text{survey} \Omega_\text{pixel} / \Omega_\text{survey}$~[s].
We assume the ``full'' (as opposed to ``pathfinder'') configuration in Table~2 of \cite{Li16}, with:
\beq
\left\{
\bal
&N_\text{feeds} = 500\\
&T_\text{sys} = 3\text{ K}\\
&\Delta\nu = 10\text{ MHz}\\
&\nu\simeq 30\text{ GHz}\\
&\Omega_\text{survey} = 6.25\text{ deg}^2\\
&\tau_\text{survey} = 2,250\text{ hr}\\
\eal
\right.
\eeq
We convert from Rayleigh-Jeans temperature [$\mu$K] to intensity units [Jy/sr] via
$\sigma_{I, \text{voxel}}
=
\frac{2\nu^2 k_B}{c^2}
\sigma_{T, \text{voxel}}$.
Finally, we obtain the noise power spectrum from the voxel intensity noise:
$N(\vk) = \sigma_{I, \text{voxel}}^2 \mathcal{V}_\text{voxel}$~[(Jy/sr)$^2$(Mpc/$h$)$^3$].

\subsection{CONCERTO}

Refs.~\cite{Serra16, Dumitru19} introduce the voxel noise temporal power spectrum
$\mathcal{S}_{I, \text{voxel}}
=
\frac{\text{NEI} / \sqrt{N_\text{voxel}}}{\Omega_\text{beam}}$ [Jy/sr$\sqrt{\text{s}}$],
where
$\text{NEI} / \sqrt{N_\text{voxel}} = 155$~mJy$\sqrt{\text{s}}$.
The voxel intensity noise is then obtained as
$\sigma_{I, \text{voxel}} = \mathcal{S}_{I, \text{voxel}} / \sqrt{t_\text{voxel}}$,
where
$t_\text{voxel} = t_\text{survey} N_\text{voxel} \Omega_\text{beam} / \Omega_\text{survey}$.
In this expression,
\beq
\left\{
\bal
&t_\text{survey} = 1,500\text{ hr}\\
&N_\text{voxel} = 1,500\\
&\Omega_\text{survey} = 2\text{ deg}^2\\
&\Omega_\text{beam} = 2\pi \left( \theta_\text{beam}/2.355 \right)^2\\
&\theta_\text{beam} = 1.22 \lambda_\text{obs}(z) / D\\
&D = 12\text{ m}\\
\eal
\right.
\eeq
Finally the voxel intensity noise is converted to noise power spectrum via
$N(\vk) = \sigma_{I, \text{voxel}}^2 \mathcal{V}_\text{voxel}$~[(Jy/sr)$^2$(Mpc/$h$)$^3$].

\subsection{HETDEX}

For HETDEX, we follow \cite{Hill08, Hill16, Cheng19}. 
We assume a $3''\times 3''$ pixel, with a spectral resolution $\mathcal{R}=800$ over $300$~deg$^2$.
We assume a flux noise from each voxel of
$5\ \sigma_{f, \text{voxel}} = 5.5\times 10^{-17}$~erg/s/cm$^2$ \cite{Hill08, Hill16}, which we convert to intensity noise via
$\sigma_{I, \text{voxel}} = \sigma_{f, \text{voxel}} \mathcal{R}/\nu(z) / \Omega_\text{pixel}$.
Finally, we infer the noise power spectrum as above:
$N(\vk) = \sigma_{I, \text{voxel}}^2 \mathcal{V}_\text{voxel}$~[(Jy/sr)$^2$(Mpc/$h$)$^3$].

\subsection{CDIM}

We follow the specifications from \cite{Cooray16, Cheng19} for CDIM.
Focusing on the deep survey, we assume a 100~deg$^2$ survey, with 1~arcsec$^2$ pixels and $\mathcal{R}=300$.
We assume a flux noise from each voxel of
$5\ \sigma_{f, \text{voxel}} = 10^{-18}$~erg/s/cm$^2$ \cite{Cooray16}.
We follow the same procedure as for HETDEX above to convert it to a noise power spectrum.

%%%%%%%%%%%%%%%%%%%%%%%%%%%%%%%%%%%%%%%%%%%%%%%%%%%%%%%%%%
\section{LIM vs galaxy detection: Comparison to Ref.~\cite{Cheng19}} 
\label{app:comparison_cheng19}

% Comparison with Cheng+19
Ref.~\cite{Cheng19} compares LIM and galaxy detection as tracers of the matter density.
They provide a series of intuitive toy models, with increasing complexity, to thoroughly explore the various regimes.
They also provide for the first time a rigorous formalism, based on the Fisher information, to derive an optimal estimator of the matter density field from a given intensity map pixel.
In this formalism, LIM and galaxy detection appear as two limiting cases of the optimal estimator.
In this section, we have built upon this study and extended it in a number of ways.

First, we compared LIM and galaxy detection not only as cosmological probes, but as astrophysical ones.
We showed in which cases LIM allows to probe fainter galaxies than are individually detectable.

Turning to the cosmological information, ref.~\cite{Cheng19} focuses on a given pixel. This ignores the correlations across pixels, and gives a scale-independent answer to the comparison of LIM and galaxy detection.
Here, we instead focus on individual Fourier scales. This automatically accounts for pixel-to-pixel correlations, and allows us to compare LIM and galaxy detection as a function of the wavevector $k$.
In our formalism, the pixel size disappears as long as the PSF is well sampled. For an undersampled PSF (as for SPHEREx), the pixel geometry appears through the pixel-convolved PSF.

Ref.~\cite{Cheng19} assumes a fixed matter overdensity $\delta$ in the pixel of interest. In other words, the cosmic variance of the matter density field is not included.
Here, we instead include the cosmic variance in the comparison, as relevant when reconstructing the linear matter power spectrum from the data.
This introduces a new regime, where cosmic variance dominates, 
and the SNR per Fourier mode saturates to unity.
In this regime, LIM and galaxy detection tend to be equivalent, regardless of their differing clustering bias and number density of sources, since the shot noise is negligible.
Without cosmic variance, either LIM or galaxy detection would be preferred over the other.
This would be relevant when relying on sample variance cancellation methods \cite{Seljak09}, in the very low detector noise and shot noise regime.

In ref.~\cite{Cheng19}, the clustering bias does not enter in the comparison between LIM and galaxy detection, and is assumed luminosity-independent for simplicity.
In our study, the bias enters for two reasons. 
First, it determines the comparison between cosmic variance and shot noise, through the usual dimensionless number $\bar{n}_\text{eff}b^2 P_\text{lin}$.
Second, we model its luminosity dependence, which matters when comparing LIM and galaxy detection, since the detected galaxies are bright and generally have a high bias.

Our underlying halo models are slightly different.
The model in ref.~\cite{Cheng19} does not include halos for simplicity, only galaxies.
We include halos, which simply adds the 1-halo term as an additional source of noise.
Otherwise, both models include galaxies as a Poisson sampling.
Finally, ref.~\cite{Cheng19} works at the map level, rather than the power spectrum level here.
We believe that the two approaches, field and power level, should give the same answer, given the same halo model.

Finally, ref.~\cite{Cheng19} tackled the question of the optimal estimator, beyond the simple LIM or galaxy detection.
We have not attempted to answer this question here.
Instead, we simply point out two other approaches than LIM and galaxy detection.
First, we explored masking the detected point sources in the LIM. 
This keeps only the faint, undetected sources in the intensity map, lowering its bias but potentially increasing the effective number density. 
For the experiments considered here, this did not significantly change the results in Fig.~\ref{fig:np_lim_vs_galaxies}.
Again, these are the point sources at the target redshift, not foreground point sources, which we did not model.
Finally, the detected galaxies do not have to be number-weighted. 
Since their individual luminosities are measured, they can be weighted by any function of luminosity.
An interesting weighting to explore would for example approximate a mass weighting, to recover the matter density field \cite{Hamaus10}.
We leave this to future work.

%%%%%%%%%%%%%%%%%%%%%%%%%%%%%%%%%%%%%%%%%%%%%%%%%%%%%%%%%%

\bibliographystyle{JHEP}
\bibliography{refs}

\providecommand{\href}[2]{#2}\begingroup\raggedright\begin{thebibliography}{10}

\bibitem{Cheng19}
Y.-T. {Cheng}, R.~{de Putter}, T.-C. {Chang} and O.~{Dor{\'e}},
  \emph{{Optimally Mapping Large-scale Structures with Luminous Sources}},
  \href{https://doi.org/10.3847/1538-4357/ab1b2b}{\emph{\apj} {\bfseries 877}
  (2019) 86} [\href{https://arxiv.org/abs/1809.06384}{{\ttfamily 1809.06384}}].

\bibitem{Breysse16}
P.~C. {Breysse}, E.~D. {Kovetz} and M.~{Kamionkowski}, \emph{{The high-redshift
  star formation history from carbon-monoxide intensity maps}},
  \href{https://doi.org/10.1093/mnrasl/slw005}{\emph{\mnras} {\bfseries 457}
  (2016) L127} [\href{https://arxiv.org/abs/1507.06304}{{\ttfamily
  1507.06304}}].

\bibitem{Yue15}
B.~{Yue}, A.~{Ferrara}, A.~{Pallottini}, S.~{Gallerani} and L.~{Vallini},
  \emph{{Intensity mapping of [C II] emission from early galaxies}},
  \href{https://doi.org/10.1093/mnras/stv933}{\emph{\mnras} {\bfseries 450}
  (2015) 3829} [\href{https://arxiv.org/abs/1504.06530}{{\ttfamily
  1504.06530}}].

\bibitem{Silva17}
M.~B. {Silva}, S.~{Zaroubi}, R.~{Kooistra} and A.~{Cooray}, \emph{{Tomographic
  Intensity Mapping versus Galaxy Surveys: Observing the Universe in H-alpha
  emission with new generation instruments}}, {\emph{arXiv e-prints} (2017)
  arXiv:1711.09902} [\href{https://arxiv.org/abs/1711.09902}{{\ttfamily
  1711.09902}}].

\bibitem{Lidz09}
A.~{Lidz}, O.~{Zahn}, S.~R. {Furlanetto}, M.~{McQuinn}, L.~{Hernquist} and
  M.~{Zaldarriaga}, \emph{{Probing Reionization with the 21 cm Galaxy
  Cross-Power Spectrum}},
  \href{https://doi.org/10.1088/0004-637X/690/1/252}{\emph{\apj} {\bfseries
  690} (2009) 252} [\href{https://arxiv.org/abs/0806.1055}{{\ttfamily
  0806.1055}}].

\bibitem{Gong12}
Y.~{Gong}, A.~{Cooray}, M.~{Silva}, M.~G. {Santos}, J.~{Bock}, C.~M. {Bradford}
  et~al., \emph{{Intensity Mapping of the [C II] Fine Structure Line during the
  Epoch of Reionization}},
  \href{https://doi.org/10.1088/0004-637X/745/1/49}{\emph{\apj} {\bfseries 745}
  (2012) 49} [\href{https://arxiv.org/abs/1107.3553}{{\ttfamily 1107.3553}}].

\bibitem{Lidz11}
A.~{Lidz}, S.~R. {Furlanetto}, S.~P. {Oh}, J.~{Aguirre}, T.-C. {Chang},
  O.~{Dor{\'e}} et~al., \emph{{Intensity Mapping with Carbon Monoxide Emission
  Lines and the Redshifted 21 cm Line}},
  \href{https://doi.org/10.1088/0004-637X/741/2/70}{\emph{\apj} {\bfseries 741}
  (2011) 70} [\href{https://arxiv.org/abs/1104.4800}{{\ttfamily 1104.4800}}].

\bibitem{Dinda18}
B.~R. {Dinda}, A.~A. {Sen} and T.~R. {Choudhury}, \emph{{Dark energy
  constraints from the 21\raisebox{-0.5ex}cm intensity mapping surveys with
  SKA1}}, {\emph{arXiv e-prints} (2018) arXiv:1804.11137}
  [\href{https://arxiv.org/abs/1804.11137}{{\ttfamily 1804.11137}}].

\bibitem{Carucci17}
I.~P. {Carucci}, P.-S. {Corasaniti} and M.~{Viel}, \emph{{Imprints of
  non-standard dark energy and dark matter models on the 21cm intensity map
  power spectrum}},
  \href{https://doi.org/10.1088/1475-7516/2017/12/018}{\emph{\jcap} {\bfseries
  2017} (2017) 018} [\href{https://arxiv.org/abs/1706.09462}{{\ttfamily
  1706.09462}}].

\bibitem{Moradinezhad19}
A.~{Moradinezhad Dizgah}, G.~K. {Keating} and A.~{Fialkov}, \emph{{Probing
  Cosmic Origins with CO and [C II] Emission Lines}},
  \href{https://doi.org/10.3847/2041-8213/aaf813}{\emph{\apjl} {\bfseries 870}
  (2019) L4} [\href{https://arxiv.org/abs/1801.10178}{{\ttfamily 1801.10178}}].

\bibitem{Fonseca18}
J.~{Fonseca}, R.~{Maartens} and M.~G. {Santos}, \emph{{Synergies between
  intensity maps of hydrogen lines}},
  \href{https://doi.org/10.1093/mnras/sty1702}{\emph{\mnras} {\bfseries 479}
  (2018) 3490} [\href{https://arxiv.org/abs/1803.07077}{{\ttfamily
  1803.07077}}].

\bibitem{Munoz15}
J.~B. {Mu{\~n}oz}, Y.~{Ali-Ha{\"\i}moud} and M.~{Kamionkowski},
  \emph{{Primordial non-gaussianity from the bispectrum of 21-cm fluctuations
  in the dark ages}},
  \href{https://doi.org/10.1103/PhysRevD.92.083508}{\emph{\prd} {\bfseries 92}
  (2015) 083508} [\href{https://arxiv.org/abs/1506.04152}{{\ttfamily
  1506.04152}}].

\bibitem{Kovetz17}
E.~D. {Kovetz}, M.~P. {Viero}, A.~{Lidz}, L.~{Newburgh}, M.~{Rahman},
  E.~{Switzer} et~al., \emph{{Line-Intensity Mapping: 2017 Status Report}},
  {\emph{ArXiv e-prints} (2017) }
  [\href{https://arxiv.org/abs/1709.09066}{{\ttfamily 1709.09066}}].

\bibitem{Kovetz19}
E.~{Kovetz}, P.~C. {Breysse}, A.~{Lidz}, J.~{Bock}, C.~M. {Bradford}, T.-C.
  {Chang} et~al., \emph{{Astrophysics and Cosmology with Line-Intensity
  Mapping}},  in \emph{\baas}, vol.~51, p.~101, May, 2019,
  \href{https://arxiv.org/abs/1903.04496}{{\ttfamily 1903.04496}}.

\bibitem{Keating16}
G.~K. {Keating}, D.~P. {Marrone}, G.~C. {Bower}, E.~{Leitch}, J.~E. {Carlstrom}
  and D.~R. {DeBoer}, \emph{{COPSS II: The Molecular Gas Content of Ten Million
  Cubic Megaparsecs at Redshift z$\sim$3}},
  \href{https://doi.org/10.3847/0004-637X/830/1/34}{\emph{\apj} {\bfseries 830}
  (2016) 34} [\href{https://arxiv.org/abs/1605.03971}{{\ttfamily 1605.03971}}].

\bibitem{Pullen18}
A.~R. {Pullen}, P.~{Serra}, T.-C. {Chang}, O.~{Dor{\'e}} and S.~{Ho},
  \emph{{Search for C II emission on cosmological scales at redshift Z
  {\ensuremath{\sim}} 2.6}},
  \href{https://doi.org/10.1093/mnras/sty1243}{\emph{\mnras} {\bfseries 478}
  (2018) 1911} [\href{https://arxiv.org/abs/1707.06172}{{\ttfamily
  1707.06172}}].

\bibitem{paper1}
E.~{Schaan} and M.~{White}, \emph{{Multi-tracer intensity mapping:
  Cross-correlations, Line noise \& Decorrelation}}, {\emph{arXiv e-prints}
  (2021) arXiv:2103.01964} [\href{https://arxiv.org/abs/2103.01964}{{\ttfamily
  2103.01964}}].

\bibitem{Mao08}
Y.~{Mao}, M.~{Tegmark}, M.~{McQuinn}, M.~{Zaldarriaga} and O.~{Zahn},
  \emph{{How accurately can 21cm tomography constrain cosmology?}},
  \href{https://doi.org/10.1103/PhysRevD.78.023529}{\emph{\prd} {\bfseries 78}
  (2008) 023529} [\href{https://arxiv.org/abs/0802.1710}{{\ttfamily
  0802.1710}}].

\bibitem{Bernal19}
J.~L. {Bernal}, P.~C. {Breysse}, H.~{Gil-Mar{\'\i}n} and E.~D. {Kovetz},
  \emph{{User's guide to extracting cosmological information from
  line-intensity maps}},
  \href{https://doi.org/10.1103/PhysRevD.100.123522}{\emph{\prd} {\bfseries
  100} (2019) 123522} [\href{https://arxiv.org/abs/1907.10067}{{\ttfamily
  1907.10067}}].

\bibitem{Castorina19}
E.~{Castorina} and M.~{White}, \emph{{Measuring the growth of structure with
  intensity mapping surveys}},
  \href{https://doi.org/10.1088/1475-7516/2019/06/025}{\emph{\jcap} {\bfseries
  2019} (2019) 025} [\href{https://arxiv.org/abs/1902.07147}{{\ttfamily
  1902.07147}}].

\bibitem{Beane18}
A.~{Beane} and A.~{Lidz}, \emph{{Extracting Bias Using the Cross-bispectrum: An
  EoR and 21 cm-[C II]-[C II] Case Study}},
  \href{https://doi.org/10.3847/1538-4357/aae388}{\emph{\apj} {\bfseries 867}
  (2018) 26} [\href{https://arxiv.org/abs/1806.02796}{{\ttfamily 1806.02796}}].

\bibitem{Dore14}
O.~{Dor{\'e}}, J.~{Bock}, M.~{Ashby}, P.~{Capak}, A.~{Cooray}, R.~{de Putter}
  et~al., \emph{{Cosmology with the SPHEREX All-Sky Spectral Survey}},
  {\emph{arXiv e-prints} (2014) arXiv:1412.4872}
  [\href{https://arxiv.org/abs/1412.4872}{{\ttfamily 1412.4872}}].

\bibitem{Dore16}
O.~{Dor{\'e}}, M.~W. {Werner}, M.~{Ashby}, P.~{Banerjee}, N.~{Battaglia},
  J.~{Bauer} et~al., \emph{{Science Impacts of the SPHEREx All-Sky Optical to
  Near-Infrared Spectral Survey: Report of a Community Workshop Examining
  Extragalactic, Galactic, Stellar and Planetary Science}}, {\emph{arXiv
  e-prints} (2016) arXiv:1606.07039}
  [\href{https://arxiv.org/abs/1606.07039}{{\ttfamily 1606.07039}}].

\bibitem{Dore18}
O.~{Dor{\'e}}, M.~W. {Werner}, M.~L.~N. {Ashby}, L.~E. {Bleem}, J.~{Bock},
  J.~{Burt} et~al., \emph{{Science Impacts of the SPHEREx All-Sky Optical to
  Near-Infrared Spectral Survey II: Report of a Community Workshop on the
  Scientific Synergies Between the SPHEREx Survey and Other Astronomy
  Observatories}}, {\emph{arXiv e-prints} (2018) arXiv:1805.05489}
  [\href{https://arxiv.org/abs/1805.05489}{{\ttfamily 1805.05489}}].

\bibitem{Lagache18}
G.~{Lagache}, \emph{{Exploring the dusty star-formation in the early Universe
  using intensity mapping}},  in \emph{Peering towards Cosmic Dawn},
  V.~{Jeli{\'c}} and T.~{van der Hulst}, eds., vol.~333 of \emph{IAU
  Symposium}, pp.~228--233, May, 2018,
  \href{https://doi.org/10.1017/S1743921318000558}{DOI}
  [\href{https://arxiv.org/abs/1801.08054}{{\ttfamily 1801.08054}}].

\bibitem{Parshley18}
S.~C. {Parshley}, J.~{Kronshage}, J.~{Blair}, T.~{Herter}, M.~{Nolta}, G.~J.
  {Stacey} et~al., \emph{{CCAT-prime: a novel telescope for sub-millimeter
  astronomy}},  in \emph{\procspie}, vol.~10700 of \emph{Society of
  Photo-Optical Instrumentation Engineers (SPIE) Conference Series},
  p.~107005X, July, 2018, \href{https://doi.org/10.1117/12.2314046}{DOI}
  [\href{https://arxiv.org/abs/1807.06675}{{\ttfamily 1807.06675}}].

\bibitem{Cleary16}
K.~{Cleary}, M.-A. {Bigot-Sazy}, D.~{Chung}, S.~E. {Church}, C.~{Dickinson},
  H.~{Eriksen} et~al., \emph{{The CO Mapping Array Pathfinder (COMAP)}},  in
  \emph{American Astronomical Society Meeting Abstracts \#227}, vol.~227 of
  \emph{American Astronomical Society Meeting Abstracts}, p.~426.06, Jan.,
  2016.

\bibitem{Bower16}
G.~C. {Bower}, G.~K. {Keating}, D.~P. {Marrone} and S.~T. {YT Lee Array Team},
  \emph{{Cosmic Structure and Galaxy Evolution through Intensity Mapping of
  Molecular Gas}},  in \emph{American Astronomical Society Meeting Abstracts
  \#227}, vol.~227 of \emph{American Astronomical Society Meeting Abstracts},
  p.~426.04, Jan., 2016.

\bibitem{Crites17}
A.~{Crites}, J.~{Bock}, M.~{Bradford}, B.~{Bumble}, T.-C. {Chang}, Y.-T.
  {Cheng} et~al., \emph{{Measuring the Epoch of Reionization using [CII]
  Intensity Mapping with TIME-Pilot}},  in \emph{American Astronomical Society
  Meeting Abstracts \#229}, vol.~229 of \emph{American Astronomical Society
  Meeting Abstracts}, p.~125.01, Jan., 2017.

\bibitem{Li16}
T.~Y. {Li}, R.~H. {Wechsler}, K.~{Devaraj} and S.~E. {Church},
  \emph{{Connecting CO Intensity Mapping to Molecular Gas and Star Formation in
  the Epoch of Galaxy Assembly}},
  \href{https://doi.org/10.3847/0004-637X/817/2/169}{\emph{\apj} {\bfseries
  817} (2016) 169} [\href{https://arxiv.org/abs/1503.08833}{{\ttfamily
  1503.08833}}].

\bibitem{Dumitru19}
S.~{Dumitru}, G.~{Kulkarni}, G.~{Lagache} and M.~G. {Haehnelt},
  \emph{{Predictions and sensitivity forecasts for reionization-era [C II] line
  intensity mapping}},
  \href{https://doi.org/10.1093/mnras/stz617}{\emph{\mnras} {\bfseries 485}
  (2019) 3486} [\href{https://arxiv.org/abs/1802.04804}{{\ttfamily
  1802.04804}}].

\bibitem{Serra16}
P.~{Serra}, O.~{Dor{\'e}} and G.~{Lagache}, \emph{{Dissecting the High-z
  Interstellar Medium through Intensity Mapping Cross-correlations}},
  \href{https://doi.org/10.3847/1538-4357/833/2/153}{\emph{\apj} {\bfseries
  833} (2016) 153} [\href{https://arxiv.org/abs/1608.00585}{{\ttfamily
  1608.00585}}].

\bibitem{Concerto20}
{The CONCERTO collaboration}, P.~{Ade}, M.~{Aravena}, E.~{Barria}, A.~{Beelen},
  A.~{Benoit} et~al., \emph{{A wide field-of-view low-resolution spectrometer
  at APEX: instrument design and science forecast}}, {\emph{arXiv e-prints}
  (2020) arXiv:2007.14246} [\href{https://arxiv.org/abs/2007.14246}{{\ttfamily
  2007.14246}}].

\bibitem{PlanckParams16}
{Planck Collaboration}, P.~A.~R. {Ade}, N.~{Aghanim}, M.~{Arnaud},
  M.~{Ashdown}, J.~{Aumont} et~al., \emph{{Planck 2015 results. XIII.
  Cosmological parameters}},
  \href{https://doi.org/10.1051/0004-6361/201525830}{\emph{\aap} {\bfseries
  594} (2016) A13} [\href{https://arxiv.org/abs/1502.01589}{{\ttfamily
  1502.01589}}].

\bibitem{Bryan98}
G.~L. {Bryan} and M.~L. {Norman}, \emph{{Statistical Properties of X-Ray
  Clusters: Analytic and Numerical Comparisons}},
  \href{https://doi.org/10.1086/305262}{\emph{\apj} {\bfseries 495} (1998) 80}
  [\href{https://arxiv.org/abs/astro-ph/9710107}{{\ttfamily
  astro-ph/9710107}}].

\bibitem{Yang03}
X.~{Yang}, H.~J. {Mo} and F.~C. {van den Bosch}, \emph{{Constraining galaxy
  formation and cosmology with the conditional luminosity function of
  galaxies}},
  \href{https://doi.org/10.1046/j.1365-8711.2003.06254.x}{\emph{\mnras}
  {\bfseries 339} (2003) 1057}
  [\href{https://arxiv.org/abs/astro-ph/0207019}{{\ttfamily
  astro-ph/0207019}}].

\bibitem{Cooray16}
A.~{Cooray}, J.~{Bock}, D.~{Burgarella}, R.~{Chary}, T.-C. {Chang},
  O.~{Dor{\'e}} et~al., \emph{{Cosmic Dawn Intensity Mapper}}, {\emph{arXiv
  e-prints} (2016) arXiv:1602.05178}
  [\href{https://arxiv.org/abs/1602.05178}{{\ttfamily 1602.05178}}].

\bibitem{Hill08}
G.~J. {Hill}, K.~{Gebhardt}, E.~{Komatsu}, N.~{Drory}, P.~J. {MacQueen},
  J.~{Adams} et~al., \emph{{The Hobby-Eberly Telescope Dark Energy Experiment
  (HETDEX): Description and Early Pilot Survey Results}},  in \emph{Panoramic
  Views of Galaxy Formation and Evolution}, T.~{Kodama}, T.~{Yamada} and
  K.~{Aoki}, eds., vol.~399 of \emph{Astronomical Society of the Pacific
  Conference Series}, p.~115, Oct., 2008,
  \href{https://arxiv.org/abs/0806.0183}{{\ttfamily 0806.0183}}.

\bibitem{Hill16}
G.~J. {Hill} and {HETDEX Consortium}, \emph{{HETDEX and VIRUS: Panoramic
  Integral Field Spectroscopy with 35k Fibers}},  in \emph{Multi-Object
  Spectroscopy in the Next Decade: Big Questions, Large Surveys, and Wide
  Fields}, I.~{Skillen}, M.~{Balcells} and S.~{Trager}, eds., vol.~507 of
  \emph{Astronomical Society of the Pacific Conference Series}, p.~393, Oct.,
  2016.

\bibitem{Sobral13}
D.~{Sobral}, I.~{Smail}, P.~N. {Best}, J.~E. {Geach}, Y.~{Matsuda}, J.~P.
  {Stott} et~al., \emph{{A large H{\ensuremath{\alpha}} survey at z = 2.23,
  1.47, 0.84 and 0.40: the 11 Gyr evolution of star-forming galaxies from
  HiZELS}}, \href{https://doi.org/10.1093/mnras/sts096}{\emph{\mnras}
  {\bfseries 428} (2013) 1128}
  [\href{https://arxiv.org/abs/1202.3436}{{\ttfamily 1202.3436}}].

\bibitem{Colbert13}
J.~W. {Colbert}, H.~{Teplitz}, H.~{Atek}, A.~{Bunker}, M.~{Rafelski}, N.~{Ross}
  et~al., \emph{{Predicting Future Space Near-IR Grism Surveys Using the WFC3
  Infrared Spectroscopic Parallels Survey}},
  \href{https://doi.org/10.1088/0004-637X/779/1/34}{\emph{\apj} {\bfseries 779}
  (2013) 34} [\href{https://arxiv.org/abs/1305.1399}{{\ttfamily 1305.1399}}].

\bibitem{Mehta15}
V.~{Mehta}, C.~{Scarlata}, J.~W. {Colbert}, Y.~S. {Dai}, A.~{Dressler},
  A.~{Henry} et~al., \emph{{Predicting the Redshift 2 H{\ensuremath{\alpha}}
  Luminosity Function Using [OIII] Emission Line Galaxies}},
  \href{https://doi.org/10.1088/0004-637X/811/2/141}{\emph{\apj} {\bfseries
  811} (2015) 141} [\href{https://arxiv.org/abs/1505.07843}{{\ttfamily
  1505.07843}}].

\bibitem{Cochrane17}
R.~K. {Cochrane}, P.~N. {Best}, D.~{Sobral}, I.~{Smail}, D.~A. {Wake}, J.~P.
  {Stott} et~al., \emph{{The H {\ensuremath{\alpha}} luminosity-dependent
  clustering of star-forming galaxies from z {\ensuremath{\sim}} 0.8 to
  {\ensuremath{\sim}}2.2 with HiZELS}},
  \href{https://doi.org/10.1093/mnras/stx957}{\emph{\mnras} {\bfseries 469}
  (2017) 2913} [\href{https://arxiv.org/abs/1704.05472}{{\ttfamily
  1704.05472}}].

\bibitem{Popping16}
G.~{Popping}, E.~{van Kampen}, R.~{Decarli}, M.~{Spaans}, R.~S. {Somerville}
  and S.~C. {Trager}, \emph{{Sub-mm emission line deep fields: CO and [C II]
  luminosity functions out to z = 6}},
  \href{https://doi.org/10.1093/mnras/stw1323}{\emph{\mnras} {\bfseries 461}
  (2016) 93} [\href{https://arxiv.org/abs/1602.02761}{{\ttfamily 1602.02761}}].

\bibitem{Cassata11}
P.~{Cassata}, O.~{Le F{\`e}vre}, B.~{Garilli}, D.~{Maccagni}, V.~{Le Brun},
  M.~{Scodeggio} et~al., \emph{{The VIMOS VLT Deep Survey: star formation rate
  density of Ly{\ensuremath{\alpha}} emitters from a sample of 217 galaxies
  with spectroscopic redshifts 2 {\ensuremath{\leq}} z {\ensuremath{\leq}}
  6.6}}, \href{https://doi.org/10.1051/0004-6361/201014410}{\emph{\aap}
  {\bfseries 525} (2011) A143}
  [\href{https://arxiv.org/abs/1003.3480}{{\ttfamily 1003.3480}}].

\bibitem{Kaiser87}
N.~{Kaiser}, \emph{{Clustering in real space and in redshift space}},
  \href{https://doi.org/10.1093/mnras/227.1.1}{\emph{\mnras} {\bfseries 227}
  (1987) 1}.

\bibitem{H98}
A.~J.~S. {Hamilton}, \emph{{Linear Redshift Distortions: a Review}},  in
  \emph{The Evolving Universe}, D.~{Hamilton}, ed., vol.~231 of
  \emph{Astrophysics and Space Science Library}, p.~185, 1998,
  \href{https://doi.org/10.1007/978-94-011-4960-0\_17}{DOI}
  [\href{https://arxiv.org/abs/astro-ph/9708102}{{\ttfamily
  astro-ph/9708102}}].

\bibitem{LSST12}
{LSST Dark Energy Science Collaboration}, \emph{{Large Synoptic Survey
  Telescope: Dark Energy Science Collaboration}}, {\emph{arXiv e-prints} (2012)
  arXiv:1211.0310} [\href{https://arxiv.org/abs/1211.0310}{{\ttfamily
  1211.0310}}].

\bibitem{Ade19}
P.~{Ade}, J.~{Aguirre}, Z.~{Ahmed}, S.~{Aiola}, A.~{Ali}, D.~{Alonso} et~al.,
  \emph{{The Simons Observatory: science goals and forecasts}},
  \href{https://doi.org/10.1088/1475-7516/2019/02/056}{\emph{\jcap} {\bfseries
  2019} (2019) 056} [\href{https://arxiv.org/abs/1808.07445}{{\ttfamily
  1808.07445}}].

\bibitem{Abazajian16}
K.~N. {Abazajian}, P.~{Adshead}, Z.~{Ahmed}, S.~W. {Allen}, D.~{Alonso}, K.~S.
  {Arnold} et~al., \emph{{CMB-S4 Science Book, First Edition}}, {\emph{arXiv
  e-prints} (2016) arXiv:1610.02743}
  [\href{https://arxiv.org/abs/1610.02743}{{\ttfamily 1610.02743}}].

\bibitem{Modi19}
C.~{Modi}, M.~{White}, A.~{Slosar} and E.~{Castorina}, \emph{{Reconstructing
  large-scale structure with neutral hydrogen surveys}},
  \href{https://doi.org/10.1088/1475-7516/2019/11/023}{\emph{\jcap} {\bfseries
  2019} (2019) 023} [\href{https://arxiv.org/abs/1907.02330}{{\ttfamily
  1907.02330}}].

\bibitem{Doux16}
C.~{Doux}, E.~{Schaan}, E.~{Aubourg}, K.~{Ganga}, K.-G. {Lee}, D.~N. {Spergel}
  et~al., \emph{{First detection of cosmic microwave background lensing and
  Lyman-{\ensuremath{\alpha}} forest bispectrum}},
  \href{https://doi.org/10.1103/PhysRevD.94.103506}{\emph{\prd} {\bfseries 94}
  (2016) 103506} [\href{https://arxiv.org/abs/1607.03625}{{\ttfamily
  1607.03625}}].

\bibitem{Zhu16}
H.-M. {Zhu}, U.-L. {Pen}, Y.~{Yu}, X.~{Er} and X.~{Chen}, \emph{{Cosmic tidal
  reconstruction}},
  \href{https://doi.org/10.1103/PhysRevD.93.103504}{\emph{\prd} {\bfseries 93}
  (2016) 103504} [\href{https://arxiv.org/abs/1511.04680}{{\ttfamily
  1511.04680}}].

\bibitem{Li19}
D.~{Li}, H.-M. {Zhu} and U.-L. {Pen}, \emph{{Cross-correlation of the kinematic
  Sunyaev-Zel'dovich effect and 21 cm intensity mapping with tidal
  reconstruction}},
  \href{https://doi.org/10.1103/PhysRevD.100.023517}{\emph{\prd} {\bfseries
  100} (2019) 023517} [\href{https://arxiv.org/abs/1811.05012}{{\ttfamily
  1811.05012}}].

\bibitem{Schlegel19}
D.~{Schlegel}, J.~A. {Kollmeier} and S.~{Ferraro}, \emph{{The MegaMapper: a
  z\&gt;2 spectroscopic instrument for the study of Inflation and Dark
  Energy}},  in \emph{\baas}, vol.~51, p.~229, Sep, 2019,
  \href{https://arxiv.org/abs/1907.11171}{{\ttfamily 1907.11171}}.

\bibitem{Heneka18}
C.~{Heneka} and L.~{Amendola}, \emph{{General modified gravity with 21cm
  intensity mapping: simulations and forecast}},
  \href{https://doi.org/10.1088/1475-7516/2018/10/004}{\emph{\jcap} {\bfseries
  2018} (2018) 004} [\href{https://arxiv.org/abs/1805.03629}{{\ttfamily
  1805.03629}}].

\bibitem{Wilson19}
M.~J. {Wilson} and M.~{White}, \emph{{Cosmology with dropout selection:
  straw-man surveys \& CMB lensing}},
  \href{https://doi.org/10.1088/1475-7516/2019/10/015}{\emph{\jcap} {\bfseries
  2019} (2019) 015} [\href{https://arxiv.org/abs/1904.13378}{{\ttfamily
  1904.13378}}].

\bibitem{Tegmark97}
M.~{Tegmark}, \emph{{Measuring Cosmological Parameters with Galaxy Surveys}},
  \href{https://doi.org/10.1103/PhysRevLett.79.3806}{\emph{\prl} {\bfseries 79}
  (1997) 3806} [\href{https://arxiv.org/abs/astro-ph/9706198}{{\ttfamily
  astro-ph/9706198}}].

\bibitem{White09}
M.~{White}, Y.-S. {Song} and W.~J. {Percival}, \emph{{Forecasting cosmological
  constraints from redshift surveys}},
  \href{https://doi.org/10.1111/j.1365-2966.2008.14379.x}{\emph{\mnras}
  {\bfseries 397} (2009) 1348}
  [\href{https://arxiv.org/abs/0810.1518}{{\ttfamily 0810.1518}}].

\bibitem{Cheng16}
Y.-T. {Cheng}, T.-C. {Chang}, J.~{Bock}, C.~M. {Bradford} and A.~{Cooray},
  \emph{{Spectral Line De-confusion in an Intensity Mapping Survey}},
  \href{https://doi.org/10.3847/0004-637X/832/2/165}{\emph{\apj} {\bfseries
  832} (2016) 165} [\href{https://arxiv.org/abs/1604.07833}{{\ttfamily
  1604.07833}}].

\bibitem{Fonseca17}
J.~{Fonseca}, M.~B. {Silva}, M.~G. {Santos} and A.~{Cooray}, \emph{{Cosmology
  with intensity mapping techniques using atomic and molecular lines}},
  \href{https://doi.org/10.1093/mnras/stw2470}{\emph{\mnras} {\bfseries 464}
  (2017) 1948} [\href{https://arxiv.org/abs/1607.05288}{{\ttfamily
  1607.05288}}].

\bibitem{Lidz16}
A.~{Lidz} and J.~{Taylor}, \emph{{On Removing Interloper Contamination from
  Intensity Mapping Power Spectrum Measurements}},
  \href{https://doi.org/10.3847/0004-637X/825/2/143}{\emph{\apj} {\bfseries
  825} (2016) 143} [\href{https://arxiv.org/abs/1604.05737}{{\ttfamily
  1604.05737}}].

\bibitem{Gong17}
Y.~{Gong}, A.~{Cooray}, M.~B. {Silva}, M.~{Zemcov}, C.~{Feng}, M.~G. {Santos}
  et~al., \emph{{Intensity Mapping of H{\ensuremath{\alpha}},
  H{\ensuremath{\beta}}, [OII], and [OIII] Lines at z \&lt; 5}},
  \href{https://doi.org/10.3847/1538-4357/835/2/273}{\emph{\apj} {\bfseries
  835} (2017) 273} [\href{https://arxiv.org/abs/1610.09060}{{\ttfamily
  1610.09060}}].

\bibitem{Gong20}
Y.~{Gong}, X.~{Chen} and A.~{Cooray}, \emph{{Cosmological constraints from line
  intensity mapping with interlopers}}, {\emph{arXiv e-prints} (2020)
  arXiv:2001.10792} [\href{https://arxiv.org/abs/2001.10792}{{\ttfamily
  2001.10792}}].

\bibitem{Cheng20}
Y.-T. {Cheng}, T.-C. {Chang} and J.~J. {Bock}, \emph{{Phase-space Spectral Line
  Deconfusion in Intensity Mapping}},
  \href{https://doi.org/10.3847/1538-4357/abb023}{\emph{\apj} {\bfseries 901}
  (2020) 142} [\href{https://arxiv.org/abs/2005.05341}{{\ttfamily
  2005.05341}}].

\bibitem{Pullen13}
A.~R. {Pullen}, O.~{Dor{\'e}} and J.~{Bock}, \emph{{Intensity Mapping across
  Cosmic Times with the Ly{\ensuremath{\alpha}} Line}},
  \href{https://doi.org/10.1088/0004-637X/786/2/111}{\emph{\apj} {\bfseries
  786} (2014) 111} [\href{https://arxiv.org/abs/1309.2295}{{\ttfamily
  1309.2295}}].

\bibitem{Breysse15}
P.~C. {Breysse}, E.~D. {Kovetz} and M.~{Kamionkowski}, \emph{{Masking line
  foregrounds in intensity-mapping surveys}},
  \href{https://doi.org/10.1093/mnras/stv1476}{\emph{\mnras} {\bfseries 452}
  (2015) 3408} [\href{https://arxiv.org/abs/1503.05202}{{\ttfamily
  1503.05202}}].

\bibitem{Silva15}
M.~{Silva}, M.~G. {Santos}, A.~{Cooray} and Y.~{Gong}, \emph{{Prospects for
  Detecting C II Emission during the Epoch of Reionization}},
  \href{https://doi.org/10.1088/0004-637X/806/2/209}{\emph{\apj} {\bfseries
  806} (2015) 209} [\href{https://arxiv.org/abs/1410.4808}{{\ttfamily
  1410.4808}}].

\bibitem{Kogut15}
A.~{Kogut}, E.~{Dwek} and S.~H. {Moseley}, \emph{{Spectral Confusion for
  Cosmological Surveys of Redshifted C II Emission}},
  \href{https://doi.org/10.1088/0004-637X/806/2/234}{\emph{\apj} {\bfseries
  806} (2015) 234} [\href{https://arxiv.org/abs/1505.00266}{{\ttfamily
  1505.00266}}].

\bibitem{Liu16}
A.~{Liu}, Y.~{Zhang} and A.~R. {Parsons}, \emph{{Spherical Harmonic Analyses of
  Intensity Mapping Power Spectra}},
  \href{https://doi.org/10.3847/1538-4357/833/2/242}{\emph{\apj} {\bfseries
  833} (2016) 242} [\href{https://arxiv.org/abs/1609.04401}{{\ttfamily
  1609.04401}}].

\bibitem{Visbal10}
E.~{Visbal} and A.~{Loeb}, \emph{{Measuring the 3D clustering of undetected
  galaxies through cross correlation of their cumulative flux fluctuations from
  multiple spectral lines}},
  \href{https://doi.org/10.1088/1475-7516/2010/11/016}{\emph{\jcap} {\bfseries
  2010} (2010) 016} [\href{https://arxiv.org/abs/1008.3178}{{\ttfamily
  1008.3178}}].

\bibitem{Visbal11}
E.~{Visbal}, H.~{Trac} and A.~{Loeb}, \emph{{Demonstrating the feasibility of
  line intensity mapping using mock data of galaxy clustering from
  simulations}},
  \href{https://doi.org/10.1088/1475-7516/2011/08/010}{\emph{\jcap} {\bfseries
  2011} (2011) 010} [\href{https://arxiv.org/abs/1104.4809}{{\ttfamily
  1104.4809}}].

\bibitem{Gong14}
Y.~{Gong}, M.~{Silva}, A.~{Cooray} and M.~G. {Santos}, \emph{{Foreground
  Contamination in Ly{\ensuremath{\alpha}} Intensity Mapping during the Epoch
  of Reionization}},
  \href{https://doi.org/10.1088/0004-637X/785/1/72}{\emph{\apj} {\bfseries 785}
  (2014) 72} [\href{https://arxiv.org/abs/1312.2035}{{\ttfamily 1312.2035}}].

\bibitem{Switzer17}
E.~{Switzer}, ``{Measuring the Cosmological Evolution of Gas and Galaxies with
  the EXperiment for Cryogenic Large-aperture Intensity Mapping (EXCLAIM)}.''
  NASA APRA Proposal, Jan., 2017.

\bibitem{Switzer19}
E.~R. {Switzer}, C.~J. {Anderson}, A.~R. {Pullen} and S.~{Yang},
  \emph{{Intensity Mapping in the Presence of Foregrounds and Correlated
  Continuum Emission}},
  \href{https://doi.org/10.3847/1538-4357/aaf9ab}{\emph{\apj} {\bfseries 872}
  (2019) 82} [\href{https://arxiv.org/abs/1812.06223}{{\ttfamily 1812.06223}}].

\bibitem{White17}
M.~{White} and N.~{Padmanabhan}, \emph{{Matched filtering with interferometric
  21 cm experiments}},
  \href{https://doi.org/10.1093/mnras/stx1682}{\emph{\mnras} {\bfseries 471}
  (2017) 1167} [\href{https://arxiv.org/abs/1705.09669}{{\ttfamily
  1705.09669}}].

\bibitem{Seljak09}
U.~{Seljak}, \emph{{Extracting Primordial Non-Gaussianity without Cosmic
  Variance}}, \href{https://doi.org/10.1103/PhysRevLett.102.021302}{\emph{\prl}
  {\bfseries 102} (2009) 021302}
  [\href{https://arxiv.org/abs/0807.1770}{{\ttfamily 0807.1770}}].

\bibitem{McDonald09}
P.~{McDonald} and U.~{Seljak}, \emph{{How to evade the sample variance limit on
  measurements of redshift-space distortions}},
  \href{https://doi.org/10.1088/1475-7516/2009/10/007}{\emph{\jcap} {\bfseries
  2009} (2009) 007} [\href{https://arxiv.org/abs/0810.0323}{{\ttfamily
  0810.0323}}].

\bibitem{Moradinezhad20}
A.~{Moradinezhad Dizgah}, M.~{Biagetti}, E.~{Sefusatti}, V.~{Desjacques} and
  J.~{Nore{\~n}a}, \emph{{Primordial Non-Gaussianity from Biased Tracers:
  Likelihood Analysis of Real-Space Power Spectrum and Bispectrum}},
  {\emph{arXiv e-prints} (2020) arXiv:2010.14523}
  [\href{https://arxiv.org/abs/2010.14523}{{\ttfamily 2010.14523}}].

\bibitem{Hamaus10}
N.~{Hamaus}, U.~{Seljak}, V.~{Desjacques}, R.~E. {Smith} and T.~{Baldauf},
  \emph{{Minimizing the stochasticity of halos in large-scale structure
  surveys}}, \href{https://doi.org/10.1103/PhysRevD.82.043515}{\emph{\prd}
  {\bfseries 82} (2010) 043515}
  [\href{https://arxiv.org/abs/1004.5377}{{\ttfamily 1004.5377}}].

\end{thebibliography}\endgroup

\end{document}